\documentclass[10pt,journal,compsoc]{IEEEtran}
\IEEEoverridecommandlockouts
\usepackage{cite}
\usepackage[most]{tcolorbox}
\usepackage{amsmath}
\usepackage{amsthm}
\usepackage{xurl}
\usepackage{tikz}
\usepackage{multirow}
\usepackage[normalem]{ulem}
\usepackage{colortbl}
\usepackage{algorithm}
\usepackage{algpseudocode}
\usepackage{graphicx}
\usepackage{xcolor}
\usepackage{color}

\usepackage{varwidth}
\usepackage{rotating}

\usepackage[nolist]{acronym}

\usepackage{nameref}
\usepackage{zref-xr,zref-user}
\zxrsetup{toltxlabel=true, tozreflabel=false}
\usepackage{xcite}

\usepackage{tikz-3dplot}

\usepackage{subfigure}

\usepackage[resetlabels,labeled]{multibib}
\newcites{A}{References}

\usetikzlibrary{decorations.pathreplacing,decorations.markings,decorations.text}

\tikzset{arrow data/.style 2 args={%
      decoration={%
         markings,
         mark=at position #1 with \arrow{#2}},
         postaction=decorate}
      }%

\usetikzlibrary{shapes,backgrounds,calc}

\tikzstyle{dummy} = [rectangle, text width=0.1em, draw=white, white,
                      minimum width=0.1em, minimum height=3em, opacity=0.0]

\makeatletter
\newcommand*{\Strut}[1][0.1em]{\vrule\@width\z@\@height#1\@depth\z@\relax}
\makeatother

\definecolor{Linen}{rgb}{0.9803,0.9411,0.9019}
\definecolor{White}{rgb}{1,1,1}
\definecolor{Green}{rgb}{0.5,1,0.5}
\definecolor{Red}{rgb}{1,0.4,0.4}
\definecolor{Coral}{rgb}{1,0.4980,0.3137}
\definecolor{Grayblue}{rgb}{0.9411,0.9411,0.9803}
\definecolor{DarkLinen}{rgb}{0.729,0.7176,0.635}

\usepackage{mdframed}
\begin{document}

\begin{acronym}

\acro{5G-NR}{5G New Radio}
\acro{3GPP}{3rd Generation Partnership Project}
\acro{AC}{address coding}
\acro{ACF}{autocorrelation function}
\acro{ACR}{autocorrelation receiver}
\acro{ADC}{Analog-to-Digital Converter}
\acrodef{aic}[AIC]{Analog-to-Information Converter}     
\acro{AIC}[AIC]{Akaike information criterion}
\acro{aric}[ARIC]{asymmetric restricted isometry constant}
\acro{arip}[ARIP]{asymmetric restricted isometry property}

\acro{ARQ}{automatic repeat request}
\acro{AUB}{asymptotic union bound}
\acrodef{awgn}[AWGN]{Additive White Gaussian Noise}     
\acro{AWGN}{additive white Gaussian noise}

\acro{APSK}[PSK]{asymmetric PSK} 

\acro{waric}[AWRICs]{asymmetric weak restricted isometry constants}
\acro{warip}[AWRIP]{asymmetric weak restricted isometry property}
\acro{BCH}{Bose, Chaudhuri, and Hocquenghem}        
\acro{BCHC}[BCHSC]{BCH based source coding}
\acro{BEP}{bit error probability}
\acro{BFC}{block fading channel}
\acro{BG}[BG]{Bernoulli-Gaussian}
\acro{BGG}{Bernoulli-Generalized Gaussian}
\acro{BPAM}{binary pulse amplitude modulation}
\acro{BPDN}{Basis Pursuit Denoising}
\acro{BPPM}{binary pulse position modulation}
\acro{BPSK}{binary phase shift keying}
\acro{BPZF}{bandpass zonal filter}
\acro{BSC}{binary symmetric channels}              
\acro{BU}[BU]{Bernoulli-uniform}
\acro{BER}{bit error rate}
\acro{BS}{base station}

\acro{CP}{Cyclic Prefix}
\acrodef{cdf}[CDF]{cumulative distribution function}   
\acro{CDF}{cumulative distribution function}
\acrodef{c.d.f.}[CDF]{cumulative distribution function}
\acro{CCDF}{complementary cumulative distribution function}
\acrodef{ccdf}[CCDF]{complementary CDF}               
\acrodef{c.c.d.f.}[CCDF]{complementary cumulative distribution function}
\acro{CD}{cooperative diversity}

\acro{CDMA}{Code Division Multiple Access}
\acro{ch.f.}{characteristic function}
\acro{CIR}{channel impulse response}
\acro{cosamp}[CoSaMP]{compressive sampling matching pursuit}
\acro{CR}{cognitive radio}
\acro{cs}[CS]{compressed sensing}                   
\acrodef{cscapital}[CS]{Compressed sensing} 
\acrodef{CS}[CS]{compressed sensing}
\acro{CSI}{channel state information}
\acro{CCSDS}{consultative committee for space data systems}
\acro{CC}{convolutional coding}
\acro{Covid19}[COVID-19]{Coronavirus disease}
\acro{CAPEX}{CAPital EXpenditures}

\acro{DAA}{detect and avoid}
\acro{DAB}{digital audio broadcasting}
\acro{DCT}{discrete cosine transform}
\acro{dft}[DFT]{discrete Fourier transform}
\acro{DR}{distortion-rate}
\acro{DS}{direct sequence}
\acro{DS-SS}{direct-sequence spread-spectrum}
\acro{DTR}{differential transmitted-reference}
\acro{DVB-H}{digital video broadcasting\,--\,handheld}
\acro{DVB-T}{digital video broadcasting\,--\,terrestrial}
\acro{DL}{downlink}
\acro{DSSS}{Direct Sequence Spread Spectrum}
\acro{DFT-s-OFDM}{Discrete Fourier Transform-spread-Orthogonal Frequency Division Multiplexing}
\acro{DAS}{distributed antenna system}
\acro{DNA}{Deoxyribonucleic Acid}

\acro{EC}{European Commission}
\acro{EED}[EED]{exact eigenvalues distribution}
\acro{EIRP}{Equivalent Isotropically Radiated Power}
\acro{ELP}{equivalent low-pass}
\acro{eMBB}{Enhanced Mobile Broadband}
\acro{EMF}{Electro-Magnetic Field}
\acro{EU}{European union}
\acro{ELP}{Exposure Limit-based Power}

\acro{FC}[FC]{fusion center}
\acro{FCC}{Federal Communications Commission}
\acro{FEC}{forward error correction}
\acro{FFT}{fast Fourier transform}
\acro{FH}{frequency-hopping}
\acro{FH-SS}{frequency-hopping spread-spectrum}
\acrodef{FS}{Frame synchronization}
\acro{FSsmall}[FS]{frame synchronization}  
\acro{FDMA}{Frequency Division Multiple Access}  
\acro{FSPL}{Free Space Path Loss}  
\acro{FWA}{Fixed Wireless Access}
\acro{FDD}{Frequency Division Duplexing}

\acro{GA}{Gaussian approximation}
\acro{GF}{Galois field }
\acro{GG}{Generalized-Gaussian}
\acro{GIC}[GIC]{generalized information criterion}
\acro{GLRT}{generalized likelihood ratio test}
\acro{GPS}{Global Positioning System}
\acro{GMSK}{Gaussian minimum shift keying}
\acro{GSMA}{Global System for Mobile communications Association}

\acro{HAP}{high altitude platform}
\acro{HW}{HardWare}

\acro{IDR}{information distortion-rate}
\acro{IFFT}{inverse fast Fourier transform}
\acro{iht}[IHT]{iterative hard thresholding}
\acro{i.i.d.}{independent, identically distributed}
\acro{IoT}{Internet of Things}                      
\acro{IR}{impulse radio}
\acro{lric}[LRIC]{lower restricted isometry constant}
\acro{lrict}[LRICt]{lower restricted isometry constant threshold}
\acro{ISI}{intersymbol interference}
\acro{ITU}{International Telecommunication Union}
\acro{ICNIRP}{International Commission on Non-Ionizing Radiation Protection}
\acro{IEEE}{Institute of Electrical and Electronics Engineers}
\acro{ICES}{IEEE international committee on electromagnetic safety}
\acro{IEC}{International Electrotechnical Commission}
\acro{IARC}{International Agency on Research on Cancer}
\acro{IS-95}{Interim Standard 95}
\acro{ITU}{International Telecommunication Union}
\acro{IP}{Internet Protocol}

\acro{LEO}{low earth orbit}
\acro{LF}{likelihood function}
\acro{LLF}{log-likelihood function}
\acro{LLR}{log-likelihood ratio}
\acro{LLRT}{log-likelihood ratio test}
\acro{LOS}{Line-of-Sight}
\acro{LRT}{likelihood ratio test}
\acro{wlric}[LWRIC]{lower weak restricted isometry constant}
\acro{wlrict}[LWRICt]{LWRIC threshold}
\acro{LPWAN}{low power wide area network}
\acro{LoRaWAN}{Low power long Range Wide Area Network}
\acro{LB}{Lower Bound}

\acro{MB}{multiband}
\acro{MC}{multicarrier}
\acro{MDS}{mixed distributed source}
\acro{MF}{matched filter}
\acro{m.g.f.}{moment generating function}
\acro{MI}{mutual information}
\acro{MIMO}{multiple-input multiple-output}
\acro{MISO}{multiple-input single-output}
\acrodef{maxs}[MJSO]{maximum joint support cardinality}                       
\acro{ML}[ML]{maximum likelihood}
\acro{MMSE}{minimum mean-square error}
\acro{MMV}{multiple measurement vectors}
\acrodef{MOS}{model order selection}
\acro{M-PSK}[${M}$-PSK]{$M$-ary phase shift keying}                       
\acro{M-APSK}[${M}$-PSK]{$M$-ary asymmetric PSK} 
\acro{MSP}{Minimum Sensitivity-based Power}

\acro{M-QAM}[$M$-QAM]{$M$-ary quadrature amplitude modulation}
\acro{MRC}{maximal ratio combiner}                  
\acro{maxs}[MSO]{maximum sparsity order}                                      
\acro{M2M}{machine to machine}                                                
\acro{MUI}{multi-user interference}
\acro{mMTC}{massive Machine Type Communications}      
\acro{mm-Wave}{millimeter-wave}
\acro{MP}{mobile phone}
\acro{MPE}{maximum permissible exposure}
\acro{MAC}{media access control}
\acro{NB}{narrowband}
\acro{NBI}{narrowband interference}
\acro{NLA}{nonlinear sparse approximation}
\acro{NLOS}{Non-Line-of-Sight}
\acro{NTIA}{National Telecommunications and Information Administration}
\acro{NTP}{National Toxicology Program}
\acro{NHS}{National Health Service}
\acro{NSA}{Non Stand Alone}

\acro{OC}{optimum combining}                             
\acro{OC}{optimum combining}
\acro{ODE}{operational distortion-energy}
\acro{ODR}{operational distortion-rate}
\acro{OFDM}{orthogonal frequency-division multiplexing}
\acro{omp}[OMP]{orthogonal matching pursuit}
\acro{OSMP}[OSMP]{orthogonal subspace matching pursuit}
\acro{OQAM}{offset quadrature amplitude modulation}
\acro{OQPSK}{offset QPSK}
\acro{OFDMA}{Orthogonal Frequency-division Multiple Access}
\acro{OPEX}{OPerating EXpenditures}
\acro{OQPSK/PM}{OQPSK with phase modulation}

\acro{PAM}{pulse amplitude modulation}
\acro{PAR}{peak-to-average ratio}
\acrodef{pdf}[PDF]{probability density function}                      
\acro{PDF}{probability density function}
\acrodef{p.d.f.}[PDF]{probability distribution function}
\acro{PDP}{power dispersion profile}
\acro{PMF}{probability mass function}                             
\acrodef{p.m.f.}[PMF]{probability mass function}
\acro{PN}{pseudo-noise}
\acro{PPM}{pulse position modulation}
\acro{PRake}{Partial Rake}
\acro{PSD}{power spectral density}
\acro{PSEP}{pairwise synchronization error probability}
\acro{PSK}{phase shift keying}
\acro{PD}{Power Density}
\acro{8-PSK}[$8$-PSK]{$8$-phase shift keying}

\acro{FSK}{frequency shift keying}

\acro{QAM}{Quadrature Amplitude Modulation}
\acro{QPSK}{quadrature phase shift keying}
\acro{OQPSK/PM}{OQPSK with phase modulator }

\acro{RD}[RD]{raw data}
\acro{RDL}{"random data limit"}
\acro{ric}[RIC]{restricted isometry constant}
\acro{rict}[RICt]{restricted isometry constant threshold}
\acro{rip}[RIP]{restricted isometry property}
\acro{ROC}{receiver operating characteristic}
\acro{rq}[RQ]{Raleigh quotient}
\acro{RS}[RS]{Reed-Solomon}
\acro{RSC}[RSSC]{RS based source coding}
\acro{RFP}{Radio Frequency ``Pollution''}
\acro{r.v.}{random variable}                               
\acro{R.V.}{random vector}
\acro{RMS}{root mean square}
\acro{RFR}{radiofrequency radiation}
\acro{RIS}{Reconfigurable Intelligent Surface}
\acro{RNA}{RiboNucleic Acid}
\acro{RBS}{Radio Base Station}
\acro{RSRP}{Reference Signal Received Power}
\acro{RSRQ}{Reference Signal Received Quality}

\acro{SA}{Stand Alone}
\acro{SAN}{Spectrum ANalyzer}
\acro{SCBSES}[SCBSES]{Source Compression Based Syndrome Encoding Scheme}
\acro{SCM}{sample covariance matrix}
\acro{SEP}{symbol error probability}
\acro{SG}[SG]{sparse-land Gaussian model}
\acro{SIMO}{single-input multiple-output}
\acro{SINR}{Signal-to-Interference plus Noise Ratio}
\acro{SIR}{signal-to-interference ratio}
\acro{SISO}{single-input single-output}
\acro{SMV}{single measurement vector}
\acro{SNR}[\textrm{SNR}]{signal-to-noise ratio} 
\acro{sp}[SP]{subspace pursuit}
\acro{SS}{spread spectrum}
\acro{SW}{sync word}
\acro{SAR}{Specific Absorption Rate}
\acro{SSB}{synchronization signal block}
\acro{SCPI}{Standard Commands for Programmable Instruments}
\acro{SS-RSRP}{Synchronization Signal Reference Signal Received Power}

\acro{TH}{time-hopping}
\acro{ToA}{time-of-arrival}
\acro{TR}{transmitted-reference}
\acro{TW}{Tracy-Widom}
\acro{TWDT}{TW Distribution Tail}
\acro{TCM}{trellis coded modulation}
\acro{TDD}{Time Division Duplexing}
\acro{TDMA}{Time Division Multiple Access}
\acro{TCP}{Transmission Control Protocol}

\acro{UAV}{unmanned aerial vehicle}
\acro{uric}[URIC]{upper restricted isometry constant}
\acro{urict}[URICt]{upper restricted isometry constant threshold}
\acro{UWB}{ultrawide band}
\acro{UWBcap}[UWB]{Ultrawide band}   
\acro{URLLC}{Ultra Reliable Low Latency Communications}
         
\acro{wuric}[UWRIC]{upper weak restricted isometry constant}
\acro{wurict}[UWRICt]{UWRIC threshold}                
\acro{UE}{User Equipment}
\acro{UL}{uplink}
\acro{UB}{Upper Bound}

\acro{WiM}[WiM]{weigh-in-motion}
\acro{WLAN}{wireless local area network}
\acro{wm}[WM]{Wishart matrix}                               
\acroplural{wm}[WM]{Wishart matrices}
\acro{WMAN}{wireless metropolitan area network}
\acro{WPAN}{wireless personal area network}
\acro{wric}[WRIC]{weak restricted isometry constant}
\acro{wrict}[WRICt]{weak restricted isometry constant thresholds}
\acro{wrip}[WRIP]{weak restricted isometry property}
\acro{WSN}{wireless sensor network}                        
\acro{WSS}{wide-sense stationary}
\acro{WHO}{World Health Organization}
\acro{Wi-Fi}{wireless fidelity}

\acro{sss}[SpaSoSEnc]{sparse source syndrome encoding}

\acro{VLC}{visible light communication}
\acro{VPN}{virtual private network} 
\acro{RF}{Radio-Frequency}
\acro{FSO}{free space optics}
\acro{IoST}{Internet of space things}

\acro{GSM}{Global System for Mobile Communications}
\acro{2G}{second-generation cellular network}
\acro{3G}{third-generation cellular network}
\acro{4G}{fourth-generation cellular network}
\acro{5G}{5th-generation cellular network}	
\acro{gNB}{next-generation Node-B}
\acro{NR}{New Radio}
\acro{UMTS}{Universal Mobile Telecommunications Service}
\acro{LTE}{Long Term Evolution}

\acro{QoS}{Quality of Service}
\end{acronym}

\title{Dominance of Smartphone Exposure \\in 5G Mobile Networks}
\author{\small  Luca Chiaraviglio,$^{(1,2)}$ Chiara Lodovisi,$^{(1,2)}$ Stefania Bartoletti,$^{(1,2)}$ Ahmed Elzanaty,$^{(3)}$ Mohamed Slim-Alouini,$^{(4)}$\\
(1) EE Department,  University of Rome Tor Vergata, Rome, Italy, email \{luca.chiaraviglio@uniroma2.it, Lodovisi@ing.uniroma2.it, stefania.bartoletti@uniroma2.it\}\\
(2) Consorzio Nazionale Interuniversitario per le Telecomunicazioni, Italy\\
(3) Institute for Communication Systems (ICS), University of Surrey, Guildford, United Kingdom,  email a.elzanaty@surrey.ac.uk\\
(4) CEMSE Division, King Abdullah University of Science and Technology, Thuwal, Saudi Arabia,  email slim.alouini@kaust.edu.sa
}
\maketitle
\IEEEpeerreviewmaketitle

\begin{abstract}
The deployment of 5G networks is sometimes questioned due to the impact of ElectroMagnetic Field (EMF) generated by Radio Base Station (RBS) on users. The goal of this work is to analyze such issue from a novel perspective,  by comparing RBS EMF against exposure generated by 5G smartphones in commercial deployments. The measurement of exposure from 5G is hampered by several implementation aspects,  such as dual connectivity between 4G and 5G, spectrum fragmentation, and carrier aggregation. To face such issues, we deploy a novel framework, called \textsc{5G-EA}, tailored to the assessment of smartphone and RBS exposure through an innovative measurement algorithm, able to remotely control a programmable spectrum analyzer. Results, obtained in both outdoor and indoor locations, reveal that smartphone exposure (upon generation of uplink traffic) dominates over the RBS one.  Moreover,   Line-of-Sight locations experience a reduction of around one order of magnitude on the overall exposure compared to Non-Line-of-Sight ones. In addition,  5G exposure always represents a small share (up to {38}\%) compared to {the total one radiated by the smartphone}.  
\end{abstract}

\section{Introduction}
\label{sec:intro}

According to recent reports \cite{smartphonediffusion}, more than 80\% of the world population own a smartphone. The diffusion of such equipment is so pervasive in the daily activities that it is almost impossible to imagine a future without a smartphone in our hands. One of the key drivers for the ever-increasing smartphone adoption is the ubiquitous Internet service, generally offered by mobile networks. To this purpose, 5G aims at delivering a true broadband connectivity service, especially in urban areas and densely populated zones. The sales of smartphone equipped with 5G interfaces are constantly rising, with more than 700 millions of units sold during 2022 \cite{5Gsmartphonediffusion}, in parallel with the deployment of 5G networks across the world \cite{gsmalaunch}. Therefore, 5G networks will (likely) become the main provider for smartphone connectivity in the near future.

In this scenario, the \ac{EMF} exposure from 5G networks is a hot topic in several communities (e.g., government, local committees, environmental protection agencies and academia), especially when considering the (possible, yet still not proven) implications of 5G exposure on the human health \cite{chiaraviglio2021health}. To this aim, \ac{EMF} working groups of \ac{WHO} \cite{who}, \ac{ICNIRP} \cite{international2020guidelines}, \ac{IEEE} committees \cite{bushberg2020ieee} and \ac{IEEE} standards \cite{ieee2019ieee} periodically evaluate the scientific literature, including the assessment of biological effects from \ac{EMF} exposure generated by telecommunication equipment. At present time, there is a consensus among such authoritative organizations that a clear causal correlation between exposure from mobile networks adhering to international exposure guidelines and emergence of long-term health diseases has not been observed so far. Consequently, 5G exposure does not pose any evident risk on the population health. Very frequently, however, the dispute about 5G exposure is dominated by the bias of non-scientific communities \cite{elzanaty20215g}, who associate the exposure of 5G \acp{RBS} with severe health diseases - a connection that is not (presently) proven by science.  As a result, the installation of new 5G \acp{RBS} over the territory is (sometimes) fiercely opposed by local communities and advocacy groups, who act against the (supposed) increase of exposure generated by the newly installed \acp{RBS} in their neighborhood.
 
Despite the exposure from 5G \ac{RBS} is a matter of debate - at the extent that the presence of a 5G antenna over a real estate has an impact on the property value - little or no concerns are associated with 5G smartphones, which are another (and important) source of exposure \cite{chiaraviglio2021health}. Part of the population promptly reacts against the presence of 5G towers in proximity to their living and working spaces, while almost nobody cares about the exposure that is radiated by the own smartphone when uploading/downloading hundreds of Megabytes of data through a mobile network connection. Therefore, the total exposure levels, resulting from the combination of 5G smartphones \textit{and} 5G \acp{RBS}, are almost overlooked.

The goal of this work is {twofold}. On one side, we assess in a scientific way the exposure generated by smartphones in a commercial 5G deployment. On the other one, we compare the observed smartphone exposure levels against the ones radiated by the serving \ac{RBS},  showing that the increase of signal coverage from 5G \ac{RBS} (and consequently the exposure) is highly beneficial in reducing the \ac{EMF} from the smartphone. The measurement of smartphone vs. \ac{RBS} exposure has been preliminary investigated in the context of 4G (see e.g., the very interesting paper of Schilling \textit{et al.} in \cite{schilling2022impact}), but, to the best of our knowledge, none of the previous works have conducted an in-depth measurement analysis tailored to a 5G commercial deployment. We point out, however, that our purpose is not to spread worries or alarms - as both smartphone and \ac{RBS} exposure naturally adhere to \ac{EMF} regulations and are therefore legally \textit{safe} - but rather to scientifically position the exposure from 5G \acp{RBS} in a wide picture that include the contribution of 5G smartphones, the effect of propagation conditions and the amount of traffic that is generated by \ac{UE}.

More concretely, we target the following questions: What is the amount of exposure generated by a 5G smartphone and a 5G \ac{RBS} in a commercial deployment? What is the impact of \ac{UL}/\ac{DL} traffic generated by the smartphone on the exposure levels? How do propagation conditions (like \ac{RBS} proximity/remoteness, presence/absence of buildings on the radio link towards the \ac{RBS}) influence 5G exposure levels? How does the dual connectivity  between 4G and 5G affect the exposure? The answer to these intriguing questions is the technical goal of this paper. More specifically, our original contributions include: \textit{i}) a ground-truth overview of 5G implementation features that are relevant for smartphone and \ac{RBS} exposure assessments,  with a focus on the Italian country; \textit{ii}) the definition of the measurement requirements to achieve our goal,  based on the technological features outlined in \textit{i}); \textit{iii}) the design of an innovative measurement framework, called \textsc{5G Exposure Assessment (5G-EA)}, which strongly leverages networking features (e.g., traffic generation \& monitoring, and remote programmability of spectrum analyzers) to satisfy the requirements in \textit{ii}); \textit{iv}) the application of \textsc{5G-EA} in a real 5G deployment to collect an extensive campaign of exposure measurements.

Our results demonstrate that the smartphone exposure dominates over the \ac{RBS} one upon generation of \ac{UL} traffic, especially when the \ac{UE} is in \ac{NLOS} with respect to the \ac{RBS}. On the contrary, both smartphone exposure and total \ac{EMF} are reduced up to one order of magnitude when the smartphone \ac{UL} traffic traverses a radio link in \ac{LOS} with respect to the serving \ac{RBS}. Interestingly, the exploitation of dual connectivity feature between 4G and 5G reveals that only a small {smartphone} exposure share (at most equal to {38}\%) is due to 5G, while the largest exposure levels are derived from the carrier aggregation over 4G bands. Moreover, both total and smartphone exposure-per-bit metrics are inversely proportional to the maximum amount of \ac{UL} traffic generated by the smartphone in the measurement location, thus suggesting that innovative exposure estimators, based on the reporting of maximum \ac{UL} traffic from the smartphone, can be designed.

Last but not least, we demonstrate that the complexity of the measurement procedures (which need to track spatial/temporal variations of 4G carrier aggregation and dual connectivity between 4G and 5G) can be efficiently tackled by a framework encompassing a softwarized measurement algorithm, like the one developed in this work. The design of softwarized-based \ac{EMF} measurement procedures, running on general purpose machines and able to remotely control spectrum analyzers, indicate the potentials of a new market, in which the \ac{EMF} measurement algorithms are designed, shared and adopted by a community of experts, while the manufacturers ``open'' the interfaces of the measurement equipment to support the remote programmability from non-proprietary software.

The rest of the paper  is organized as follows. Related works are analyzed in Sec.~\ref{sec:rel_work}. Sec.~\ref{sec:5G_primer} includes a primer about the implementation aspects of 5G networks that are relevant to \ac{EMF} monitoring, with a focus on the Italian country - useful for the layman in the field. Sec.~\ref{sec:emf_requirements} defines the measurement requirements, taking into account our goals \textit{and} the 5G implementation aspects of Sec.~\ref{sec:5G_primer}. The design of the \textsc{5G-EA} measurement framework is described in Sec.~\ref{sec:5G_SE}.  Results, retrieved from a real 5G deployment, are detailed in Sec.~\ref{sec:results}. Finally, Sec.~\ref{sec:conclusions} concludes our work and reports possible future directions.

\section{Related Work}
\label{sec:rel_work}

\begin{figure}
\centering
 	\subfigure[Environmental Exposure]
	{
		\includegraphics[width=4.2cm]{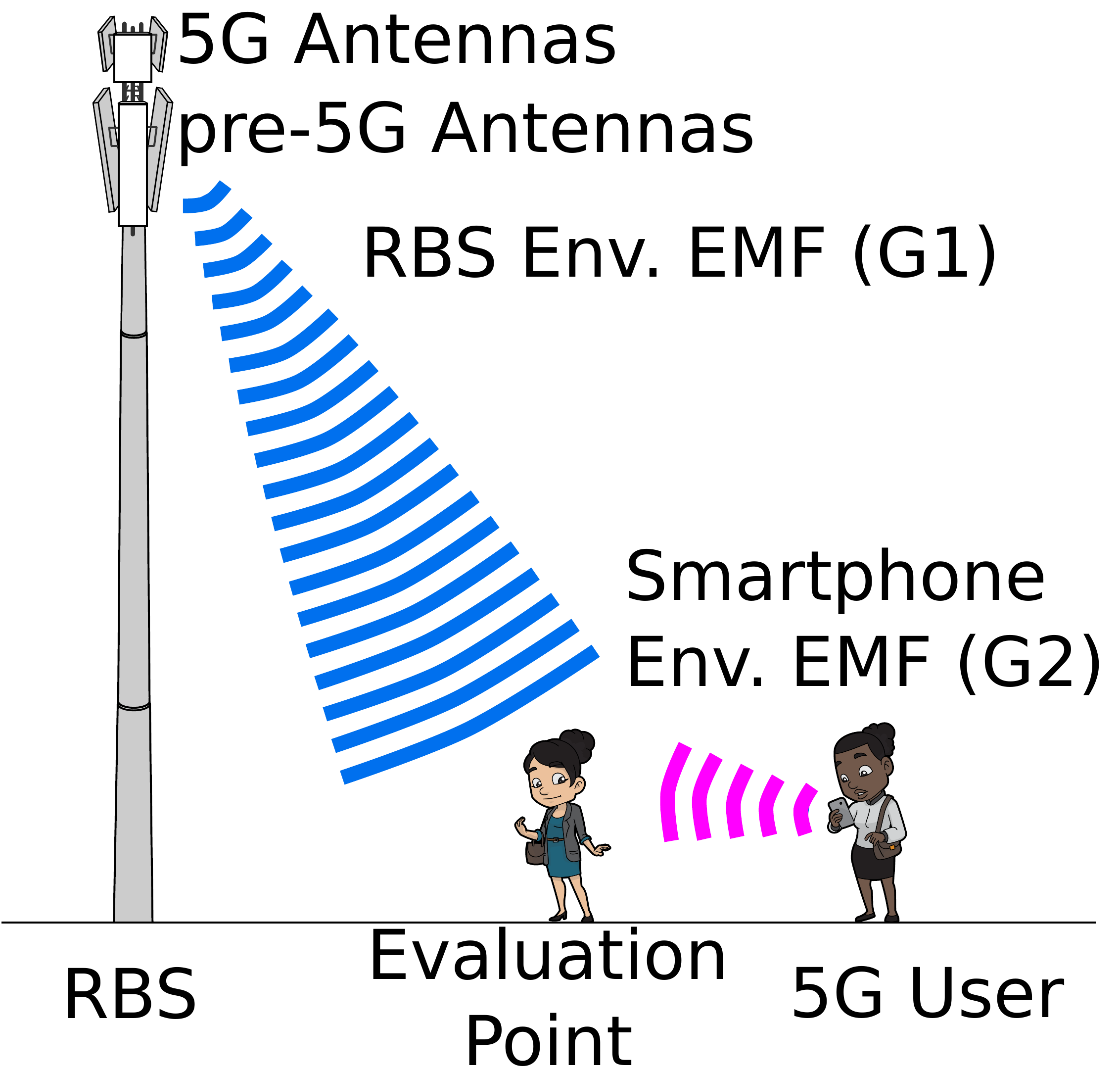}
		\label{fig:env_exposure}

	}
 	\subfigure[Active Traffic Exposure]
	{
		\includegraphics[width=4.2cm]{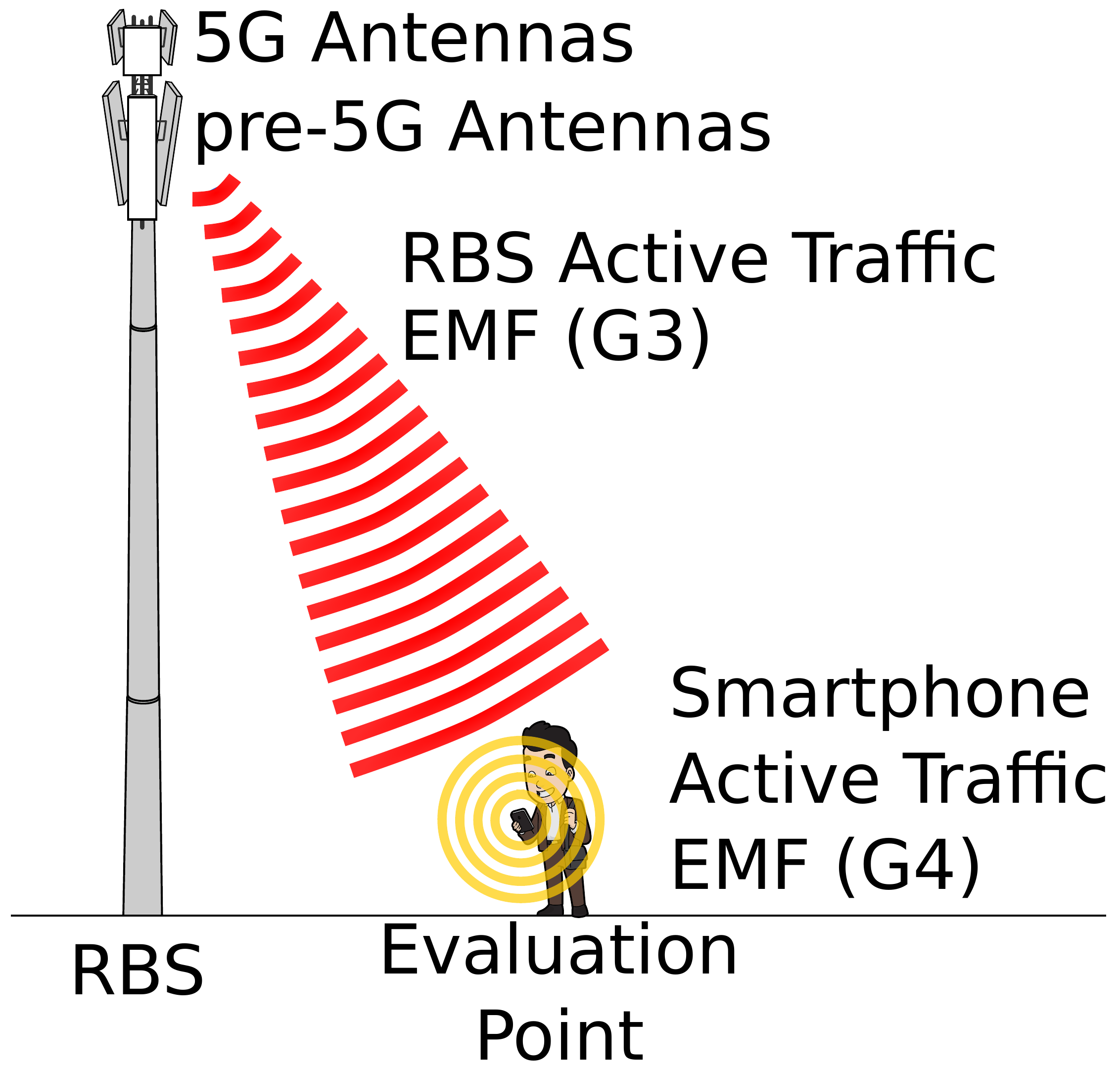}
		\label{fig:active_exposure}
	}
	\caption{Taxonomy of exposure measurements (figure best viewed in colors).}
	\label{fig:taxonomy_exposure}
\end{figure}

Fig.~\ref{fig:taxonomy_exposure} sketches the main taxonomy of \ac{RBS} and smartphone exposure measurements over an evaluation point. More in depth, we identify the following groups: $G1$) environmental exposure from 5G \ac{RBS}, $G2$) environmental exposure from nearby 5G smartphones, $G3$) exposure generated by the 5G \ac{RBS} when a 5G smartphone is used to inject active traffic in the evaluation point,  $G4$) exposure generated by the 5G smartphone in the same condition of $G3$. Intuitively, groups $G1$-$G2$ identify measurements taken without injecting any traffic in the measurement positions, and resulting in general lower exposure levels than groups $G3$-$G4$. 

In the following, we initially focus on the works tailored to the \ac{RBS} side,  i.e.,  covering groups $G1$ and/or $G3$. Then, we focus on the works investigating the active traffic exposure from 5G smartphone (group $G4$). Finally, we consider the works that integrate joint measurement of environmental/active traffic exposure from \ac{RBS} \textit{and} active traffic exposure from smartphone (groups $G1$ + $G3$ + $G4$) - although we did not find any previous work tailored to 5G. As a side comment, we intentionally leave apart group $G2$, as the exposure contributions from nearby terminals rapidly decrease to negligible levels when they are not in close proximity to the evaluation point.

\subsection{Exposure Measurements from 5G RBS}
\label{subsec:bs}

We initially focus on the works targeting: \textit{i}) measurement of environmental exposure from 5G \ac{RBS} (group $G1$) \cite{chiaraviglio2021massive,betta20225g,9887762,elbasheir2022measurement,hausl2022mobile} and \textit{ii}) measurement of active traffic exposure from 5G \ac{RBS} (group $G3$) \cite{adda2020theoretical,aerts2021situ,migliore2021new,bornkessel2021determination,schilling2022analysis,chiaraviglio2022emf,liu2021field,heliot2022empirical,aerts2019situ,chountala2021radio,wali2022rf}.

Focusing on $G1$ \cite{chiaraviglio2021massive,betta20225g,9887762,elbasheir2022measurement,hausl2022mobile}, Chiaraviglio \textit{et al.} \cite{chiaraviglio2021massive} perform a massive evaluation of a 5G \ac{RBS} covering a town. Betta \textit{et al.} \cite{betta20225g,9887762} and Elbasheir \textit{et al.} \cite{elbasheir2022measurement} collect \ac{RBS} exposure information through the measurement of the pilot signals. Hausl \textit{et al.} \cite{hausl2022mobile} analyze the received power over the control channels in a 5G network by employing a code-selective measurement methodology. 

Focusing on $G3$ \cite{adda2020theoretical,aerts2021situ,migliore2021new,bornkessel2021determination,schilling2022analysis,chiaraviglio2022emf,liu2021field,heliot2022empirical,aerts2019situ,chountala2021radio,wali2022rf}, Adda \textit{et al.} \cite{adda2020theoretical}, Aerts \textit{et al.} \cite{aerts2021situ}, Migliore \textit{et al.} \cite{migliore2021new}, Bornkessel \textit{et al.} \cite{bornkessel2021determination}, Schilling \textit{et al.} \cite{schilling2022analysis}, Chiaraviglio \textit{et al.} \cite{chiaraviglio2022emf}, Liu \textit{et al.} \cite{liu2021field}, Heliot \textit{et al.} \cite{heliot2022empirical} share the idea of measuring the exposure from 5G \acp{RBS} by forcing traffic with a terminal in the \ac{DL} direction from the \ac{RBS}. The works of Aerts \textit{et al.} \cite{aerts2019situ}, Chountala \textit{et al.} \cite{chountala2021radio} and Wali \textit{et al.} \cite{wali2022rf} complement the previous ones by adding the evaluation of 5G \ac{RBS} exposure when \ac{UL} traffic is forced with a terminal. In general, such works demonstrate that the exposure from 5G \ac{RBS} depends on the amount of traffic that is injected towards the measurement location. Moreover, the active traffic contribution from the 5G \ac{RBS} is generally higher than the environmental one. 

Compared to \cite{chiaraviglio2021massive,betta20225g,9887762,elbasheir2022measurement,hausl2022mobile,adda2020theoretical,aerts2021situ,migliore2021new,bornkessel2021determination,schilling2022analysis,chiaraviglio2022emf,liu2021field,heliot2022empirical,aerts2019situ,chountala2021radio,wali2022rf}, we tackle the 5G \ac{EMF} measurement from a novel perspective, by including the contribution of the smartphone in the \ac{EMF} assessments (groups $G1$ + $G3$ + $G4$). Although we exploit some findings/intuitions of the  literature (like the idea of forcing \ac{DL}/\ac{UL} traffic towards the measurement location), our work presents an innovative  measurement framework, called \textsc{5G-EA}, tailored to the assessment of both smartphone and \ac{RBS} \ac{EMF}.

\subsection{Exposure Measurement from 5G UE}
\label{subsec:ue}

We focus hereafter on the literature addressing active traffic exposure measurements from 5G \ac{UE} (group $G4$) \cite{xu2017power,nedelcu2021uplink,joshi2020actual,lee2021emf,deaconescu2022dynamics,miclaus2022peculiarities}. Xu \textit{et al.} \cite{xu2017power} perform measurements of 5G \ac{UE} power density in a semi-anechoic chamber. Nedelcu \textit{et al.} \cite{nedelcu2021uplink} analyze the \ac{UL} contribution from 5G \ac{UE} in terms of radiated power. Joshi \textit{et al.} \cite{joshi2020actual} and Lee \textit{et al.} \cite{lee2021emf} analyze 5G \ac{UE} output power levels that are collected from measurements in commercial networks. Deaconescu \textit{et al.} \cite{deaconescu2022dynamics} and Miclaus \textit{et al.} \cite{miclaus2022peculiarities} collect \ac{EMF} measurement from a 5G \ac{UE} in an indoor controlled environment, with and without generating \ac{UL}/\ac{DL} traffic. Overall, such works indicate that the exposure from 5G smartphones is non-negligible, and that a huge variation in the exposure levels can be observed. In contrast to \cite{xu2017power,nedelcu2021uplink,joshi2020actual,lee2021emf,deaconescu2022dynamics,miclaus2022peculiarities}, in this work we go two steps further by: \textit{i}) integrating the exposure from 5G \ac{RBS} and \textit{ii}) performing exposure assessments of groups $G1$ + $G3$ + $G4$ both in indoor controlled environments and \textit{into the wild}, i.e., several outdoor locations covered by a commercial 5G \ac{RBS}.

\subsection{Joint Measurement of Smartphone \textit{and} RBS Exposure}
\label{sec:joint_measurements}

The last category relevant to our study is focused on the joint assessment of smartphone and \ac{RBS} exposure. In this case, we did not find any work covering groups $G1$ + $G3$ + $G4$ in the 5G domain. Focusing instead on pre-5G technologies, the most relevant work to ours is the one of Schilling \textit{et al.} \cite{schilling2022impact}, in which the authors propose a method based on \ac{EMF} measurements to evaluate the combined exposure from both smartphone and \ac{RBS} in 4G deployments. Interestingly, a strong reduction in the \ac{UL} transmission power is observed when the link conditions are improved. Moreover, the total exposure in a macro cell scenario is dominated by the smartphone contribution. Eventually, the authors advocate the need for a balance between \ac{RBS} and smartphone exposure. 

In line with \cite{schilling2022impact}, our work is also focused on the joint assessment of smartphone \textit{and} \ac{RBS} exposure. However, differently from \cite{schilling2022impact}, we focus on a novel domain: the 5G exposure assessment of groups $G1$ + $G3$ + $G4$, which requires a different exposure framework than the one used by \cite{schilling2022impact} for the 4G evaluations. In addition, 5G smartphones currently employ a dual 4G/5G connectivity to support the data transfers. Therefore, our innovative framework evaluates the exposure over both 4G and 5G bands. This last aspects further complicates the exposure assessment compared to \cite{schilling2022impact}, since multiple 4G/5G carriers are dynamically used for the data transfer.

\section{5G Implementation Aspects Relevant to EMF Monitoring}
\label{sec:5G_primer}

\begin{figure}[t]
\centering
\includegraphics[width=8.8cm]{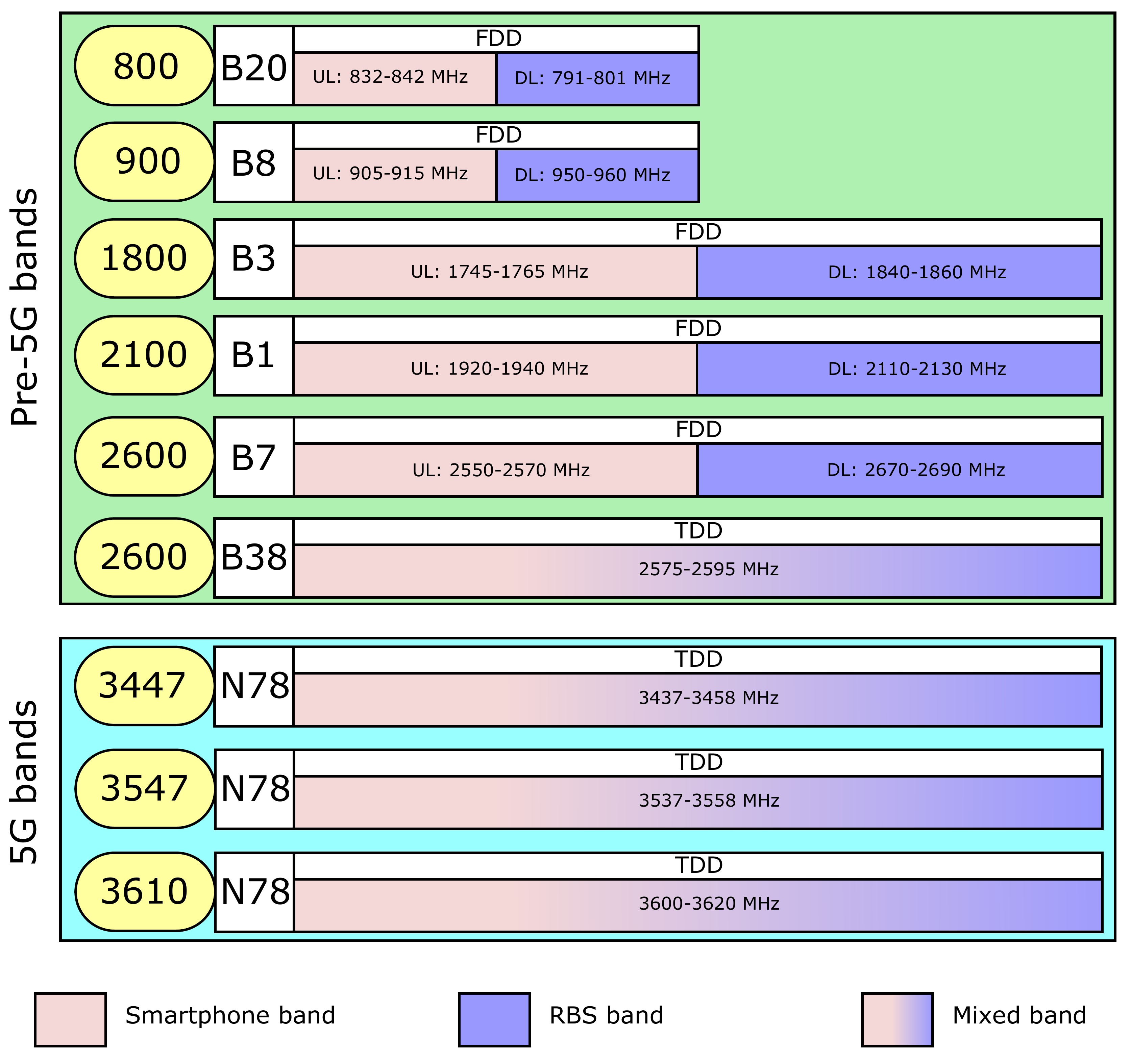}
\caption{Pre-5G and 5G bands in use by the W3 mobile operator.}
\label{fig:frequencies_overview}
\end{figure}

In this section, we provide a brief overview of the key 5G implementation aspects that have an influence the design of our \ac{EMF} assessments, with a focus on the Italian country. 

\textbf{Spectrum Fragmentation.} 5G encompasses a wide set of spectrum portions, including frequencies lower than 1 GHz, frequencies between 1 and 6 GHz - a.k.a. the mid-band - and close-to-mm-Wave frequencies at around 26-27 GHz. The most widespread option to provide 5G mobile service in Italy is the mid-band, thanks to the fact that the adopted frequencies can guarantee the mixture of coverage and capacity that is required during the current 5G early-adoption phase.  

The mid-band spectrum, spanning over 3.4-3.8 GHz is rather a crowded space. Historically, the 3.4-3.6 GHz portion of the spectrum was allocated to \ac{FWA} operators \cite{agcommidband}, which provided access to household customers over legacy technologies (pre-5G). The 3.6-3.8 GHz portion was instead allocated with the purpose of providing 5G service for mobile operators, with licensed spectrum blocks including both 80 MHz and 20 MHz portions \cite{agcom5g}. Clearly, the operators that were licensed 20 MHz of 5G mid-band spectrum  (like W3) could not support the same level of service as the one provided by providers operating on wider bandwidth, e.g., 80 MHz. To overcome this issue, W3 has recently signed an agreement with the \ac{FWA} operator Fastweb to lease some portions of the 3.4-3.6 GHz spectrum for the 5G mobile service \cite{agcomw3}.  Despite the total allocation of licensed and leased bandwidth is non-negligible (typically equal to 60 MHz for W3), the spectrum blocks for delivering 5G in the mid-band are not contiguous. Up to this point, a natural question is: How do such spectrum allocations affect the considered measurements? The answer is that, for some operators (like W3), the 5G \ac{EMF} monitoring (even focusing solely on the mid-band) has to be done over multiple not-contiguous spectrum portions. Such feature generally complicates the \ac{EMF} measurement procedure, as it is necessarly (in principle) to iterate over the different 5G bands in use by the same operator to evaluate the total 5G exposure.

\begin{figure}[t]
\centering
\includegraphics[width=8cm]{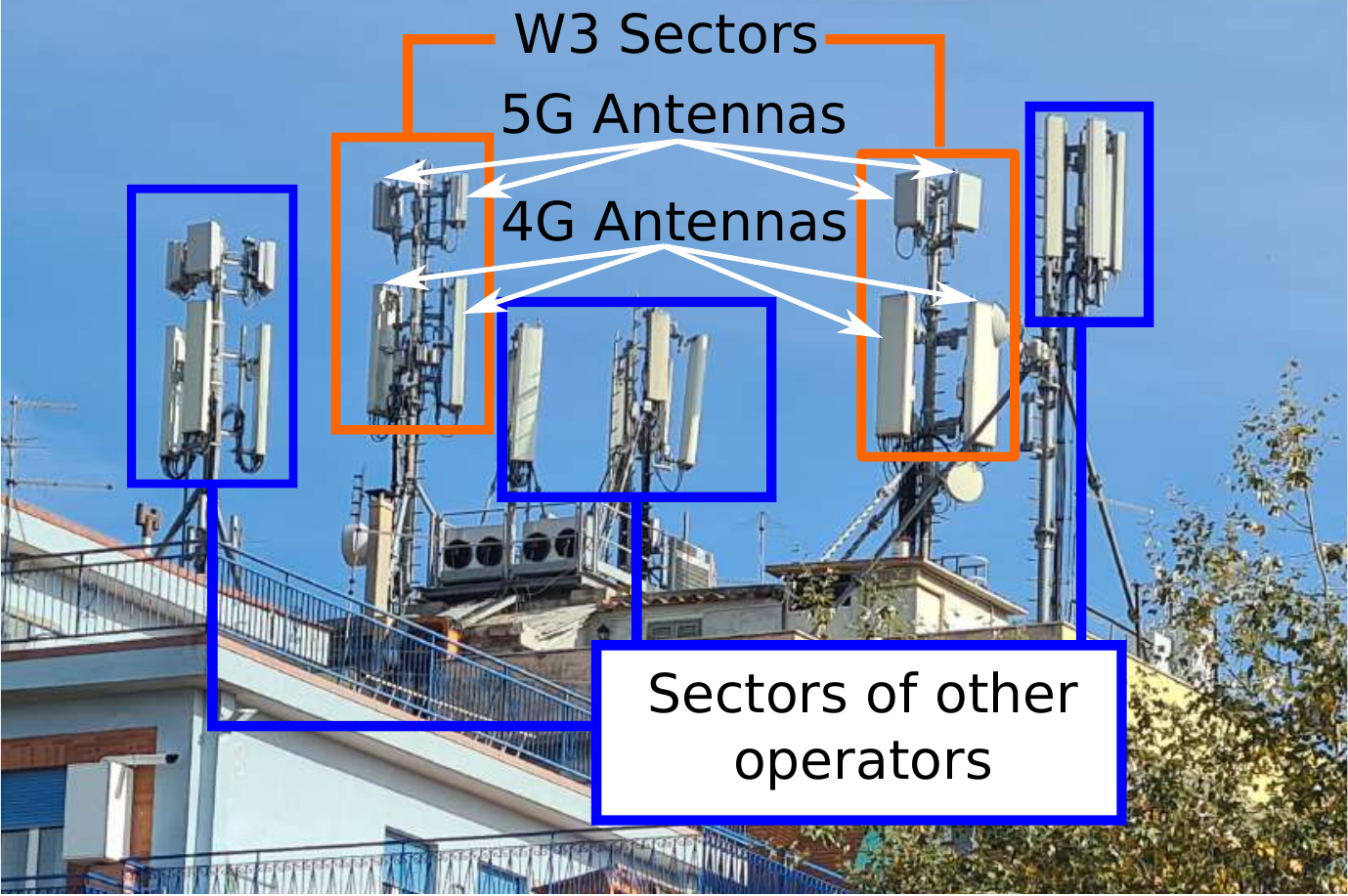}
\caption{Roof-top installation of W3 operator encompassing multiple sectors and distinct panels for 4G and 5G technologies.}
\label{fig:tower_installation}
\end{figure}

\textbf{Interviewing of 5G and 4G networks.} At the time of preparing this work, the \ac{NSA} option, in which the 5G radio access network is supported by a 4G core, is still the most widespread way to implement 5G networks in the country. Compared to a full \ac{SA} deployment, \ac{NSA} requires a strong dependability of the 5G service with respect to the 4G network. In particular, a 5G connection is always provided in parallel to a 4G one, which acts as the anchor for the dual 4G/5G connectivity \cite{agiwal2021survey}. Therefore, our \ac{EMF} assessments have to include the monitoring of the 4G bands that are used in parallel to the 5G ones, as the injected traffic is (likely) flowing over both 4G and 5G channels. To further complicate such feature, Fig.~\ref{fig:frequencies_overview} reports the band allocations of W3 over pre-5G and 5G technologies (valid for the the city of Rome). Astonishingly, the number of possible bands that can be used by 4G is huge, as all the spectrum portions licensed to W3 over 800-2600 MHz frequencies can be potentially used for 4G services. 

\textbf{Dynamicity of carrier aggregation.} Another key implementation aspect that strongly affects the \ac{EMF} monitoring is the (possible) carrier aggregation across multiple 4G bands, which are used in parallel to the 5G ones.  As reported in relevant 3GPP standards \cite{3gpppowerclass}, there are plenty of possible carrier aggregation combinations, ranging from sets composed of 1-2 bands up to ones including several pre-5G spectrum portions in use by the operator. The selected combination of aggregated carriers for a given connection is a \textit{local} decision, which depends on many features (like the propagation conditions reported by the smartphone) that are monitored by the serving \ac{RBS}. Consequently, the adopted set of carriers {cannot} be  determined \textit{a priori} and it depends on the measurement location. The dynamicity in the carrier aggregation has to be taken into account in our measurement procedures, in order to limit the exposure assessment only on those bands that are used for the transfer of the injected traffic.

\textbf{Time Division Duplexing.} Fig.~\ref{fig:frequencies_overview} details the assignment of frequencies for the \ac{UL} and \ac{DL} directions. In different spectrum portions (B20, B8, B3, B1 and B7), the \ac{FDD} rigidly separates the \ac{UL} frequencies with respect the \ac{DL} ones. On the other hand, the B38 spectrum of 4G and all the 5G portions in use by W3 (covering the N78 band) are employing the \ac{TDD}, which adopts multiplexing of both \ac{UL} and \ac{DL} over the same frequencies. Intuitively, \ac{TDD} complicates the dissection of smartphone vs. \ac{RBS} exposure contributions, as 
both time-frequency domains have to be jointly analyzed to distinguish the \ac{DL} from the \ac{RBS} with respect to the \ac{UL} from the smartphone.  

\textbf{RBS co-location.} Fig.~\ref{fig:tower_installation} shows a typical roof-top \ac{RBS} installation, which includes multiple 5G and 4G sectors of W3 operator, as well as radio equipment of other operators that are co-located on the same site. Since our goal is {to} consider the impact of smartphone and \ac{RBS} exposure for a given connection, we need to distinguish the \ac{EMF} contribution of the considered operator with respect to the co-located ones that serve the same area (i.e., the sectors of other operators in the figure).

\section{5G EMF Measurement Requirements}
\label{sec:emf_requirements}

\begin{table}[t]
\caption{5G Features \& Goals vs.  EMF {Measurement} Requirements.}
\label{tab:fetures_goals_requirements}
\scriptsize
\centering
\begin{tabular}{|m{0.3cm}|m{3.2cm}|m{3.5cm}|m{0.2cm}|}
\hline
\rowcolor{Grayblue} \multicolumn{2}{|c|}{\textbf{Feature/Goal}} & \multicolumn{2}{c|}{\textbf{Requirement}}\\
\hline
\textit{FG1} & Operator and Technology Impact & Narrow-band frequency-selective measurements & \textit{R1}\\
\hline
\rowcolor{Linen} \textit{FG2} & 5G (and 4G) \ac{TDD} & Dissection of smartphone vs.  \ac{RBS} exposure  & \textit{R2}\\
\hline
\textit{FG3} & 45/5G Dual Connectivity & Multiple monitoring over 4G and 5G technologies  & \textit{R3}\\
\hline
\rowcolor{Linen} \textit{FG4} & Spectrum Fragmentation & Detection \& iteration   & \\
\cline{1-2}
\rowcolor{Linen} \textit{FG5} & Dynamic carrier aggregation & over the adopted set of carriers & \multirow{-2}{*}{\textit{R4}}\\
\hline
\textit{FG6} & Impact of traffic & Forcing smartphone traffic in the \ac{UL}/\ac{DL} directions & \textit{R5} \\
\hline
\rowcolor{Linen} \textit{FG7} & Impact of propagation & Selection of representative measurement locations & \textit{R6} \\
\hline
\end{tabular}
\end{table}

We analyze hereafter the measurement requirements that are instrumental for the definition of the \ac{EMF} monitoring framework,  starting from the goals of our work in Sec.~\ref{sec:intro} and the implementation features detailed in Sec.~\ref{sec:5G_primer}. To this aim, Tab.~\ref{tab:fetures_goals_requirements} highlights the transition from 5G features and goals (\textit{FG1}-\textit{FG7}) into concrete \ac{EMF} measurement requirements (\textit{R1}-\textit{R6}).  

More in depth, the need of distinguishing the 5G exposure contribution w.r.t. other technologies and/or other operators (\textit{FG1}) impose to adopt an approach based on narrow-band frequency-selective measurements (\textit{R1}).  This is a first and important requirement, as narrow-band measurements can be performed only by adopting more complex instrumentation tools and procedures than the ones used for wide-band approaches. Secondly, the adoption of \ac{TDD} in 5G bands (\textit{FG2}) requires to define a measurement methodology able to dissect the smartphone exposure contribution vs. the \ac{RBS} one,  which,  obviously, {cannot} be based on frequency separation (\textit{R2}). Third, the dual connectivity between 4G/5G (\textit{FG3}) requires to perform both 4G and 5G exposure assessments (\textit{R3}).  Fourth, the adoption of non-contiguous 5G portions in a fragmented spectrum (\textit{FG4}), as well as the dynamic carrier aggregation feature (\textit{FG5}), suggest that the \ac{EMF} measurements should be done only on the combination of 4G/5G carriers in use by a given connection. Therefore, rather than iterating over the whole set of spectrum portions assigned to the operator - an operation that would result in a waste of time and resources - a mechanism able to detect the carriers used for a given data transfer should be designed (\textit{R4}).  Such feature should be complemented by an \ac{EMF} measurement procedure (possibly automated) able to iterate over the selected set of carriers and measure the exposure on each carrier (\textit{R4}).  Fifth, the impact of traffic (\textit{FG6}) has to be evaluated with a procedure able to force the traffic in \ac{UL}/\ac{DL} directions (\textit{R5}). Finally, the impact of propagation (\textit{FG7}) can be solely assessed by selecting a representative set of measurement locations (\textit{R6}), subject to meaningful propagation conditions (e.g., \ac{LOS}/\ac{NLOS}, \ac{RBS} proximity/distance). It is also clear that \textit{FG6} and \textit{FG7} inherently require that the adopted measurement chain should be easily portable over the territory, as several measurements in different locations should be performed in order to retrieve a meaningful set of results.

\section{\textsc{5G-EA} Framework Description}
\label{sec:5G_SE}

We divide the presentation of the \textsc{5G-EA} framework into the following steps: \textit{i}) measured exposure metrics, \textit{ii}) adopted tools and \ac{HW} chains, \textit{iii}) description of measurement algorithm, and \textit{iv}) implementation details.

\subsection{Measured Exposure Metrics}

In principle, any exposure assessment strongly depends on the target metrics that need to be measured. In particular, the classical taxonomy defines \ac{SAR}/absorbed power density for \ac{UE} assessments vs. electric field/plane-wave power density for \ac{RBS} evaluation \cite{chiaraviglio2021health}. The \ac{SAR} and absorbed power density metrics are useful when the measurement target is the near-field assessment, in which the radiating source is (almost) attached to the body (e.g., an \ac{UE} close to the ear during a phone call). Despite such metrics are still relevant for today equipment (and for \ac{UE} {\ac{SAR}-based} limits), we point out that the typical smartphone user makes phone calls with the equipment attached to the ear only to a limited extent. In fact, recent statistics \cite{smartphoneuse} reveal that smartphones are mainly used for downloading/uploading data traffic, with the \ac{UE} hold at a non negligible distance from the head/chest in order to read/produce content on the screen. Since our goal is to evaluate the exposure in such conditions - which represent a typical 5G scenario -  in this work we always impose a minimum distance between the \ac{UE} generating traffic and the evaluation point of our measurement.

\begin{table}[t]
\caption{Frequencies, wavelengths, Fraunhofer regions and far field distances for the W3 bands with smartphone antenna length $L_{\text{A}}=0.08$~[m].}
\label{tab:far_field}
\scriptsize
\centering
\begin{tabular}{|m{1.2cm}|>{\columncolor{Linen}}m{1.5cm}|m{1.4cm}|>{\columncolor{Linen}}m{2cm}|}
\hline
\rowcolor{Grayblue} \textbf{Frequency} $f$ & \textbf{Wavelength} $\lambda_f$ & \textbf{Fraunhofer Region} $\frac{2 L_{\text{A}}^2}{\lambda_f}$ & \textbf{Far-field distance} $D^{\text{FF}}_f$ \\
\hline
800~[Mhz] & 0.37~[m] & 0.03~[m] & $>$0.37~[m]\\
900~[Mhz] & 0.33~[m] & 0.04~[m] & $>$0.33~[m]\\
1800~[Mhz] & 0.16~[m] & 0.08~[m] & $>$0.16~[m] \\
2100~[Mhz] & 0.14~[m] & 0.09~[m] & $>$0.14~[m] \\
2600~[Mhz] & 0.12~[m] & 0.11~[m] & $>$0.12~[m]\\
3447~[Mhz] & 0.09~[m] & 0.15~[m] & $>$0.15~[m] \\
3547~[Mhz] & 0.08~[m] & 0.15~[m] & $>$0.15~[m]\\
3610~[Mhz] & 0.08~[m] & 0.15~[m] & $>$0.15~[m]\\
\hline
\end{tabular}
\end{table}

Apart from better matching the actual smartphone usage, the introduction of a minimum distance between the \ac{UE} and the measurement point may allow operating in the far-field region from the \ac{UE}, which is formally defined as:
\begin{equation}
\label{eq:far_field_distance}
D^{\text{FF}}_f > \max \left(\lambda_f, L_{\text{A}}, \frac{2 L_{\text{A}}^2}{\lambda_f}\right)
\end{equation}
where $\lambda_f$ is the wavelength of frequency $f$, $L_{\text{A}}$ is the length of the radiating antenna, while the term $\frac{2 L_{\text{A}}^2}{\lambda_f}$ represents the limit of the Fraunhofer region. To give an example, Tab.~\ref{tab:far_field} reports the values of $\lambda_f$ and $D^{\text{FF}}_f$ for the bandwidth allocation of W3 and a smartphone antenna length $L_{\text{A}}$ equal to 0.08~[m]. As expected, the observed far-field distances $D^{\text{FF}}_f$ strongly depend on the considered frequencies, but, however, we can note that the minimum $D^{\text{FF}}_f$ is lower than $0.2$~[m] for 4G frequencies above or equal 1800~[MHz] and for all 5G frequencies.

By imposing the distance $D^{\text{FF}}_f$ from the \ac{UE}, we are able to operate in far-field, a condition that is also generally experienced when considering the \ac{RBS} as the source of radiation. In this way, an homogeneous set of metrics (e.g., electric field and/or plane-wave power density) can be used to measure both \ac{UE} and \ac{RBS} exposure. This is in turn beneficial for adopting the same measurement equipment when assessing the \ac{UE}/\ac{RBS} exposure, as detailed in the following subsection.


\begin{figure}[t]
\centering
 	\subfigure[Exposure Assessment and Traffic Generation Chains (UE side)]
	{
		\includegraphics[width=7cm]{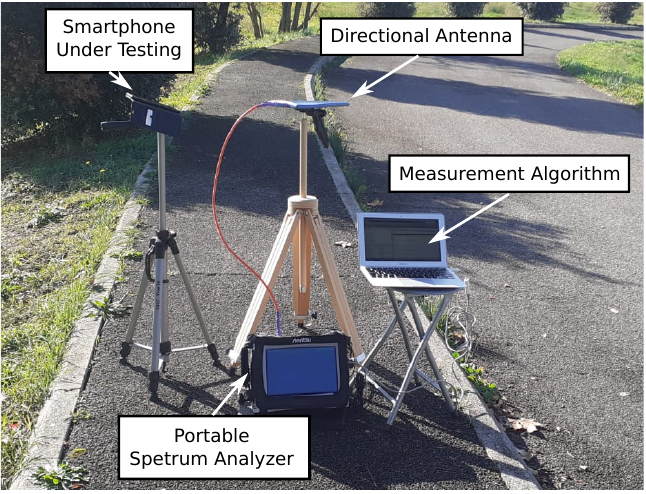}
		\label{fig:example_step_env_measuring}
	}
 	\subfigure[Isolation of \ac{RBS} contribution for a measurement point in \ac{LOS} conditions.]
	{
		\includegraphics[width=7cm]{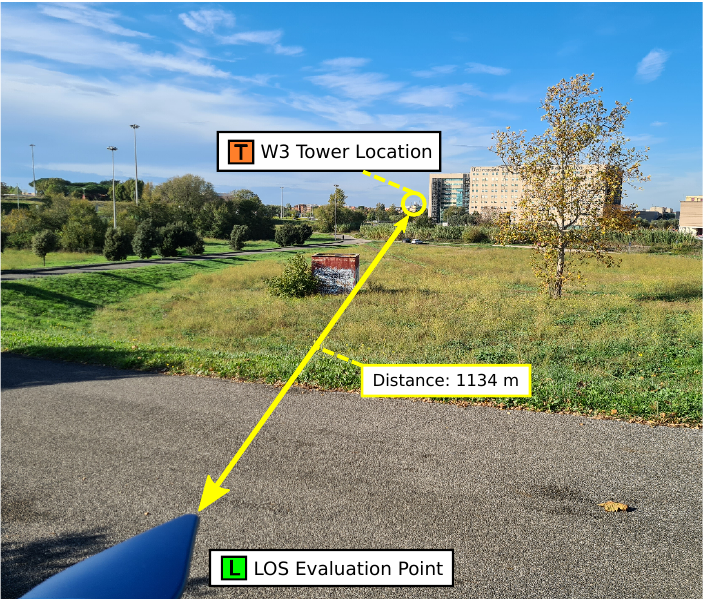}
		\label{fig:los_example}

	}
\caption{Views of the equipment tools in a given measurement location.}
\label{fig:hw_chain_direct_antenna}
\end{figure}

\subsection{Tools and HW Chains}

We describe {hereafter} the equipment tools, which are also sketched in Fig.~\ref{fig:example_step_env_measuring}. Focusing on the exposure assessment chain, we employ the following \ac{HW} ($E1$-$E5$):
\begin{enumerate}
\item[$E1$)] Hand-held \ac{SAN} Anritsu MS2090A with maximum frequency range equal to 32~[GHz] with 110~[Mhz] of maximum bandwidth analysis, equipped with one battery plus another one of backup;
\item[$E2$)] Directive antenna Aaronia 6080, with frequency range 680~[MHz]-8~[GHz], maximum gain equal to 6~[dBi], nominal impedance of 50~[Ohm];
\item[$E3$)] Coaxial cable Anritsu flexible RF 1~[m] Cable K(f)-K(m) DC-40~[GHz], connecting the \ac{SAN} to the directive antenna;
\item[$E4$)] Laptop MacBook Air with Intel Core i5 1.3 [GHz] CPU, 4~[GB] of RAM, 256~[GB] of memory, equipped with Matlab R2017b and RSVisa 5.12.1 driver;
\item[$E5$)] Ethernet cable of 1~[m] length, Cat.5E, verified TIA-EIA-568-C.2, connecting the laptop to the \ac{SAN}.
\end{enumerate}
Focusing on $E1$, the \ac{SAN} allows implementing narrow-band measurements, and thus matching requirement $R1$. Focusing then on $E2$, the directionality of the adopted antenna allows spatially separating the contribution of the \ac{UE} and the one of the \ac{RBS}. As shown in Fig.~\ref{fig:los_example}, the antenna is oriented towards the considered source. In this way, the contribution of other sources, e.g., a \ac{UE} placed behind the measurement antenna, is not sensed. By selectively pointing the directive antenna towards the \ac{UE} or the \ac{RBS}, we can isolate their respective contributions, and thus matching requirement $R2$, even when the monitoring is performed over \ac{TDD} bands. In addition, the short coaxial cable of $E3$ guarantees almost negligible signal degradation between the directive antenna and the \ac{SAN} - a feature that is instrumental for measuring the relatively low environmental exposure values of 5G. The \ac{SAN} is then connected to the $E4$ laptop via the dedicated $E5$ cable. The core of our framework is a custom measurement algorithm written in Matlab and running on $E4$. The algorithm allows remotely programming the \ac{SAN} to perform multiple monitoring of 4G and 5G bands (requirement $R3$), as well as to implement the automatic detection and iteration over the adopted set of carriers (requirement $R4$). 

Focusing then on the traffic generation chain, we adopt the following tools:
\begin{enumerate} 
\item[$T1$)] Samsung S20+ 5G smartphone, equipped with Android 11 (1st May 2021) and Magic \texttt{Iperf} v.1.0 App client.  
\item[$T2$)] Dell Poweredge R230 server, equipped with 4 cores Intel Xeon E3-1230, 64 GB of RAM, Ubuntu 18.04.1 OS and \texttt{Iperf} v.3.1.3.
\end{enumerate}
More concretely, $T2$ is installed at the University building, and made accessible through a public \ac{IP} address. In addition, the \texttt{Iperf} program is used to generate synthetic traffic between $T1$ and $T2$ (either in the \ac{DL} or \ac{UL} direction). In this way, we accomplish requirement \textit{R5}.

Fig.~\ref{fig:example_step_env_measuring} shows the measurement setup in a given location. Both smartphone and \ac{SAN} are placed on tripods above around 1~[m] from ground, in order to mimic exposure evaluations representative of users. The required setup does not involve any electricity plug. This fact, coupled also with the overall small size of $E1$-$E5$ and $T1$ (as shown in Fig.\ref{fig:example_step_env_measuring}), as well as the availability of a second backup battery for the \ac{SAN}, allows easily repeating the measurements in different locations of the territory, and thus matching requirement $R6$.  






\subsection{Algorithm Description}

\begin{figure}[t]
\centering
\includegraphics[width=8cm]{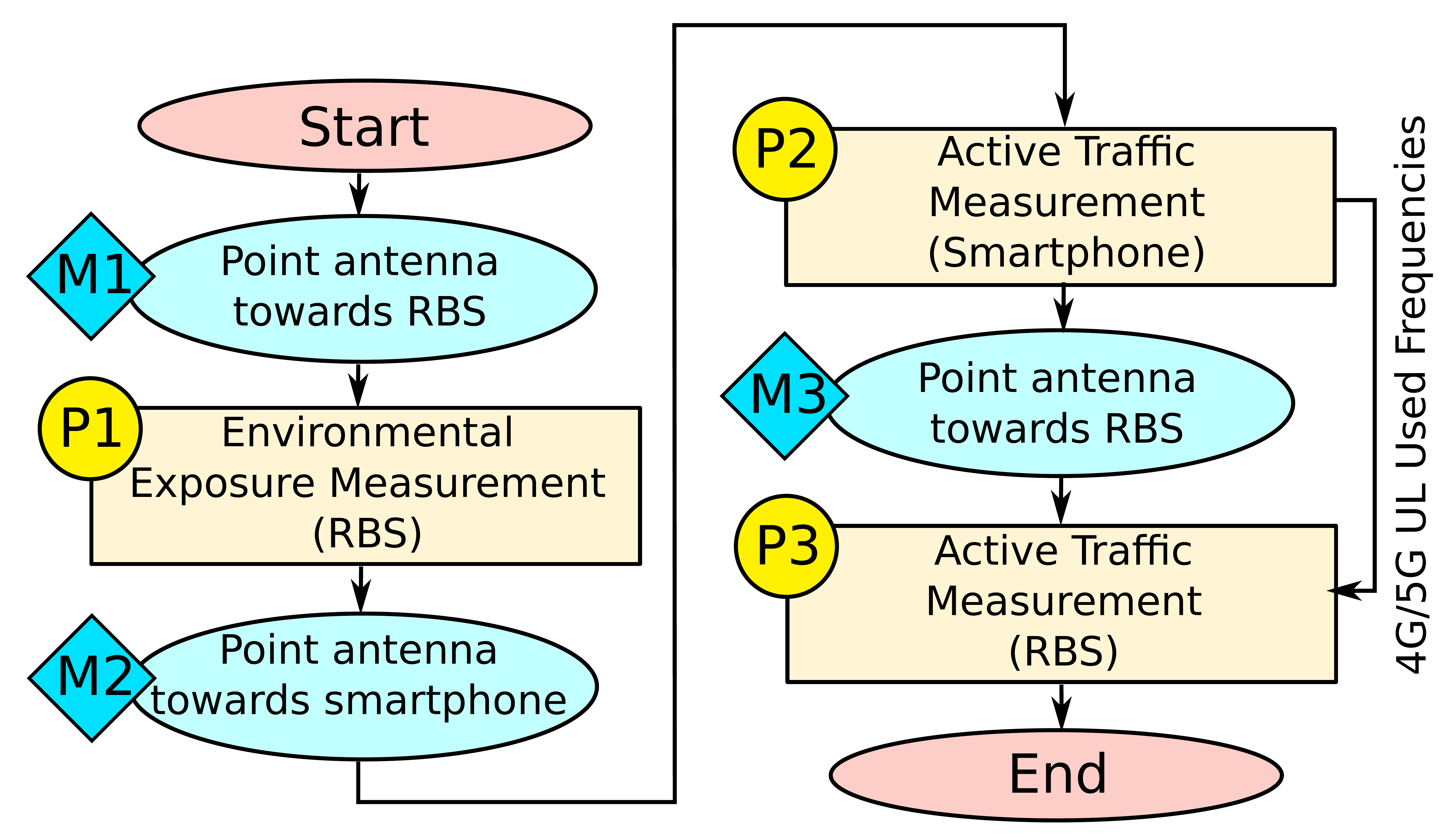}
\caption{High level steps of the measurement algorithm implemented in \textsc{5G-EA}.}
\label{fig:mm_high_level}
\end{figure}

\begin{figure*}[t]
\centering
\includegraphics[width=14cm]{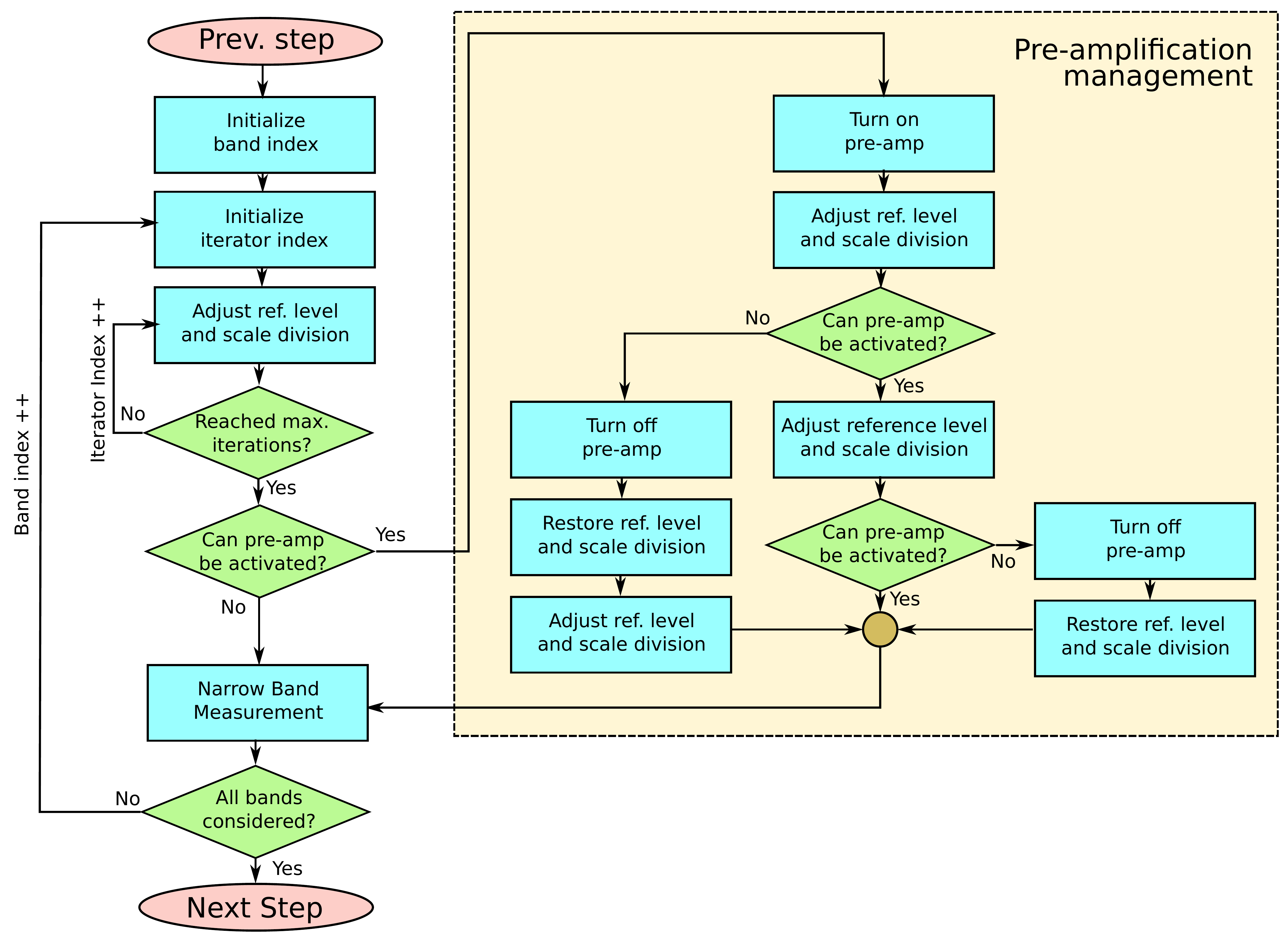}
\caption{Steps to perform $P1$ (RBS environmental exposure).}
\label{fig:env_steps_base_station}
\end{figure*}

Fig.~\ref{fig:mm_high_level} provides a high level description of the measurement algorithm implemented in the \textsc{5G-EA} framework. In more detail, we apply a \textit{divide-et-impera} approach to split the complex measurement procedure into the following sub-problems: \textit{i}) evaluation of \ac{RBS} environmental exposure (step $P1$),  \textit{ii}) evaluation of active traffic exposure from the smartphone (step $P2$), \textit{iii}) evaluation of active traffic exposure from the \ac{RBS} (step $P3$). In addition, the algorithm is complemented by three manual orientations $M1$-$M3$ of the directive antenna, which are instrumental to correctly separate \ac{RBS} vs. smartphone contributions.  More concretely, the directive antenna is pointed towards the \ac{RBS} before starting $P1$ ($M1$ block of Fig.~\ref{fig:mm_high_level}), then it is pointed towards the smartphone before running $P2$ ($M2$ block), and finally it is pointed again {towards} the \ac{RBS} before executing $P3$ ($M3$ block).

Intuitively, all the considered bands in use by the operator (\ac{TDD} and \ac{DL} \ac{FDD}) are swept during the environmental exposure assessment of $P1$. Then, the goal of $P2$ is to restrict the set of monitored bands only on those ones in use by the current active traffic connection - which is kept alive from $P2$ to $P3$. In this way, the monitoring during $P3$ is done on the same traffic conditions that are experienced in $P2$.

In the following, we describe in more details steps $P1$-$P3$.

\begin{algorithm}[t]
\small
\caption{Pseudo-Code of the \texttt{adjust\_ref\_level\_scale\_div} routine of $P1$}
\label{alg:adj_scale_div}
\textbf{Input:} set of \texttt{SAN\_settings}, current frequency start \texttt{curr\_f\_min}, current frequency stop \texttt{curr\_f\_max}, safety margin for the ref. level \texttt{safety\_margin}, maximum time (in s) for searching the maximum level \texttt{max\_time\_search}, pre-amplifier state \texttt{pre\_amp\_state}, minimum level matrix \texttt{min\_l},   number of y ticks on the screen \texttt{y\_ticks}\\  
\textbf{Output:} \texttt{ref\_level}, \texttt{scale\_div}
\begin{algorithmic}[1]	
	\State{set\_SAN(SAN\_settings);}
	\State{max\_l=-200;}
	\For{i=1:max\_time\_search}
		\State{max\_l=max\_lev\_search(max\_l,curr\_f\_min,curr\_f\_max);}
		\State{sleep(1);}
	\EndFor
	\State{ref\_level=ceil(max\_l)+safety\_margin;}
	\State{set\_SAN(ref\_level);}
	\State{scale\_div=abs(ref\_level-min\_l(curr\_f\_min,pre\_amp\_state))/ y\_ticks;}
	\State{set\_SAN(scale\_div);}
\end{algorithmic}
\end{algorithm}

\subsubsection{P1 - Environmental RBS EMF} 

Fig.~\ref{fig:env_steps_base_station} shows the high level flowchart of $P1$. Initially, the set of bands to be monitored are selected, based on the operator that is under consideration.
For example, in the case of W3 operator all \ac{TDD} and \ac{DL} \ac{FDD} bands shown in Fig.~\ref{fig:frequencies_overview} are considered for the environmental assessment of \ac{RBS} exposure. The algorithm then iterates over the set of bands. For each considered band, an automatic procedure to adjust reference level and scale division of the observed signal is implemented. Intuitively, the reference level is the upper limit of the y axis in a spectrum plot (where the x axis is the set of monitored frequencies), while the scale division allows tuning the unit of the y ticks and consequently the lower limit on the y axis. By jointly optimizing the reference level and the scale division, we can achieve a double goal: \textit{i}) the signal that is being monitored can be qualitatively checked on the screen of the \ac{SAN}, \textit{ii}) the measurement resolution is tuned to the actual signal that is observed, and thus the impact of (possible) measurement uncertainties is limited. 

More specifically, the adjustment of the reference level and scale division reported in the flow chart of Fig.~\ref{fig:env_steps_base_station} is sketched in the \texttt{adjust\_ref\_level\_scale\_div} routine of Alg.~\ref{alg:adj_scale_div}. This function requires as input a set of basic \ac{SAN} settings (whose values are going to be presented in detail in Sec. \ref{subsubsec:p1_parameters}), the starting and ending frequency for the considered band (\texttt{curr\_f\_min} and \texttt{curr\_f\_max}), the \texttt{safety\_margin} parameter that is used when setting the reference level, the \texttt{max\_time\_search} parameter to govern the maximum time for searching the  maximum reference level, the \texttt{pre\_amp\_state}  boolean variable storing the state of the \ac{SAN} pre-amplifier (active or inactive), the \texttt{min\_l} matrix including the values of the minimum sensed levels (which depend on the adopted frequency and the pre-amplifier state)  and the \texttt{y\_ticks} parameter representing the number of y ticks on the \ac{SAN} screen. 

The routine then proceeds as follows. The basic \ac{SAN} settings are implemented in line 1, which include e.g., the detector type, the measured unit, and the trace detector. The maximum signal level \texttt{max\_l} is initialized to a very low value in line 2. Then, a live searching of \texttt{max\_l}  is iteratively performed in lines 3-6, up to the maximum time \texttt{max\_time\_search}. At the end of this step, the maximum recorded signal level is stored in \texttt{max\_l}. The reference level \texttt{ref\_level} is then set by adding to \texttt{max\_l} the safety margin in line 7. In line 8, the resulting reference level is applied. In addition, the exact scale division, in order to entirely show the dynamics of the signal between \texttt{ref\_level} and \texttt{min\_l}, is computed in line 9 and then applied to the \ac{SAN} in line 10.

The execution of Alg.~\ref{alg:adj_scale_div} is then iterated up to a maximum number (iterator index in the flowchart of Fig.~\ref{fig:env_steps_base_station}), in order to improve the setting of reference level and scale division. In the following step, a check on the pre-amplification is performed.  If the current reference level is lower than a pre-amplification threshold, the signal can be pre-amplified by the \ac{SAN} (right part of Fig.~\ref{fig:env_steps_base_station}).\footnote{Reference levels higher than the pre-amplification threshold may result into an \ac{ADC} over-range after activating the pre-amplification of the signal. Therefore, this feature should be activated only for those signals whose reference level and dynamics are within the \ac{ADC} limits.} Such feature is particularly useful for the environmental assessment of 5G signals, which are normally very low and close to the equipment noise level, due to a relatively low usage of 5G on such early phase of adoption. After turning on the pre-amplifier, the \texttt{adjust\_ref\_level\_scale\_div} routine is called again, in order to adjust the amplified signal levels. However, this procedure may increase the reference level again above the maximum one allowed by the pre-amplifier. Consequently, a check on the pre-amplification threshold is done again. In case pre-amplification can be kept turned on, a further adjustment of reference level and scale division is done - and a further check on the pre-amplification is performed. In case pre-amplification is not supported, the pre-amplification is turned off, the reference level and scale division are reverted back to the last values before pre-amplification, and (eventually) a further call of the \texttt{adjust\_ref\_level\_scale\_div} routine is run.


\begin{algorithm}[t]
\small
\caption{Pseudo-Code of the \texttt{nar\_band\_meas} routine of $P1$, $P2$ and $P3$}
\label{alg:nar_band_meas}
\textbf{Input:} set of basic \texttt{SAN\_settings}, current frequency start \texttt{curr\_f\_min}, current frequency stop \texttt{curr\_f\_max}, number of samples \texttt{n\_samples}, inter sample time (in s)  \texttt{int\_sample\_time}\\  
\textbf{Output:} Array of exposure values \texttt{curr\_exp} in dBm/m$^2$ ($P1$) or V/m ($P2$ and $P3$)
\begin{algorithmic}[1]	
	\State{set\_SAN(SAN\_settings);}
	\For{i=1:n\_samples}
		\State{curr\_exp(i)=get\_SA(curr\_f\_min,curr\_f\_max);}
		\State{sleep(int\_sample\_time);}
	\EndFor
\end{algorithmic}
\end{algorithm}

After setting reference level and scale division (and eventual activation of pre-amplification), the signal is ready to be measured. To this aim, a narrow band measurement, expanded in 
Alg.~\ref{alg:nar_band_meas}, is invoked. The function takes as input a set of basic \ac{SAN} settings, the current frequency start \texttt{curr\_f\_min} and frequency stop  \texttt{curr\_f\_max},  the number of sampled channel power measurements \texttt{n\_samples}, and the inter-sample time \texttt{int\_sample\_time}. The routine then produces as output an array of exposure values \texttt{curr\_exp}.
The logic of the procedure is very simple: after setting the \ac{SAN} parameters, a channel power computation function is iteratively invoked on the \ac{SAN}. When all the samples are recorded, the algorithm returns the array of exposure measurements \texttt{curr\_exp}. At the end of $P1$, the \ac{RBS} environmental exposure is measured for the set of bands in use by the operator.

\subsubsection{P2 - Active Traffic Smartphone EMF} 

The goal of the second part of the algorithm is to perform the assessment of the exposure generated by the smartphone upon active traffic generation. To this aim, the dual 4G/5G connectivity and carrier aggregation features suggest that multiple bands (unknown \textit{a-priori}) can be used in parallel for the data transfer. On the other hand, measuring the exposure on the entire set of bands in use by the operator may result in a waste of resources, in terms of: \textit{i}) overly increase of time to perform the assessment,  \textit{ii}) waste of consumption of the \ac{SAN} battery (which is a precious resource) and \textit{iii}) excessive traffic consumption on the smartphone (which may be critical for limited data traffic plans). To face such issues altogether, $P2$ adopts the following intuition. First, the \ac{TDD} and \ac{UL} \ac{FDD} bands in use by the data transfer are detected. Then, the exposure assessment is done only on the selected subset of spectrum portions currently in use. 

\begin{figure}
\centering
\includegraphics[width=7.5cm]{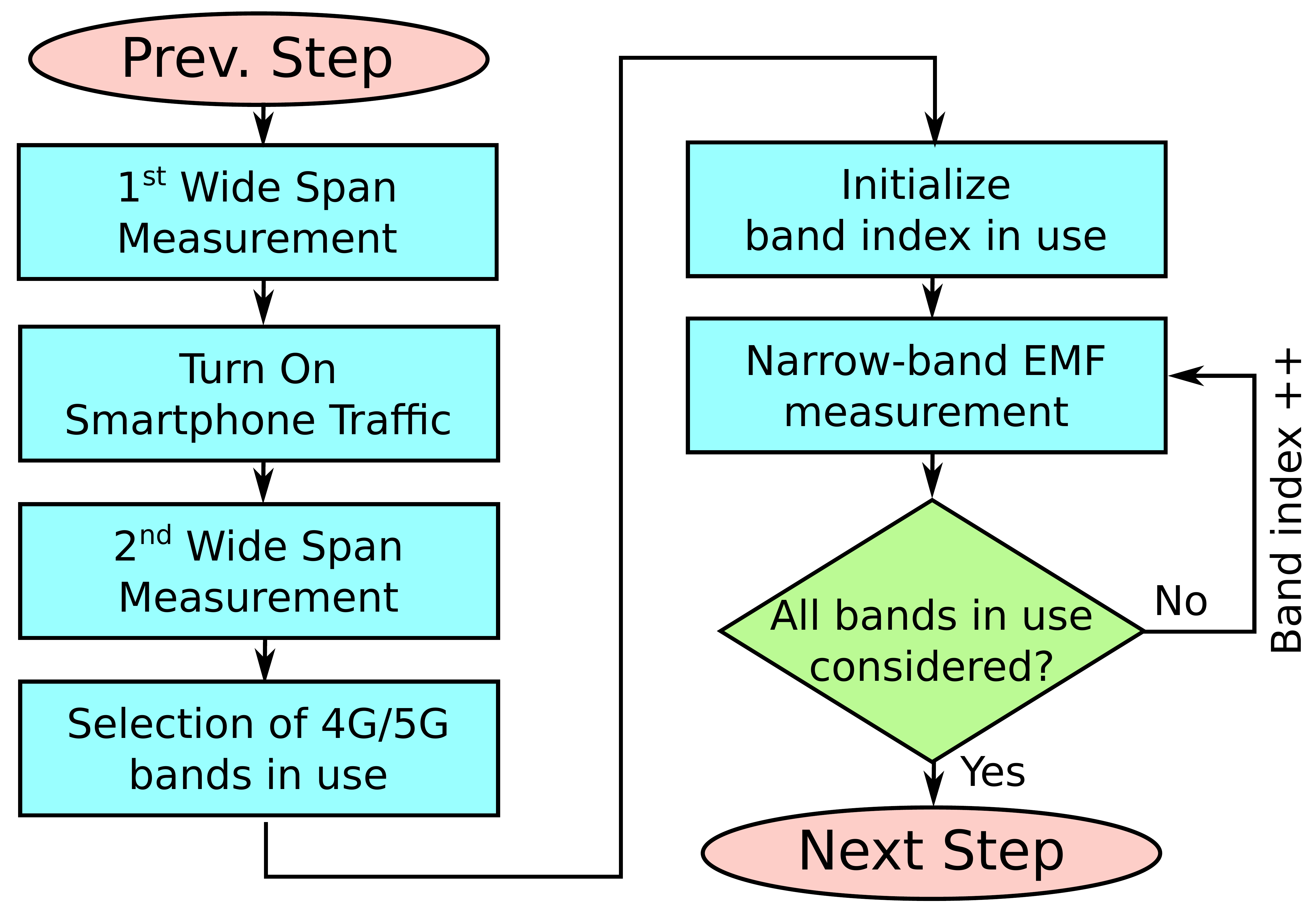}
\caption{Steps to perform $P2$ (active traffic smartphone exposure).}
\label{fig:active_traffic_steps_smartphone}
\end{figure}

To this aim, Fig.~\ref{fig:active_traffic_steps_smartphone} sketches the main operations performed during $P2$. Initially, a wide span assessment is done, in order to detect the peak(s) of the sensed signals on a very large range of frequencies (including all the ones in use by the operator under investigation). The goal of this scan is not to measure exposure, but rather to get a quick indication on the frequencies that carry most of signal power before injecting any traffic. In the following step, the active traffic is {generated towards} the smartphone, by executing the \texttt{Iperf} program. Then, a second wide span assessment is done. The detection of the subset of 4G/5G bands in use for the data transfer is done by comparing the peaks recorded during the first scan vs. the ones observed on the second one. 

\begin{algorithm}[t]
\small
\caption{Pseudo-Code of the \texttt{sel\_band\_use} routine of P2}
\label{alg:sel_band_use}
\textbf{Input:} array of \ac{EMF} values from 1st span \texttt{emf\_array\_1st\_span}, array of \ac{EMF} values from 2nd span \texttt{emf\_array\_2nd\_span}, array of frequency values \texttt{freq\_array}, threshold increase parameter \texttt {thre\_inc}\\  
\textbf{Output:} 
array of selected bands \texttt{sel\_band\_array}
\begin{algorithmic}[1]	
\State{sel\_band\_array=initialize();}
\For{f=1:size(freq\_array)}
	\If{incr\_percent(emf\_array\_2nd\_span\_rev(f), emf\_array\_1st\_span(f)) $>$ thre\_incr}
		\State{band\_index=find\_b\_index(f);}
		\State{sel\_band\_array(band\_index)=1;}
		\If{fdd\_array(band\_index)==1}
			\State{band\_index\_fdd=find\_b\_index(f,FDD);}
			\State{sel\_band\_array(band\_index\_fdd)=1;}
		\EndIf
	\EndIf
\EndFor
\end{algorithmic}
\end{algorithm}

The detection of the 4G/5G bands is expanded in Alg.~\ref{alg:sel_band_use}. The routine takes as input the \texttt{emf\_array\_1st\_span} array of \ac{EMF} values (indexed by frequency) that were sensed during the first span, the \texttt{emf\_array\_2nd\_span} array of \ac{EMF} values that were sensed during the second span, and a threshold increase parameter \texttt{thre\_inc} (in \%) to activate the detection. The algorithm then produces as output the subset of bands \texttt{sel\_band\_array} that are detected for the current data transfer. The logic of the function is quite simple: for each considered frequency, the sample in \texttt{emf\_array\_2nd\_span} is compared against the corresponding one \texttt{emf\_array\_1st\_span}, by computing the percentage variation of \ac{EMF}. If such variation is greater than the \texttt{thre\_inc} parameter, the band is included in the list of spectrum portions that are monitored for the current data transfer. The intuition here is in fact to exploit the increase of \ac{EMF} as a result of the usage of specific bands in the \ac{UL}. In case the current selected band employs \ac{FDD}, then the corresponding one in the \ac{DL} is also included in the list of selected ones. For example, let us assume that the 1745-1765~[MHz] band of Fig.~\ref{fig:frequencies_overview} is detected in the \ac{UL}. This portion of the spectrum belongs to the B3 \ac{FDD} band, which also includes the 1840-1860~[MHz] band for the \ac{DL}. This second portion will be likely used when evaluating the active traffic from the \ac{RBS}, and therefore it is included in the list of bands to be monitored - when considering \ac{RBS} active traffic exposure. At the end of the algorithm the array \texttt{sel\_band\_array} stores the list of band indexes in use for the current data transfer.

Coming back to the flowchart of Fig.~\ref{fig:active_traffic_steps_smartphone}, the blocks on the right details the steps for the \ac{EMF} assessment on the selected set of bands. In particular, the initial band is selected - the index in \texttt{sel\_band\_array} with lowest frequency and belonging to \ac{UL}. Then, the narrow-band \ac{EMF} measurement on the selected band is performed. The logic is in common with the exposure measurement of $P1$, and sketched in Alg.~\ref{alg:nar_band_meas}. In particular, the main differences rely on a different set of basic \ac{SAN} settings and on a different measurement metric (in terms of V/m). Once the measurement has been completed for the current band, $P2$ passes to the next one, until all the \ac{TDD} and \ac{UL} \ac{FDD} bands in \texttt{sel\_band\_array} are considered (band index in Fig.~\ref{fig:active_traffic_steps_smartphone}). At the end of $P2$, a set of exposure arrays, one for each considered band, is available.

\begin{figure}[t]
\centering
\includegraphics[width=4cm]{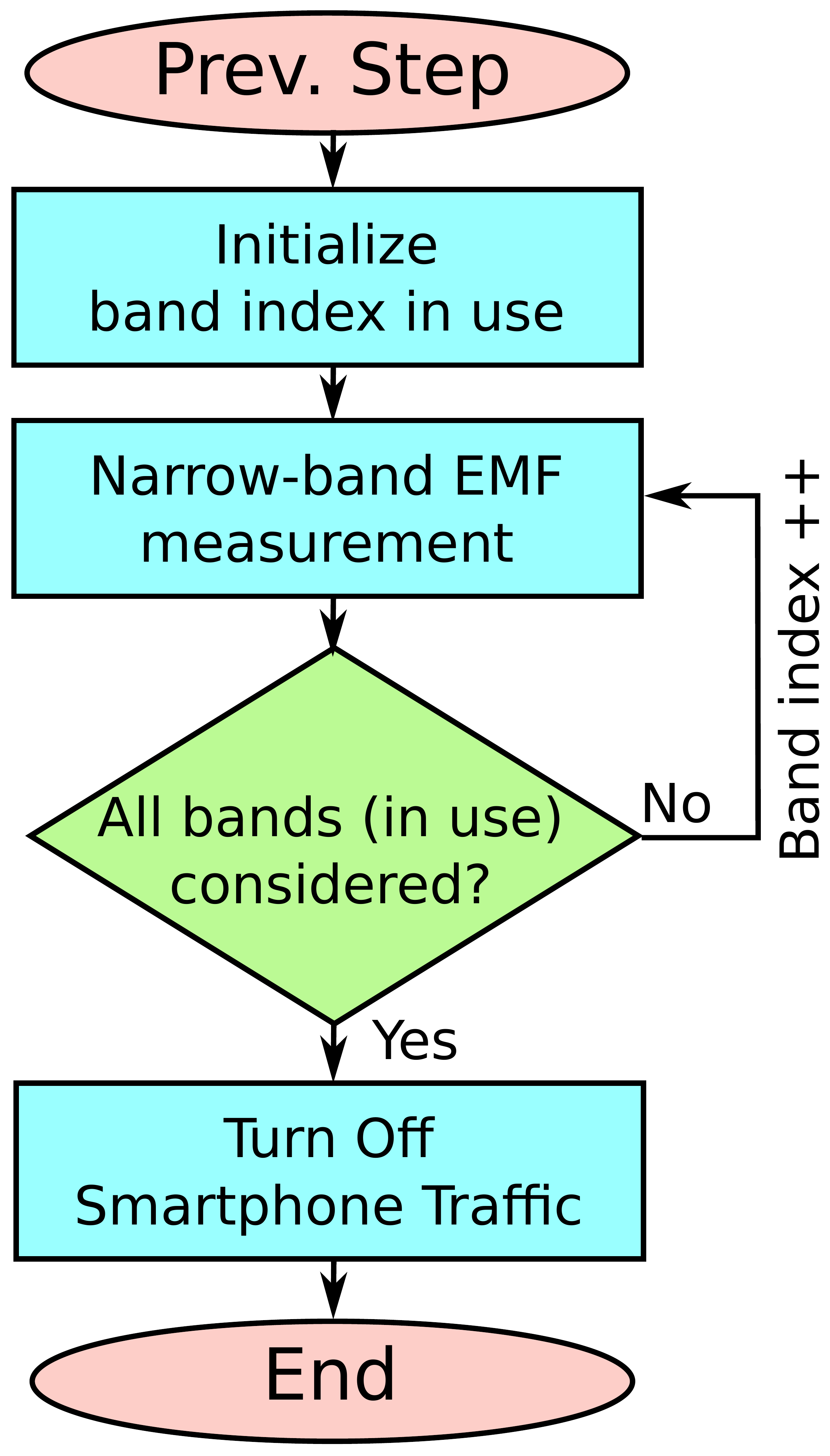}
\caption{Steps to perform $P3$ (active traffic \ac{RBS} exposure).}
\label{fig:active_traffic_steps_base_station}
\end{figure}

\subsubsection{P3 - Active Traffic RBS EMF}

The last part of our measurement technique involves the assessment of \ac{RBS} exposure while keeping active the current data transfer. Fig.~\ref{fig:active_traffic_steps_base_station} highlights the main blocks that realize this functionality. The logic is very similar to the smartphone assessment, except for the following differences: \textit{i}) the evaluation include \ac{TDD} and \ac{FDD} \ac{DL} bands (with the directive antenna pointed towards the \ac{RBS}), \textit{ii}) the smartphone traffic is turned off after completing the scan over the considered band. Similarly to $P2$, a set of exposure arrays is available at the end of the procedure.

\subsection{Implementation Details}

We implement $P1$-$P2$-$P3$ parts of \textsc{5G-EA} framework as a set of scripts written in Matlab - except from the traffic generation, which is governed by {the} \texttt{Iperf} program running on the smartphone and dedicated server. {An unique aspect} of our framework is the implementation of the measurement algorithm in software, on a general purpose machine that controls the \ac{SAN}. This is another innovation brought by our work, which opens the way for possible future investigation in the softwarization of \ac{EMF} assessments.

More technically, the high level functionalities reported in $P1$-$P2$-$P3$ are translated into a set of basic operations, coded as low-level \ac{SCPI} and transfered from/to the \ac{SAN} through a \ac{TCP} connection. The output of the \ac{SAN} (e.g., the array including the exposure values) are then sent back over the same connection in \ac{SCPI} format. In this way, the process is completely automated and the the post-analysis of the obtained data can be done directly in Matlab - in the same script running the measurement algorithm.


\section{Results}
\label{sec:results}

We present our outcomes through the following steps: \textit{i}) description of evaluation scenarios, \textit{ii}) parameter settings of \textsc{5G-EA} framework, \textit{iii}) exposure assessments.

\subsection{Evaluation Scenarios}

\begin{figure}
\centering
\includegraphics[height=8.5cm,angle=-90]{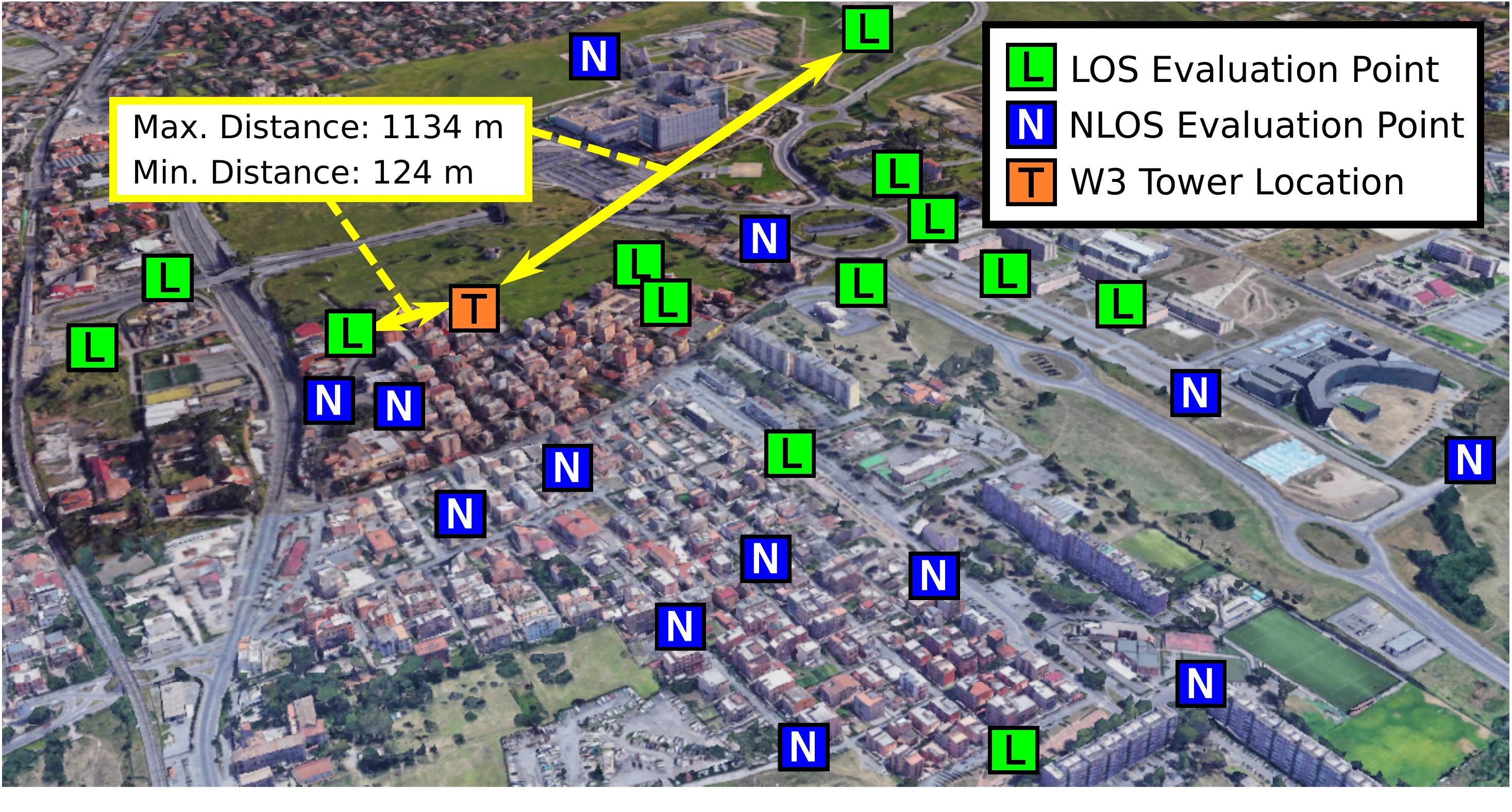}
\label{fig:tower_and_measurement_locations}
\caption{Positioning of W3 tower and outdoor \ac{LOS} (L green square) and \ac{NLOS} (N blue square) measurement locations (map source: Google Maps) - figure best viewed in colors.}
\label{fig:base_station_measurement_points}
\end{figure}

We consider a set of measurement points in the coverage area of the W3 roof-top installation shown in Fig.~\ref{fig:tower_installation}, with the frequency assignment reported in Fig.~\ref{fig:frequencies_overview}. The installation is located in the area close the University of Rome Tor Vergata in Rome (Italy). More concretely, we perform our experiments in both outdoor and indoor locations, due to the following reasons. First, we aim at massively performing measurements under different propagation conditions, which are obviously influenced by terrain parameters like the distance from the \ac{RBS}, the level of urbanization around the measurement point and the presence of buildings/obstacles on the path towards the \ac{RBS}. Second, we exploit the indoor locations to perform detailed and in-depth measurements, with the goal of corroborating the results from the outdoor locations with tests covering e.g., sensitivity analysis of the exposure vs. variation key parameters, such as throughtput, distance from the smartphone, and smartphone orientation.

Focusing on the outdoor tests, Fig.~\ref{fig:base_station_measurement_points} reports a 3D map of the measurement locations. In total, 26 measurement locations are selected for the tests, based on the following criteria: \textit{i}) spreading the tests over the territory around the W3 tower, and \textit{ii}) finding locations that are suitable for placing the instruments (e..g., avoid private streets, locations in close proximity to each other, etc.). The {3D} distance of each measurement location from the \ac{RBS} varies between a minimum of 124~[m] up to a maximum of 1134~[m],\footnote{{The percentage difference between ground (2D) distance and 3D one is always smaller than 2\%. Consequently, both distances almost overlap.}} in order to capture a wide set of exposure conditions. In addition, the measurement points are placed in the coverage area of each W3 sector shown in Fig.~\ref{fig:tower_installation}, in order to further strengthen our analysis. {We refer the reader to Appendix A, which provides detailed information about: \textit{i}) radio configuration of the W3 installation under consideration, \textit{ii}) other RBSs in the surroundings of the considered area, \textit{iii}) taxonomy of outoor locations, \textit{iv}) measurement time.}

\begin{figure}
\centering

	\subfigure[LOS Location - Engineering Parking]
	{
		\includegraphics[width=4cm]{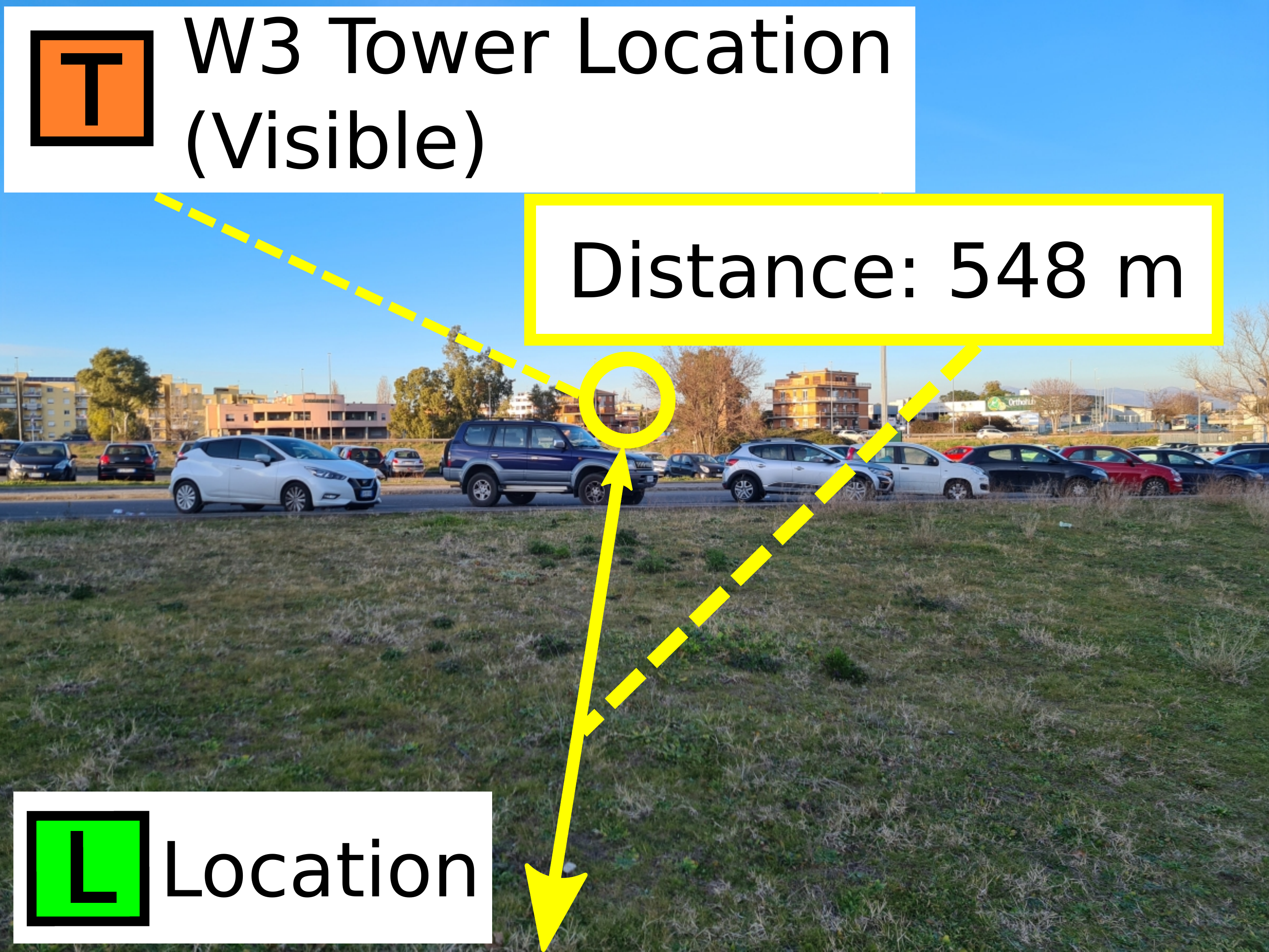}
		\label{fig:park_eng_rot}
	}
 	\subfigure[NLOS Location - Church Square]
	{
		\includegraphics[width=4cm]{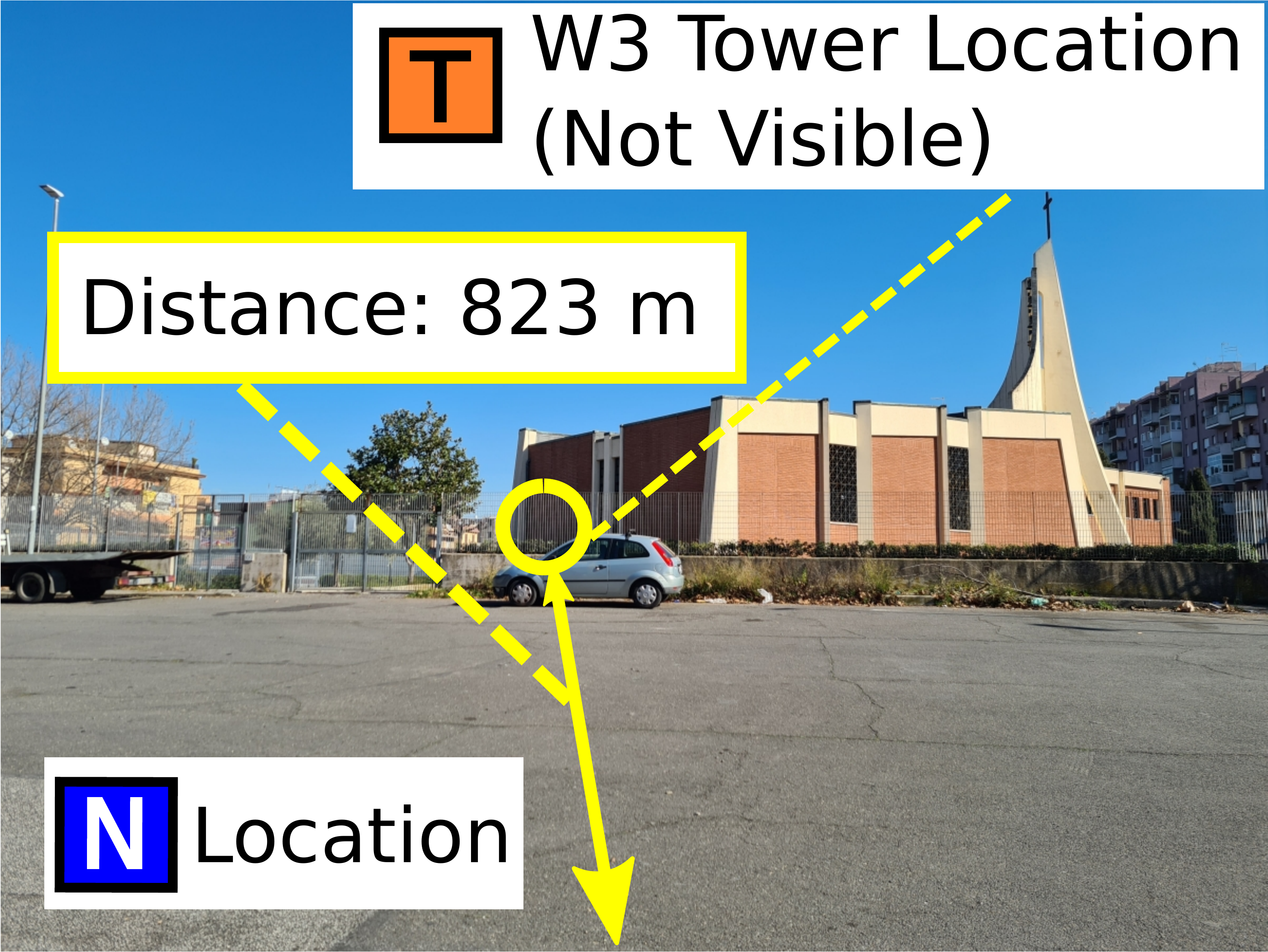}
		\label{fig:atletico_rot}

	}
\caption{Example of \ac{LOS} and \ac{NLOS} outdoor measurement locations.}
\label{fig:los_nlos_outdoor_example}
\end{figure}

To give more insights, Fig.~\ref{fig:los_nlos_outdoor_example} reports two representative examples of outdoor measurement points. When considering \ac{LOS} locations (Fig.~\ref{fig:park_eng_rot}), the \ac{RBS} is visible from the measurement point. On the contrary, the installation is not visible in \ac{NLOS} locations (Fig.~\ref{fig:atletico_rot}). In both cases, the directive antenna is pointed towards the  \ac{RBS} location during operations $M1$ and $M3$.

\begin{figure}[t]
\centering

\subfigure[Positioning of LOS and NLOS indoor locations]
{
	\includegraphics[width=4cm]{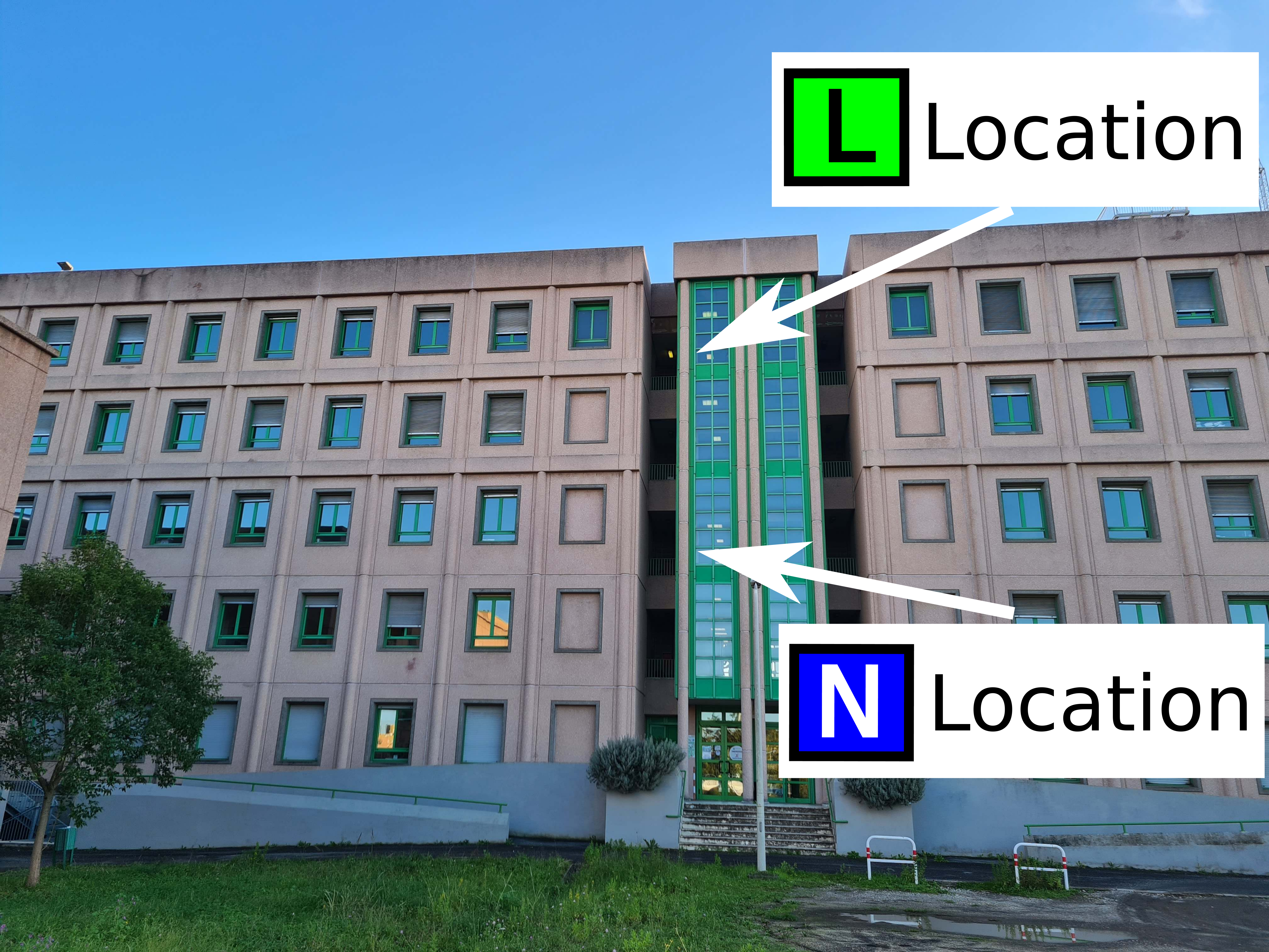}
	\label{fig:uni_test_view_external}
}
\subfigure[View of LOS indoor location]
{
	\includegraphics[width=4cm]{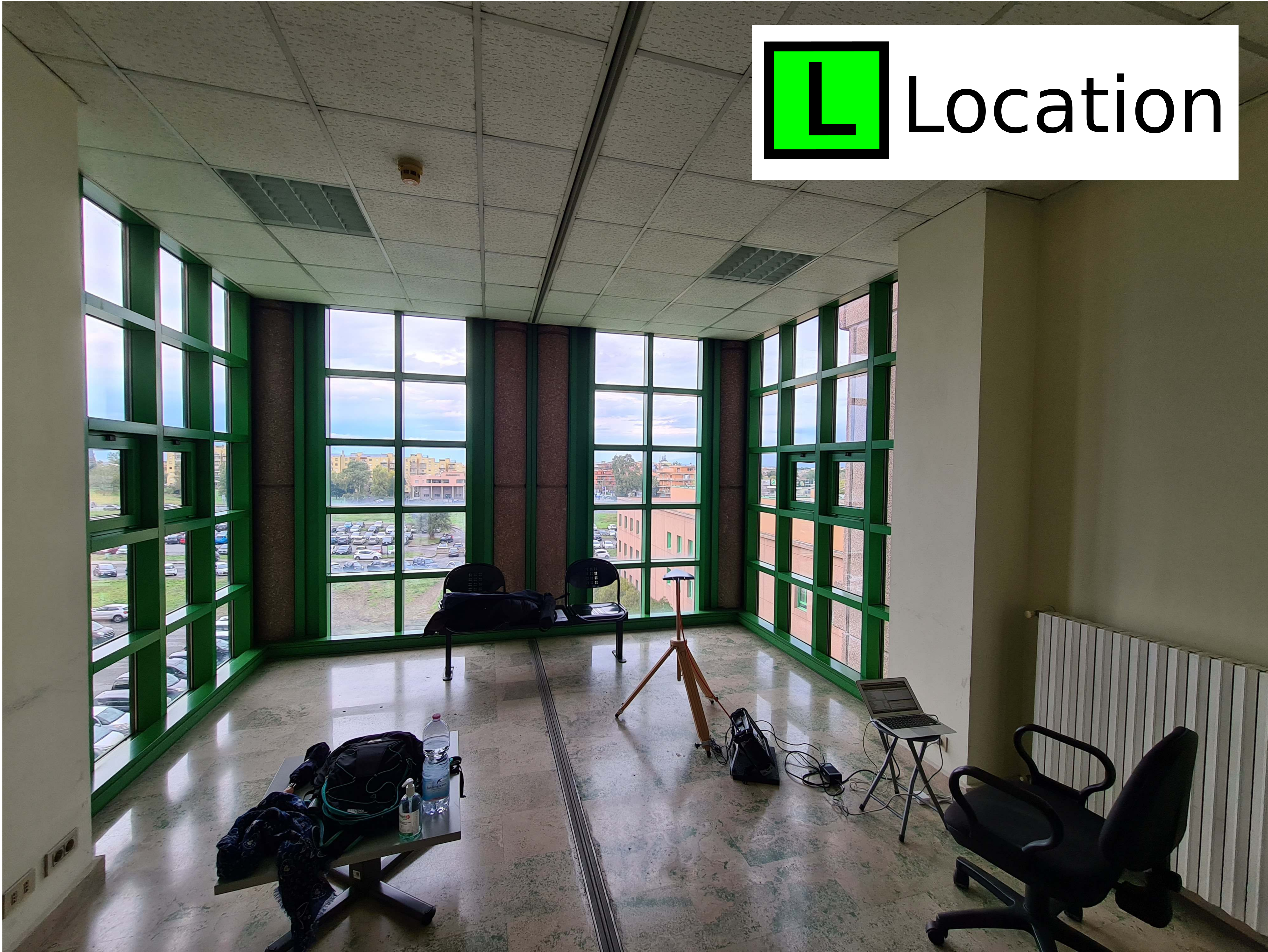}
	\label{fig:uni_test_view_internal}

}

\subfigure[Window view from LOS location]
{
	\includegraphics[width=4cm]{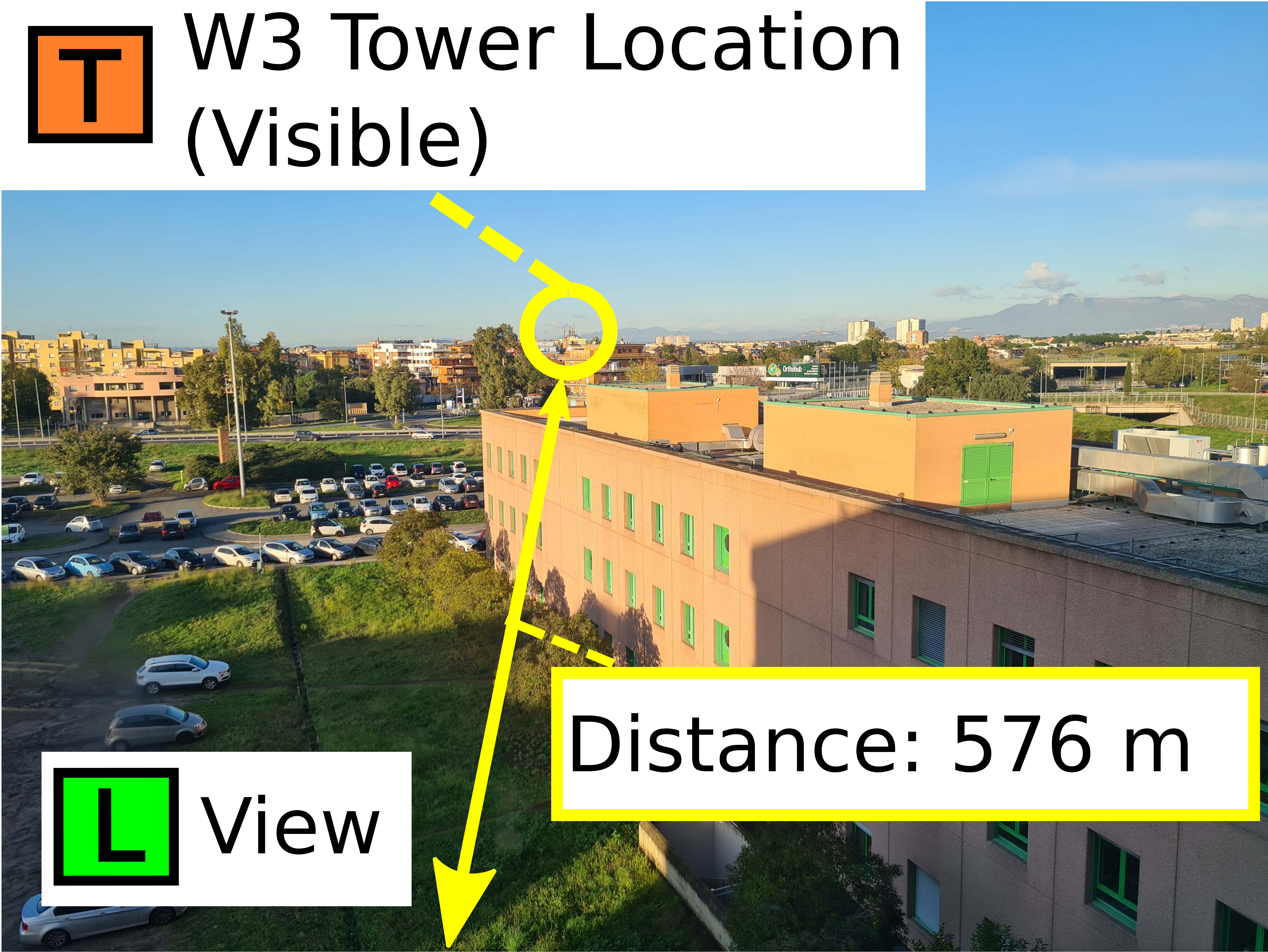}
	\label{fig:uni_test_view_LOS}
}
\subfigure[Window view from NLOS location]
{
	\includegraphics[width=4cm]{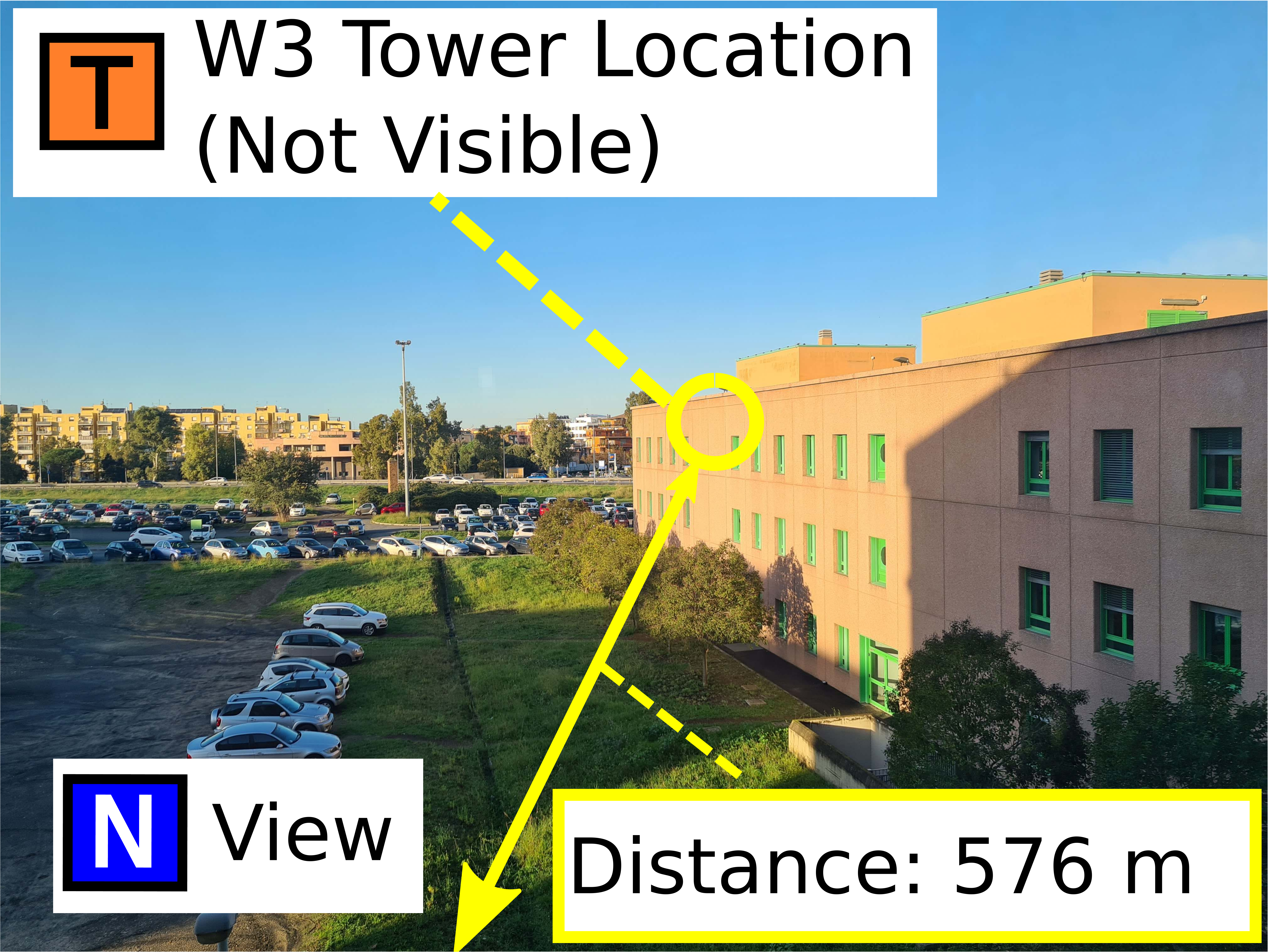}
	\label{fig:uni_test_view_NLOS}
}
\caption{LOS and NLOS indoor measurement locations.}
\label{fig:los_nlos_indoor}
\end{figure}

Focusing then on the indoor tests, we consider two locations at the Engineering building of the University, shown in Fig.~\ref{fig:uni_test_view_external}. In particular, we consider a \ac{LOS} location at the fourth floor and a \ac{NLOS} one at the second one. The room hosting the \ac{LOS} measurements is shown in Fig.~\ref{fig:uni_test_view_internal}. In addition, the environment hosting the \ac{NLOS} tests is identical to the \ac{LOS} one (not shown due to the lack of space). Interestingly, the walls are made of thin concrete pillars and big glasses that are mounted on small metallic frames. This structure provides in general good penetration of outdoor mobile signals inside the building. The window view from both locations is shown in Fig.~\ref{fig:uni_test_view_LOS}-Fig.~\ref{fig:uni_test_view_NLOS}. Focusing on the \ac{LOS} environment, the path towards the \ac{RBS} is free from obstacles and the distance is within the \ac{RBS} coverage area. Focusing instead on the \ac{NLOS} location, the distance from the \ac{RBS} is identical to the \ac{LOS} one, but, obviously, the \ac{RBS} sight is obstructed by a building, which forces the signal to follow a \ac{NLOS} path.


\subsection{Parameters Settings}

We provide herefater the settings of the main parameters of \textsc{5G-EA} framework. In particular, we shed light on the measurement antenna positioning and the algorithm parameters in $P1$, $P2$, and $P3$, respectively. {We refer the reader to Appendix B for more insights about calibration and uncertainty aspects of the measurement chain.}

\begin{figure}
\centering
\includegraphics[width=8cm]{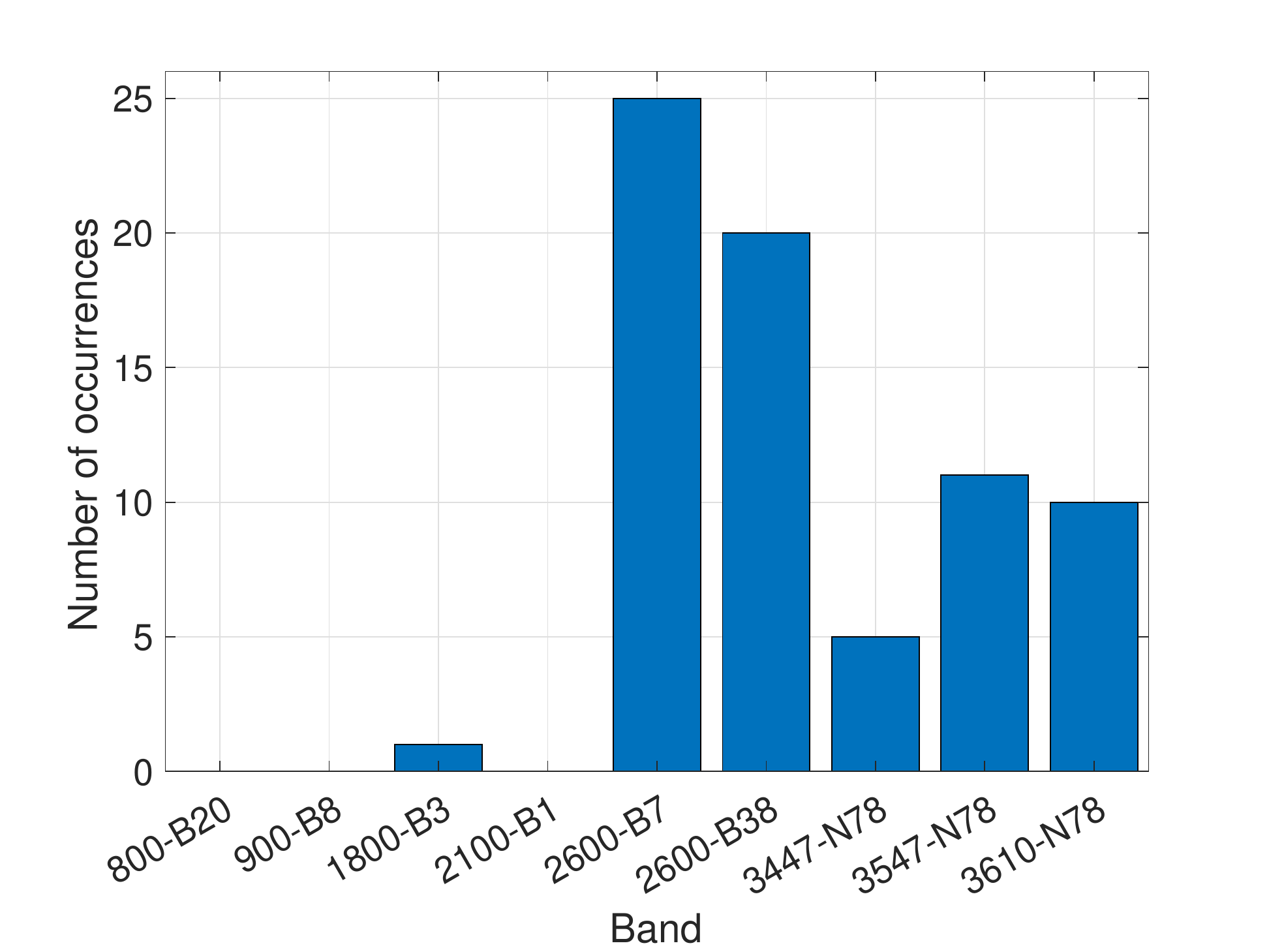}
\caption{Occurrence of bands used for the 4G/5G data transfers in the considered scenarios}
\label{fig:bw_used}
\end{figure}

\subsubsection{Measurement Antenna Placement} 

\textsc{5G-EA} requires a careful orientation and positioning of the measurement antenna, in order to properly dissect \ac{RBS} vs. smartphone exposure. Focusing on $M1$ and $M3$, the antenna is simply pointed towards the \ac{RBS} location. Focusing on $M2$, the antenna is pointed towards the smartphone. As already shown in Tab.~\ref{tab:far_field}, the far-field distance $D_f^{\text{FF}}$ has to be enforced in our experiments, in order to avoid near-field effects. On the other hand, the distance should mimic the actual exposure conditions to the head/chest that is experienced by a typical user. Therefore, there is trade-off between the (small) distance for a meaningful assessment and the (relatively large) distance that has to be enforced to preserve far-field conditions. In this work, we have found that a good compromise among such competing goals can be achieved by setting a distance from the smartphone equal to 0.25~[m]. Although this number may apparently violate the far-field conditions for the 800~[MHz] and 900~[MHz] bands (as shown in Tab.~\ref{tab:far_field}), in practice we have found that such bands are not used for 4G data transfers. 

To corroborate the previous outcome, {Fig.}~\ref{fig:bw_used} reports the occurrence of bands that are used for the data transfers in the outdoor locations of our experiments. Interestingly, most of  transfers employ 4G bands at around 2600~[MHz], while the 1800~[MHz] band is seldom used. On the other hand, the 800~[MHz] and 900~[MHz] bands are not used by the data transfers. Eventually, all the three 5G bands at 3.4-3.6~[GHz] are almost equally adopted. In this way, the minimum frequency used for the exposure assessment can be assumed to be the 1800~[MHz], band. Therefore, the distance of 0.25~[m] is sufficiently large to provide far-field conditions.

\begin{table}
\caption{Parameters of \texttt{adjust\_ref\_level\_scale\_div} routine ($P1$).}
\label{tab:adj_ref_level}
\scriptsize
\centering
\begin{tabular}{|m{0.3cm}|c|c|}
\hline
\rowcolor{Grayblue} & \textbf{Parameter} & \textbf{Value/Setting}\\
\hline
\rowcolor{Linen} & Unit & dBm/m$^2$ \\
\rowcolor{Linen}& Attenuation & Auto\\
\rowcolor{Linen}& Resolution Bandwidth & Auto\\
\rowcolor{Linen}& Video Bandwidth & Auto\\
\rowcolor{Linen}& Number of sweep points & Auto\\
\rowcolor{Linen}& Trace Detector & Root Mean Square (RMS) \\
\rowcolor{Linen}& Type Detector & Rolling Max \\
\rowcolor{Linen}& Initial Reference Level & 65 dBm/m$^2$\\
\rowcolor{Linen} \multirow{-8}{*}{\begin{sideways}\texttt{SAN\_settings}\end{sideways}} & Initial Scale/Div & 15\\
\hline
& \texttt{curr\_f\_min} & \multirow{2}{*}{Fig.~\ref{fig:frequencies_overview} (\ac{FDD} \ac{DL}, \ac{TDD})} \\ 
& \texttt{curr\_f\_max} &  \\
& \texttt{pre\_amp\_state} & Fig.~\ref{fig:env_steps_base_station}\\ 
& \texttt{min\_l} & Tab.~\ref{tab:min_level}\\
& \texttt{pre\_amp\_thre} & -48.77 dBm/m$^2$\\
& \texttt{safety\_margin} & 10 dBm/m$^2$ \\
& \texttt{max\_time\_search} & 5~s\\
\multirow{-7}{*}{\begin{sideways}Param. Settings\end{sideways}}& \texttt{y\_ticks} & 10\\
\hline
\end{tabular}
\end{table}

\begin{table}
\caption{\texttt{min\_l} values as a function of frequency and pre-amplifier state.}
\label{tab:min_level}
\scriptsize
\centering
\begin{tabular}{|c|>{\columncolor{Linen}}m{2.1cm}|m{2.1cm}|}
\hline
\rowcolor{Grayblue}  & \multicolumn{2}{c|}{\texttt{min\_l}} \\
\cline{2-3}
\rowcolor{Grayblue} \multirow{-2}{*}{\texttt{curr\_f\_min}} & \texttt{pre\_amp\_state}=0 (inactive) & \texttt{pre\_amp\_state}=1 (active)\\
\hline
791, 832, 950, 905~[MHz] & -100~[dBm/m$^2$] & -115~[dBm/m$^2$]\\
1840, 1745, 2110, 1920~[MHz] & -95~[dBm/m$^2$] & -107~[dBm/m$^2$]\\
2670, 2550, 2570~[MHz] & -90~[dBm/m$^2$] & -103~[dBm/m$^2$]\\
3437, 3537, 3600~[MHz] & -85~[dBm/m$^2$] & -97~[dBm/m$^2$]\\
\hline
\end{tabular}
\end{table}

\subsubsection{$P1$ Parameters} 
\label{subsubsec:p1_parameters}

Tab.~\ref{tab:adj_ref_level} reports the parameters of the \texttt{adjust\_ref\_level\_scale\_div} function. In more detail, the upper part of the table expands the basic \ac{SAN} parameters. In particular, we adopt a rolling max type detector, as our goal here is to first sense the maximum signal levels and then adjust accordingly reference level and scale division. Focusing  on the other routine parameters (bottom part of the table), the values of \texttt{curr\_f\_min} and \texttt{curr\_f\_max} are taken from Fig.~\ref{fig:frequencies_overview}, by considering the W3 bands over \ac{FDD} \ac{DL} and \ac{TDD} - since the exposure from \ac{RBS} is the target of $P1$. In addition, the pre-amplifier state (on/off) is governed by the logic reported in Fig.~\ref{fig:env_steps_base_station}. Obviously, the pre-amplifier is inactive when the first call of \texttt{adjust\_ref\_level\_scale\_div} is run (left part of Fig.~\ref{fig:env_steps_base_station}). However, in case the pre-amplification management branch is followed (right part of Fig.~\ref{fig:env_steps_base_station}), the \texttt{pre\_amp\_state} state that is passed to  \texttt{adjust\_ref\_level\_scale\_div} may be active. 

Focusing on the remaining parameters of \texttt{adjust\_ref\_level\_scale\_div}, the \texttt{min\_l} matrix is reported in Tab.~\ref{tab:min_level}. The values reported in the table are retrieved by visualizing the noise level on the \ac{SAN} in each considered band.
Clearly, when the pre-amplifier is turned on, the noise level can be notably reduced (right part of the table). Moreover, the \texttt{pre\_amp\_thre} threshold,  which is used in the decision block of Fig.~\ref{fig:env_steps_base_station} to compare the reference level and activatate/deactivate the pre-amplification,  is set to -48.77 dBm/m$^2$ (a setting that depends on the \ac{SAN} \ac{HW} and the features of the directive antenna). In addition, the \texttt{safety\_margin} parameter is set to 10~[dBm/m$^2$] - an empirical value that was tested to correctly work on all the considered bands. Finally, the maximum time for searching the signal peak over the considered band is set to 5~[s], while the number of y ticks is set to \texttt{y\_ticks}=10. In this way, the time required to run a single call of \texttt{adjust\_ref\_level\_scale\_div} is at least equal to 5~[s].
 
\begin{table}
\caption{SAN and parameter settings for the \texttt{nar\_band\_meas} routine of $P1$, $P2$ and $P3$.}
\label{tab:ch_pow_meas}
\scriptsize
\centering
\begin{tabular}{|m{0.05cm}|m{2.1cm}|m{2.5cm}|m{2cm}|}
\hline
\rowcolor{Grayblue} &  & \multicolumn{2}{c|}{\textbf{Value/Setting}}\\[0.1mm]
\cline{3-4}
\rowcolor{Grayblue} & \multirow{-2}{*}{\textbf{Parameter}} & & \\[-2mm]
\rowcolor{Grayblue} & & \multicolumn{1}{c|}{$P1$} & \multicolumn{1}{c|}{$P2$, $P3$} \\[0.1mm]
\hline
\rowcolor{Linen} & & & \\[-2mm]
\rowcolor{Linen}& Unit & \multicolumn{1}{c|}{dBm/m$^2$} & \multicolumn{1}{c|}{V/m} \\[0.1mm]
\cline{2-4}
\rowcolor{Linen}& & \multicolumn{2}{c|}{} \\[-2mm]
\rowcolor{Linen}& Attenuation & \multicolumn{2}{c|}{Auto}\\[0.1mm]
\cline{2-4}
\rowcolor{Linen}& & \multicolumn{2}{c|}{} \\[-2mm]
\rowcolor{Linen}& Resolution Bandwidth & \multicolumn{2}{c|}{Auto} \\[0.1mm]
\cline{2-4}
\rowcolor{Linen}& & \multicolumn{2}{c|}{} \\[-2mm]
\rowcolor{Linen}& Video Bandwidth & \multicolumn{2}{c|}{Auto} \\[0.1mm]
\cline{2-4}
\rowcolor{Linen}& & \multicolumn{2}{c|}{} \\[-2mm]
\rowcolor{Linen}& Sweep points & \multicolumn{2}{c|}{Auto}\\[0.1mm]
\cline{2-4}
\rowcolor{Linen}& & \multicolumn{2}{c|}{} \\[-2mm]
\rowcolor{Linen}& Trace Detector & \multicolumn{2}{c|}{Root Mean Square (RMS)} \\[0.1mm]
\cline{2-4}
\rowcolor{Linen}& & \multicolumn{2}{c|}{} \\[-2mm]
\rowcolor{Linen}& Type Detector & \multicolumn{2}{c|}{Rolling Average} \\[0.1mm]
\cline{2-4}
\rowcolor{Linen}& & \multicolumn{2}{c|}{} \\[-2mm]
\rowcolor{Linen}& Avg. samples & \multicolumn{2}{c|}{100} \\[0.1mm]
\cline{2-4}
\rowcolor{Linen}& & & \\[-2mm]
\rowcolor{Linen}& Reference Level & Set by the logic in Fig.~\ref{fig:env_steps_base_station} & 6~V/m \\[0.1mm]
\cline{2-4}
\rowcolor{Linen}& & & \\[-2mm]
\rowcolor{Linen}\multirow{-13}{*}{\begin{sideways}\texttt{SAN\_settings}\end{sideways}} & Scale/Div & Set by the logic in Fig.~\ref{fig:env_steps_base_station} & Automatic\\[0.1mm]
\hline
& \texttt{curr\_f\_min} & \multirow{2}{*}{Fig.~\ref{fig:frequencies_overview} (\ac{FDD} \ac{DL}, \ac{TDD})} & Extracted from  \\[0.1mm]
& \texttt{curr\_f\_max} &  & \texttt{sel\_band\_array} \\[0.1mm]
\cline{2-4}
& & \multicolumn{2}{c|}{}  \\[-2mm]
& \texttt{n\_samples} & \multicolumn{2}{c|}{12} \\[0.1mm]
\cline{2-4}
& & \multicolumn{2}{c|}{}  \\[-2mm]
 \multirow{-4.5}{*}{\begin{sideways}Parameters\end{sideways}} & \texttt{int\_sample\_time} & \multicolumn{2}{c|}{0.5~s} \\
\hline
\end{tabular}
\end{table}

Tab.~\ref{tab:ch_pow_meas} reports the parameters for the \texttt{nar\_band\_meas} function. Focusing on the $P1$ settings (central column), the set of \texttt{SAN\_parameters} this time includes a rolling average as type detector, as our primary goal during this step is to perform an exposure assessment over the considered signal. This setting is inline with relevant measurement standards in the field (see e.g., \cite{iec1}). In addition, the number of samples for computing the average is set to 100, in order to consider a meaningful range. Clearly, the reference level and the scale division are updated by the logic implemented in Fig.~\ref{fig:env_steps_base_station}. Focusing then on the remaining parameters, \texttt{curr\_f\_min} and \texttt{curr\_f\_max} are set in accordance to the set of W3 bands shown in Fig.~\ref{fig:frequencies_overview} (restricted to \ac{FDD} \ac{DL} and \ac{TDD}). Finally, the number of narrow band measurements \texttt{n\_samples} is set to 12, while the time between consecutive queries of narrow band measurements is set to 0.5~[s]. In this way, the measurement time for each band is approximately equal to 12 $\times$ 0.5~[s] = 6~[s]. 


Finally, we shed light on the remaining parameters that appear in Fig.~\ref{fig:env_steps_base_station}. Focusing on the maximum number of iterations for running the \texttt{adjust\_ref\_level\_scale\_div} function, we set it to 3 - a value that provides a good balance between overall duration of detection phase and precision in setting reference level and scale division values. Focusing on the number of bands to be monitored, we set it to 9, in accordance to the \ac{FDD} \ac{DL} and \ac{TDD} bands of W3 shown in Fig.~\ref{fig:frequencies_overview}. With such settings, the required time to run $P1$ is at least equal to (5~[s] $\times$ 3 + 6~[s]) $\times$ 9 = 189~[s]. However, the actual total time for running $P1$ may be higher, due to the following reasons: \textit{i}) the additional delay that is required when communicating with the \ac{SAN} and \textit{ii}) the eventual activation of the power amplifier, which requires further calls of the  \texttt{adjust\_ref\_level\_scale\_div} routine.


\subsubsection{$P2$ Parameters} 

\begin{table}[t]
\caption{SAN settings for wide span measurements $P2$.}
\label{tab:wide_span_meas}
\scriptsize
\centering
\begin{tabular}{|c|c|}
\hline
\rowcolor{Grayblue} \textbf{Parameter} & \textbf{Value/Setting}\\
\hline
Unit & V/m \\
Attenuation & Auto\\
Pre-amplifier & Off \\
Frequency Start & 791~MHz \\
Frequency Stop & 3620~MHz \\
Resolution Bandwidth & Auto\\
Video Bandwidth & Auto\\
Sweep Points & Auto\\
Trace Detector & Root Mean Square (RMS) \\
Type Detector & Max. Detector \\
Reference Level & 6~V/m\\
Scale/Div & Automatic\\
\hline
\end{tabular}
\end{table}

We initially consider the parameters of the wide span measurement blocks reported in Fig.~\ref{fig:active_traffic_steps_smartphone}, detailed in Tab.~\ref{tab:wide_span_meas}. More in depth, most of \ac{SAN} parameters are set to the default values (i.e., automatic setting). Focusing on the remamining parameters, the selected frequencies cover all the ones in use by W3 operator. In addition, the pre-amplifier is powered off, as the exposure from the smartphone may be potentially higher than the maximum supported signal level with pre-amplification turned on - which we remind is a very effective feature to distinguish low signals w.r.t. noise level. Moreover, the reference level is set to a large value (6~[V/m]) in order to detect possible signal peaks. Focusing on the trace detector, a \ac{RMS} setting is imposed (in line with $P1$). Eventually, the max detector is used, as we remind that the scope of the wide span measurement is to perform a quick scan over the entire frequency range and to detect the signal peaks.

The second block of $P2$ is the activation of the smartphone traffic exposure. Unless otherwise specified, the \texttt{iperf} client is run on the smartphone with the following parameters: \textit{i}) \ac{IP} address and port corresponding to the \texttt{iperf} server installed at the University, \textit{ii}) bandwidth report interval set to 1~[s], \textit{iii}) number of simultaneous connections per data transfer set to 1, \textit{iv}) maximum duration of \texttt{iperf} transfer set to 120~[s] - an amount of time sufficiently large to complete the remaining steps in $P2$ and $P3$.

In the following, we focus on the parameters for selecting the 4G/5G bands in use by the data transfer (expanded in the \texttt{sel\_band\_use} routine of Tab.~\ref{alg:sel_band_use}). Obviously, the array of \ac{EMF} measurements are set by the first and second wide span assessment. The array of frequency values \texttt{freq\_array} includes all the frequencies in use by W3 operator. Finally, the treshold increase parameter \texttt{thre\_inc} is set to 30\% - a value that guarantees a good balance between (artificial) increase of \ac{EMF} due to injection of traffic from the smartphone and (natural) variation of exposure due to other effects (e.g., nearby terminals, signal fading, etc.).

\begin{figure*}[t]
\centering
 	\subfigure[Exposure Breakdown]
	{
		\includegraphics[width=6cm,angle=-90]{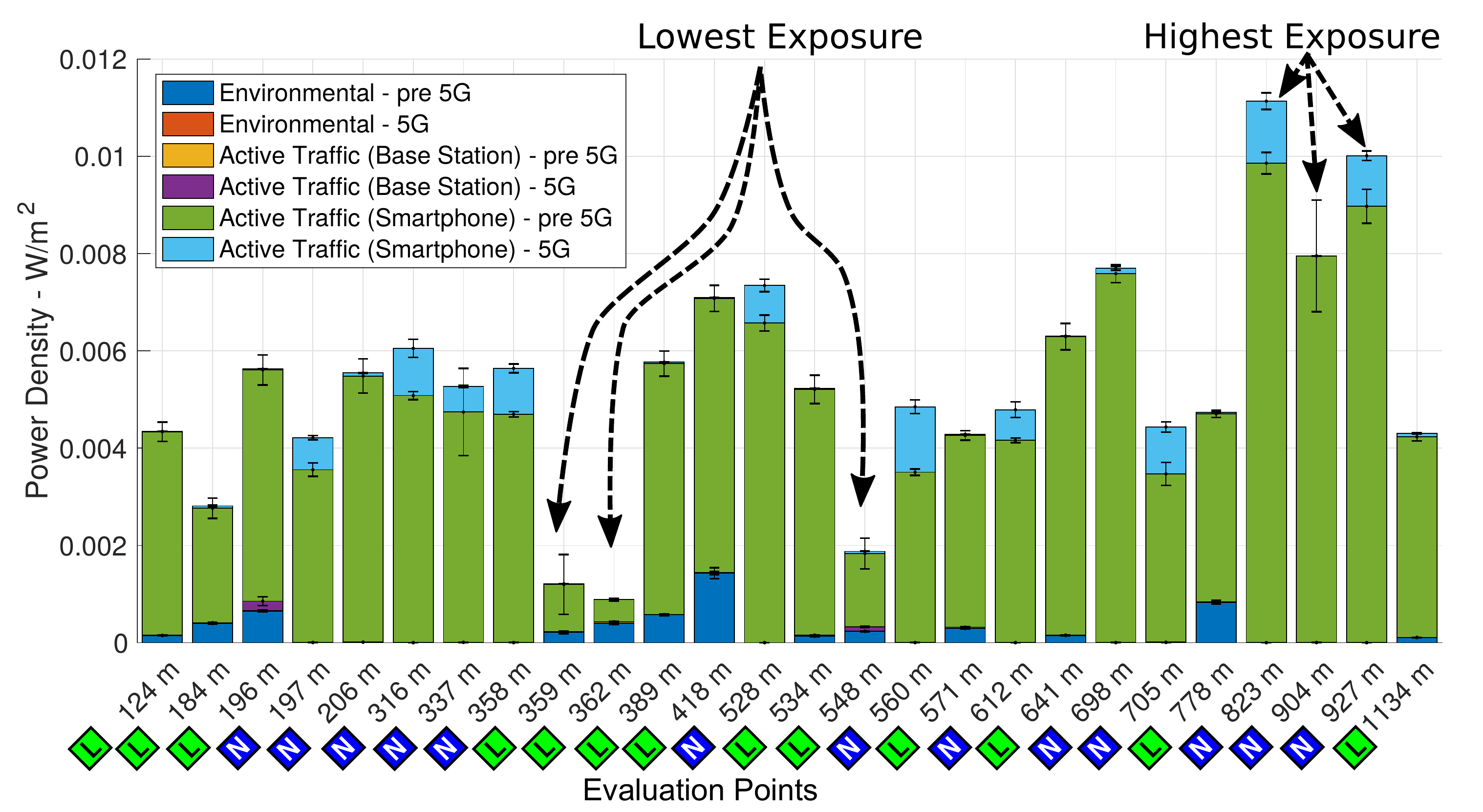}
		\label{fig:pd_details_locations}

	}
 	\subfigure[Throughput Breakdown]
	{
		\includegraphics[width=6cm,angle=-90]{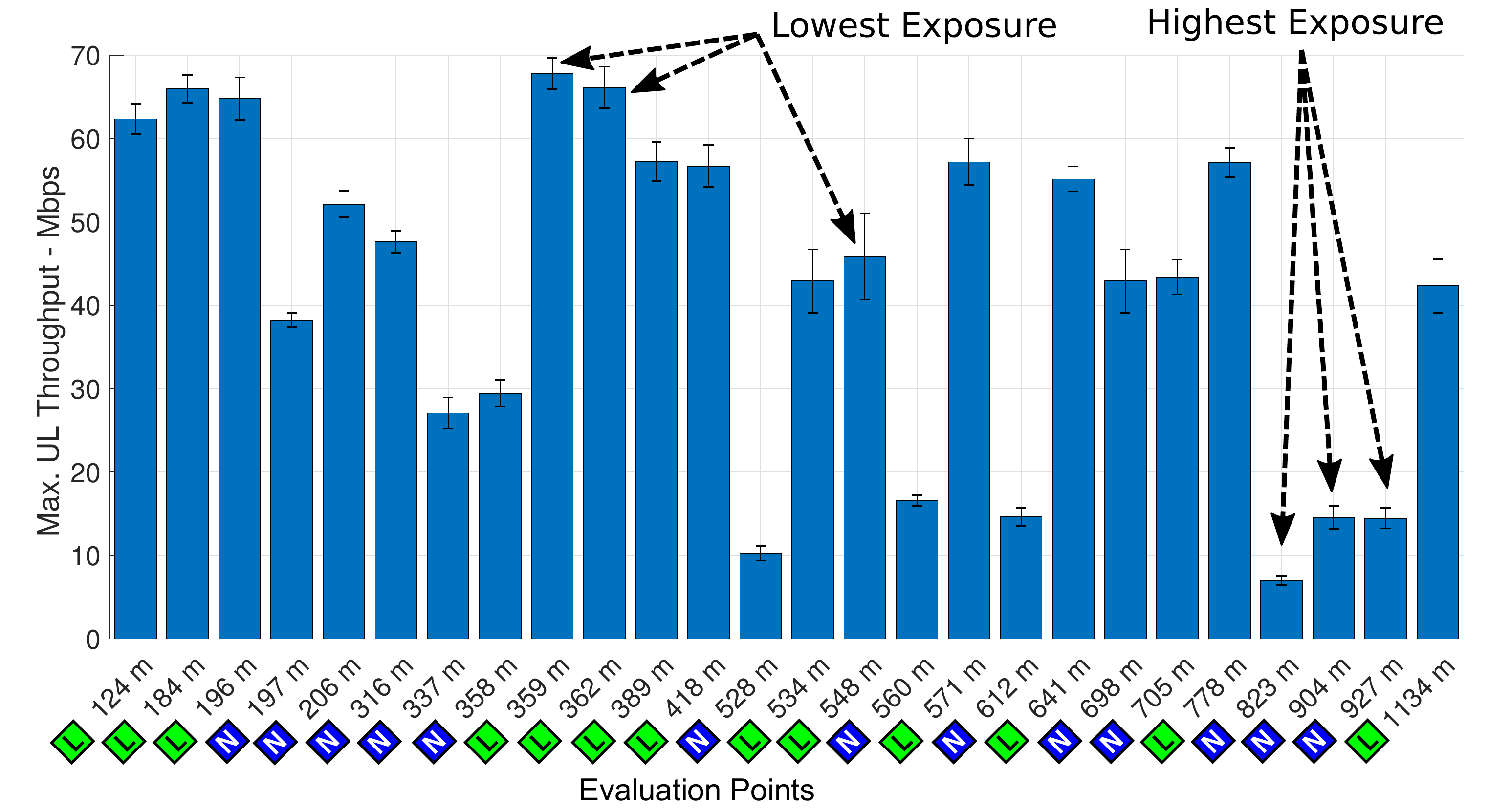}
		\label{fig:throughput_details_locations}
	}
	\caption{Characterization of exposure (top) and throughput (bottom) over the outdoor evaluation points. The points are ordered from left to right by increasing distance values with respect to the 5G W3 installation. The sight condition (L=Line-of-Sight, N=Non-Line-Of-Sight) is also reported.}
	\label{fig:exposure_throughput_variation}
\end{figure*}

Eventually, the band index in Fig.~\ref{fig:active_traffic_steps_smartphone} is initialized with the first \ac{UL} bandwidth in which an increase of exposure was detected by the \texttt{sel\_band\_use} routine. Finally, the narrow band \ac{EMF} measurement is realized through the \texttt{nar\_band\_meas} of Alg.~\ref{alg:nar_band_meas}, whose parameters for $P2$ are detailed in Tab.~\ref{tab:ch_pow_meas} on the right. The main difference w.r.t. $P1$ case relies on a different measurement metric,  expressed in terms of [V/m]. In addition, the reference level set to 6~[V/m], since the measured signal levels are expected to be non-negligible.  It is interesting to note that, when the [V/m] metric is set, the scale div are automatically tuned to show a minimum level of 0~[V/m], i.e., the minimum one. Eventually, the minimum and maximum frequency are set in accordance to the considered bandwidth, whose set is saved in the \texttt{sel\_band\_array} by the \texttt{sel\_band\_use} routine.

\subsubsection{$P3$ Parameters} 

Focusing on the $P3$ assessment shown in Fig.~\ref{fig:active_traffic_steps_base_station}, the initial band index is set with the first \ac{DL} bandwidth of W3 in use by the data transfer (detected by \texttt{sel\_band\_use} function). The narrow band assessment reported in the second block of $P3$ adopts the same parameters of $P2$ (detailed in Tab.~\ref{tab:ch_pow_meas} on the right). Finally, the \texttt{iperf} transfer is turned off on the smartphone when all the bands in use have been considered.

\subsection{Exposure Assessments}

We initially concentrate on the outcomes from outdoor measurements of Fig.~\ref{fig:base_station_measurement_points} and then we shed light on the results obtained in the indoor locations of Fig.~\ref{fig:los_nlos_indoor}.

\subsubsection{Outdoor measurements}

We run \textsc{5G-EA} over the outdoor locations, by considering the generation of \ac{UL} traffic from the smartphone to the \texttt{iperf} server. Fig.~\ref{fig:exposure_throughput_variation} reports the breakdown of exposure (top) and throughput (bottom). The exposure is expressed in terms of power density~[W/m$^2$], in order to display the different contributions (\ac{RBS} vs. smartphone, active vs. environmental, pre-5G vs. 5G) over a stacked bar. Each exposure component is expressed in terms of average value over the collected samples. Moreover, the error bars report the confidence intervals, which are computed by assuming a Gaussian distribution with a confidence level of 95\%. 

We initially focus on the collected exposure values, shown in Fig.~\ref{fig:pd_details_locations}. Several considerations hold by analyzing in detail the figure. First, the active traffic exposure from the smartphone (pre-5G and 5G) dominates over all the other ones, in all the considered locations. Second, the \ac{RBS} environmental exposure can be identified for all locations in \ac{LOS} w.r.t. the \ac{RBS}, while the same metric is negligible for all locations in \ac{NLOS}. Third, the contribution of active traffic exposure from the \ac{RBS} is almost imperceptible in all locations (except from two). Fourth, the majority of active traffic exposure from the smartphone is due to pre-5G contributions (mostly 4G), while 5G always represents a small share (at most equal to {38}\%) compared to {the total one that is radiated by the smartphone}.  Fifth, \ac{NLOS} locations generally present higher level of 5G exposure than \ac{LOS} ones. Sixth, the increase of distance generally results in an increase of exposure (left to right of the figure). However, the largest exposure variations are observed between \ac{LOS} and \ac{NLOS} evaluation points. In particular, the latters exhibit a strong increase of active traffic exposure from the smartphone compared to the formers. {As a side comment, the measured expsure levels are always orders of magnitude lower than the whole body and localized maximum power density values of \ac{ICNIRP} guidelines} \cite{international2020guidelines}.

{In the following step}, we compare the exposure of Fig.~\ref{fig:pd_details_locations} against the achieved throughput levels shown in Fig.~\ref{fig:throughput_details_locations}. Interestingly, a strong variation in the throughput levels is observed. We argue that this phenomenon is due to the different propagation conditions that are {experienced} in the measurement locations. To substantiate such observation, Fig.~\ref{fig:pd_details_locations} highlights the three locations exhibiting the lowest exposure and the other three ones providing the highest exposure levels.  Interestingly,  the formers are in \ac{LOS}, while the latters experience \ac{NLOS}. When considering the throughput metric for the same locations (Fig.~\ref{fig:throughput_details_locations}), we can note that locations with lowest exposure (\ac{LOS}) achieve very large throughput levels, typically larger than 45~[Mbps] in the \ac{UL}, while the opposite holds for locations experiencing the highest exposure levels (\ac{NLOS}), being the observed throughput lower than 16~[Mps].  Consequently,  \ac{NLOS} conditions are reflected into an increase of smartphone exposure \textit{and} a degradation of throughput levels compared to \ac{LOS} ones.\footnote{{We refer the interested reader to Appendix C for more speculations about the variations of exposure for smaller distances than the minimum one considered in this work.}}

\begin{figure}[t]
\centering
 	\subfigure[Percentage of Smartphone Exposure vs. UL Throughput]
	{
		\includegraphics[width=6.5cm]{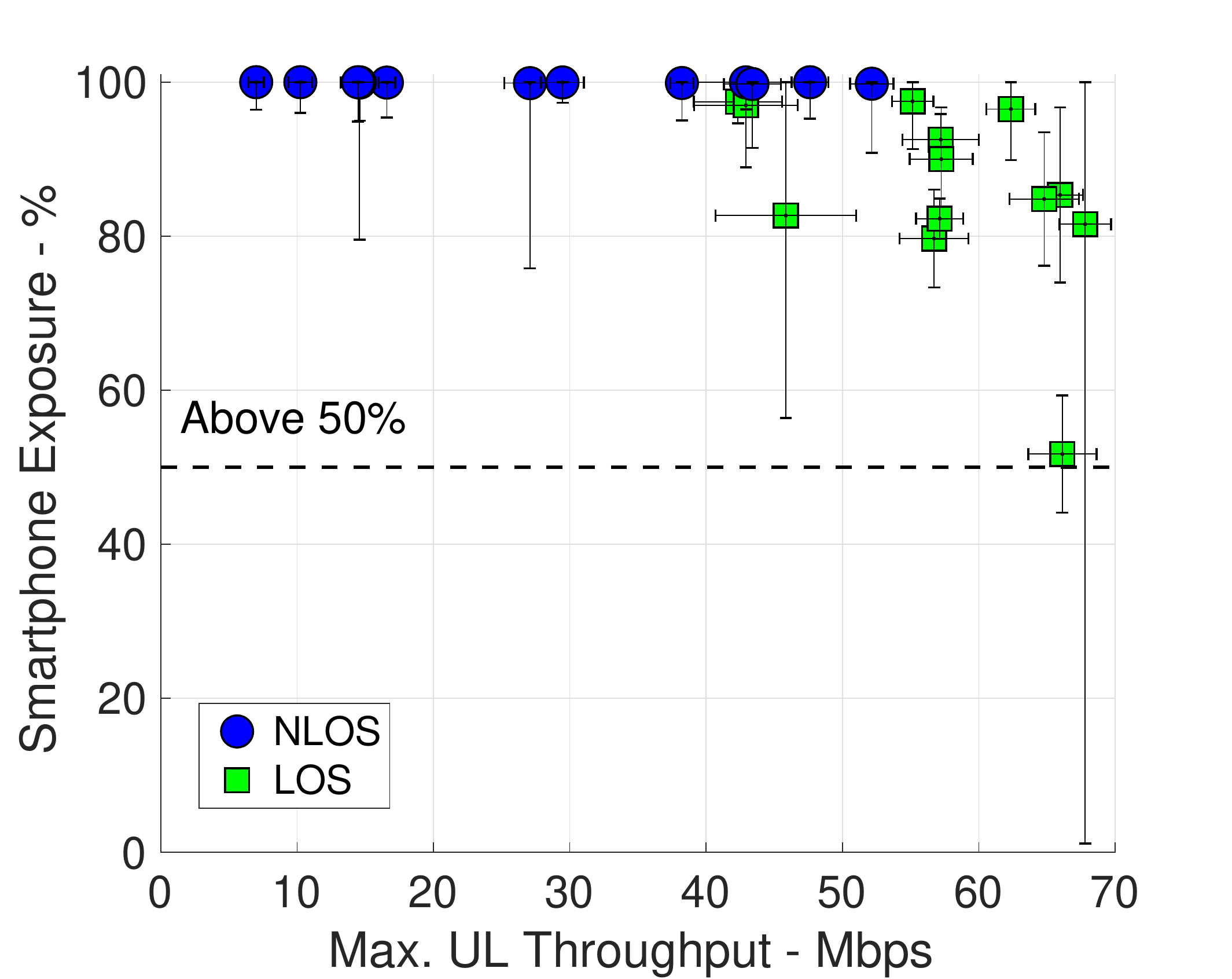}
		\label{fig:smart_exp_perc}

	}
 	\subfigure[Smartphone Exposure-per-Mbps vs. UL Throughput]
	{
		\includegraphics[width=6.5cm]{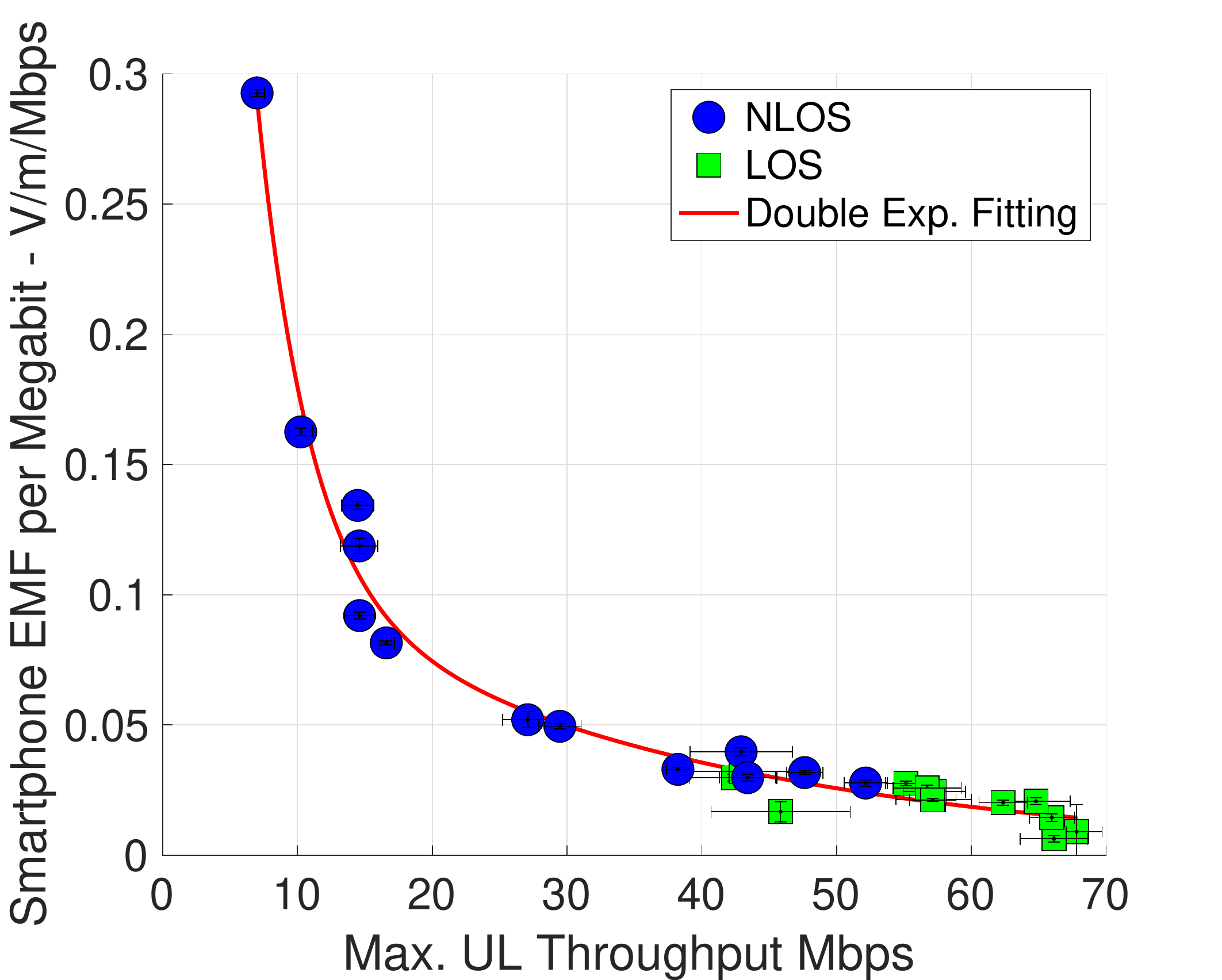}
		\label{fig:smart_exp_cost}
	}
	\caption{Incidence of \ac{UL} throughput on the smartphone active traffic exposure percentage (top) and on exposure-per-Mbps (bottom)}
	\label{fig:smart_exp_perc_costs}
\end{figure}

To provide more insights, Fig.~\ref{fig:smart_exp_perc} reports the percentage of active traffic exposure from the smartphone (w.r.t. the total one) vs. the observed {throughput} level. Each point in the figure corresponds to a measurement location (distinguished between \ac{LOS} and \ac{NLOS}), while x-y error bars are computed by assuming again 95\% of confidence levels. Interestingly, we can note that the percentage of smartphone exposure is huge (close to 100\%) for all the \ac{NLOS} measurement locations. On the contrary, the percentage of smartphone exposure tends to decrease to lower levels for the \ac{LOS} measurement locations. Moreover, a decrease is also observed when the realized \ac{UL} throughput increases. In all the cases, however, the active traffic exposure from the smartphone is always higher than 50\%, thus representing the largest source of exposure.

Having understood that there may be a strong {relationship between} the realized \ac{UL} throughput and the collected exposure levels, we compute a novel metric, called smartphone exposure-per-Mbps, which is obtained by dividing the total exposure measured in the location by the observed throughput. The metric expresses the efficiency in terms of exposure (in [V/m]) for delivering a given amount of information (in [Mbps]). When the exposure-per-Mbps is high, the system is largely inefficient, as a huge exposure is needed to transfer the information. On the contrary, when the exposure-per-Mbps is low, the efficiency of the system in delivering the same amount of information is improved. 

\begin{table}[t]
\caption{Fitting parameters for the exposure estimator.}
\label{tab:params_fitting}
\scriptsize
\centering
\begin{tabular}{|c|c|}
\hline
\rowcolor{Grayblue} \textbf{Parameter} & \textbf{Value}\\
\hline
$F_1$ & 1.146~[V/m/Mbps]\\
$E_1$ & -0.2595~[Mbps$^{-1}$]\\
$F_2$ & 0.1304~[V/m/Mbps]\\
$E_2$ & -0.0325~[Mbps$^{-1}$]\\
\hline
\end{tabular}
\vspace{-3mm}
\end{table}

Fig.~\ref{fig:smart_exp_cost} reports the smartphone exposure-per-Mbps vs. the observed throughput levels. Interestingly, the exposure-per-Mbps is inversely proportional to the throughput levels. The higher is the throughput, the flatter and {closer to zero} is the observed smartphone exposure-per-Mbps. On the contrary, the lower is the throughput, the higher is the asymptotic behavior of the exposure-per-Mbps, with the highest {values} observed for the lowest throughput levels. To better capture the aforementioned effects, we have applied the following double exponential fitting model:
\begin{equation}
\label{eq:fitting}
C^{\text{EST}}=F_1 \cdot e^{E_1 \cdot T^{\text{UL}}} + F_2 \cdot e^{E_2 \cdot T^{\text{UL}}}
\end{equation}
where $T^{\text{UL}}$ is the observed throghput level (in Mbps), $F_1$, $E_1$, $F_2$, $E_2$ are the fitting parameters (shown in Tab.~\ref{tab:params_fitting}) and $C^{\text{EST}}$ is the estimated smartphone exposure-per-Mbps.

\begin{figure*}[t]
\centering
\includegraphics[width=7.5cm,angle=-90]{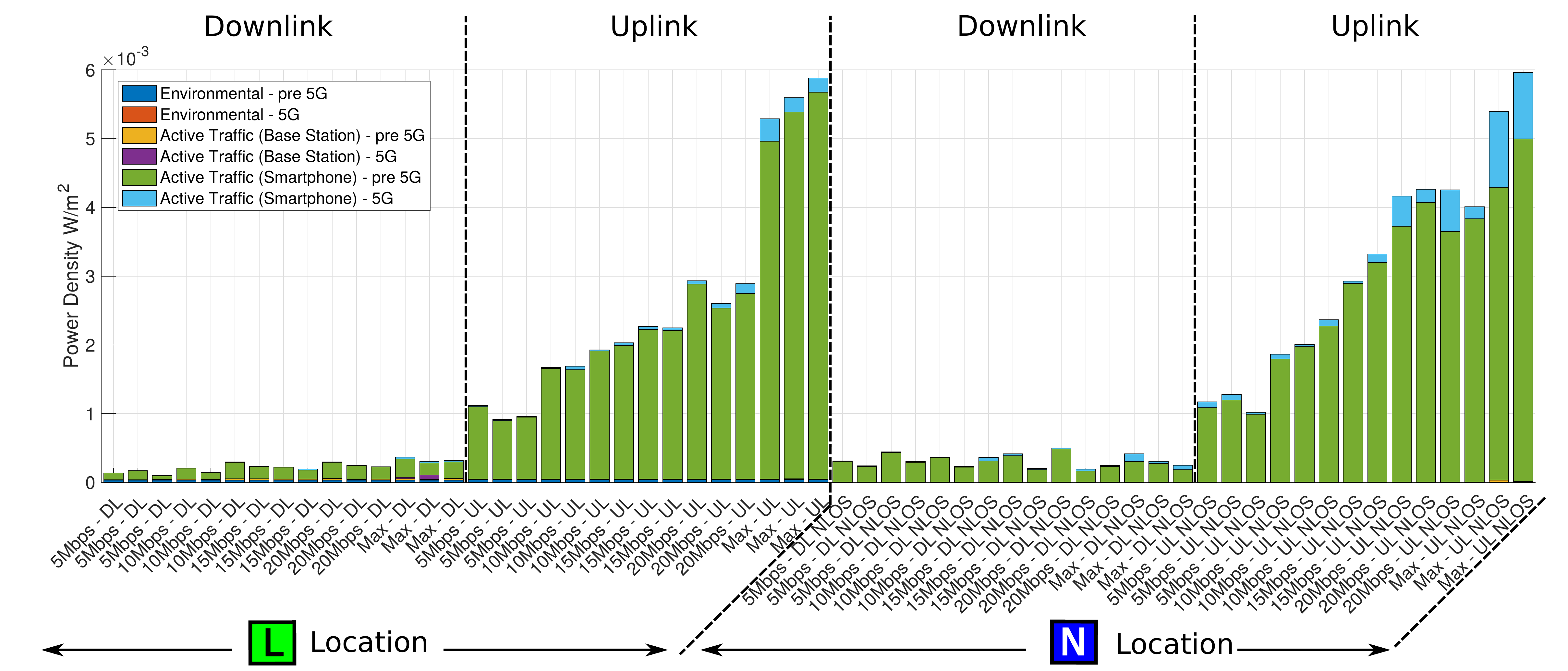}
\caption{ \ac{LOS}/\ac{NLOS} indoor locations: breakdown of exposure components vs. variation of \ac{UL}/\ac{DL} throughput.}
\label{fig:indoor_power_density}
\vspace{-3mm}
\end{figure*}

By observing in detail Fig.~\ref{fig:smart_exp_cost} we can note that the realized \ac{UL} throughput {with \texttt{iPerf} tool} can be used as an estimator of the smartphone exposure. In a practical scenario, the user could measure $T^{\text{UL}}$ {by running an \texttt{iPerf} test in the \ac{UL} direction and} a given location. Then, the smartphone exposure could be retrieved by: \textit{i}) applying the fitting model of Eq.~(\ref{eq:fitting}) to compute $C^{\text{EST}}$, \textit{ii}) computing the total estimated exposure as $C^{\text{EST}}\times T^{\text{UL}}$. Clearly, the values reported in Tab.~\ref{eq:fitting} may depend on different metrics (like the smartphone model), whose impact on the fitting model (and consequently on the exposure levels) require further investigations as future work.\footnote{{The adoption of additional metrics (derived e.g., from control channels) to predict exposure levels is discussed in Appendix D.}}

\begin{figure}
\centering
\subfigure[Total EMF vs. throughput]
{
	\includegraphics[width=6cm]{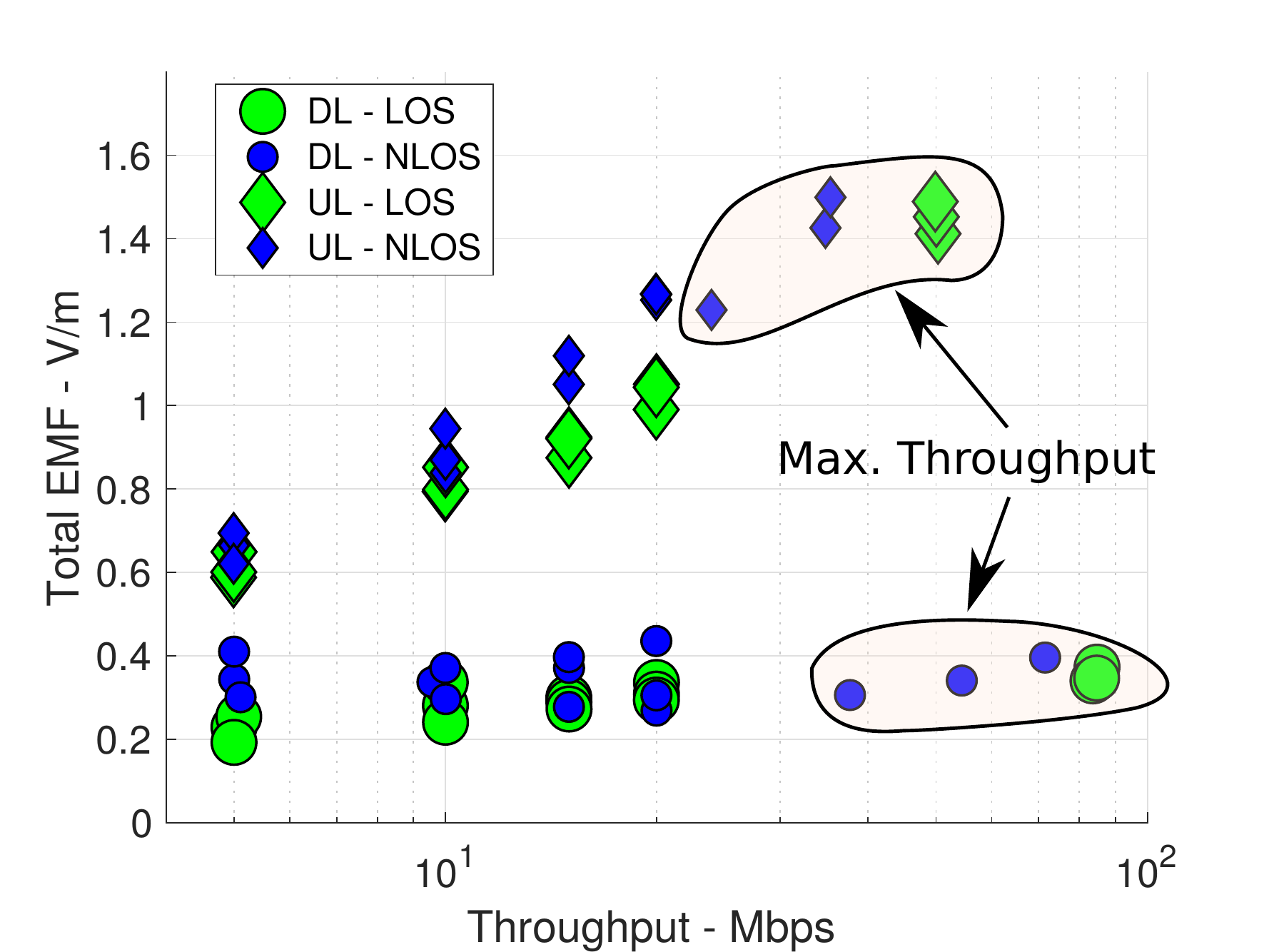}
	\label{fig:uni_test_emf_throughput}
}

\subfigure[Total Exposure-per-Mbps vs. throughput]
{
	\includegraphics[width=6cm]{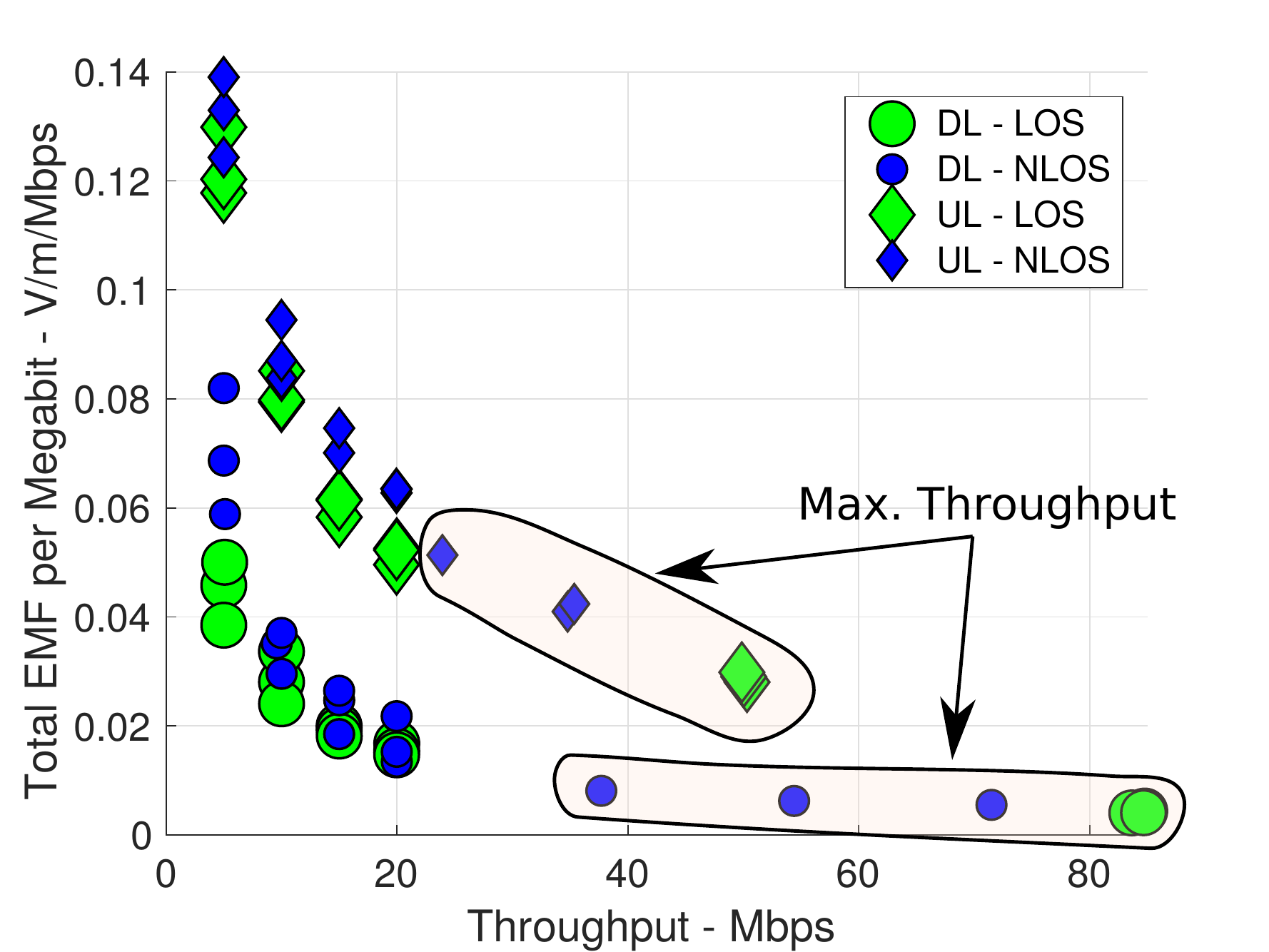}
	\label{fig:uni_test_emf_bit_throughput}
}
\caption{\ac{LOS}/\ac{NLOS} indoor locations: total EMF and total exposure-per-Mbps vs. throughput.}
\label{fig:indoor_tot_emf}
\vspace{-3mm}
\end{figure}

\subsubsection{Indoor measurements}

In the following part of our work, we extend the results of the outdoor locations by investigating the exposure in the \ac{LOS}/\ac{NLOS} indoor locations. In particular, the availability of controlled environments allows performing extensive tests, in order to deeply analyze the impact of key parameteres on the exposure levels. To this aim, we initially focus on the impact of \ac{UL} vs. \ac{DL} traffic generation. For each location, we perform a wide range of \ac{UL} and \ac{DL} tests, including the variation of the generated traffic from very low values (set to 5~[Mbps]) up to the maximum one reachable on the wireless link (several dozens of Mbps). Moreover, three independent runs are executed for each parameter setting, in order to strengthen our outcomes.

Fig.~\ref{fig:indoor_power_density} reports the breakdown of the exposure components for the considered tests. Four considerations hold by analyzing the figure. First, the exposure generated by the smartphone on pre-5G technologies dominates over the other components, both in the \ac{DL} and in the \ac{UL}. Second, the exposure in the \ac{DL} is almost one order of magnitude lower than the one recorded in the \ac{UL} tests. This is due to the fact that \ac{DL} tests generate most of traffic flows from the \texttt{Iperf} server to the client, while a very low amount of information flows on the inverse direction (e.g., segment acknowledgements and/or control information). Third, the exposure tends to increase when the \ac{UL} throughput is increased.\footnote{{This outcome, which appears apparently in contrast to the results of the outdoor locations, is instead a consequence of manual traffic setting \textit{and} stability in propagation conditions that are experienced by the indoor tests of the same location, as deeply analyzed in Appendix E.}} Fourth, the exposure tends to be higher in the \ac{NLOS} location than the \ac{LOS} one, for a given level of generated traffic. 

To better substantiate the previous findings, Fig.~\ref{fig:uni_test_emf_throughput} details the total \ac{EMF} vs. throughput over the two indoor locations. Interestingly, a strong increase of exposure is recorded when the \ac{UL} traffic is increased, easily reaching values greater than 1~[V/m] (top part of the figure). Moreover, the difference between \ac{NLOS} and \ac{LOS} exposure tends to increase with increasing \ac{UL} throughput. At last, when \texttt{Iperf} is set to generate the maximum traffic, the \ac{UL} throughput in the \ac{NLOS} location is clearly lower than the \ac{LOS} one - despite the fact that the exposure values are comparable in both locations. A similar behaviour is also observed for the maximum traffic in the \ac{DL} direction (bottom part of the figure). However, the increase of exposure due to traffic rising is less evident {than in} \ac{UL} tests.

\begin{figure}
\centering
\subfigure[Smartphone orientation variation with respect to the base station]
{
\includegraphics[width=8cm]{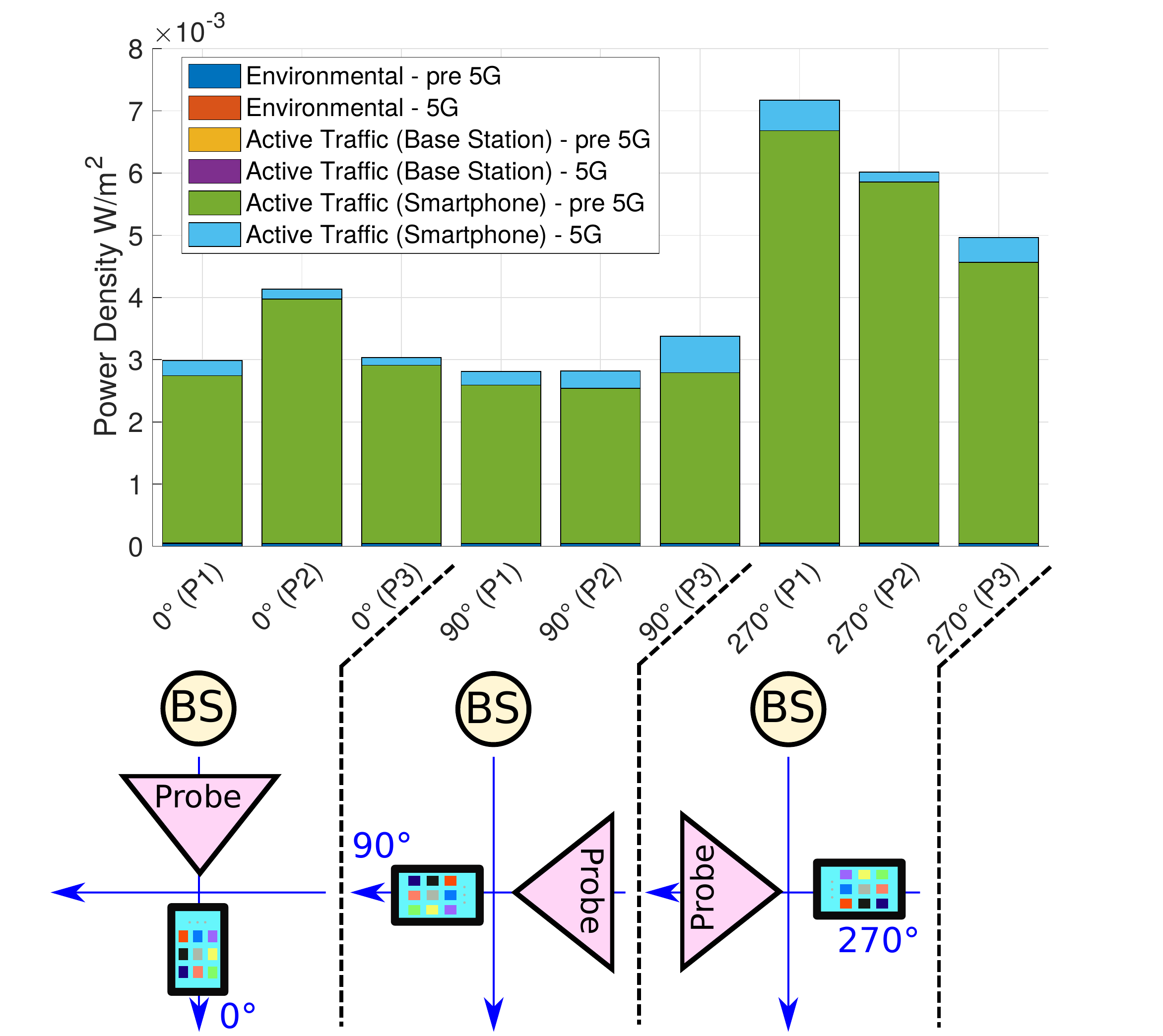}
\label{fig:var_orientation}
}

\subfigure[EMF evaluation distance variation from the smartphone]
{
\includegraphics[width=8cm]{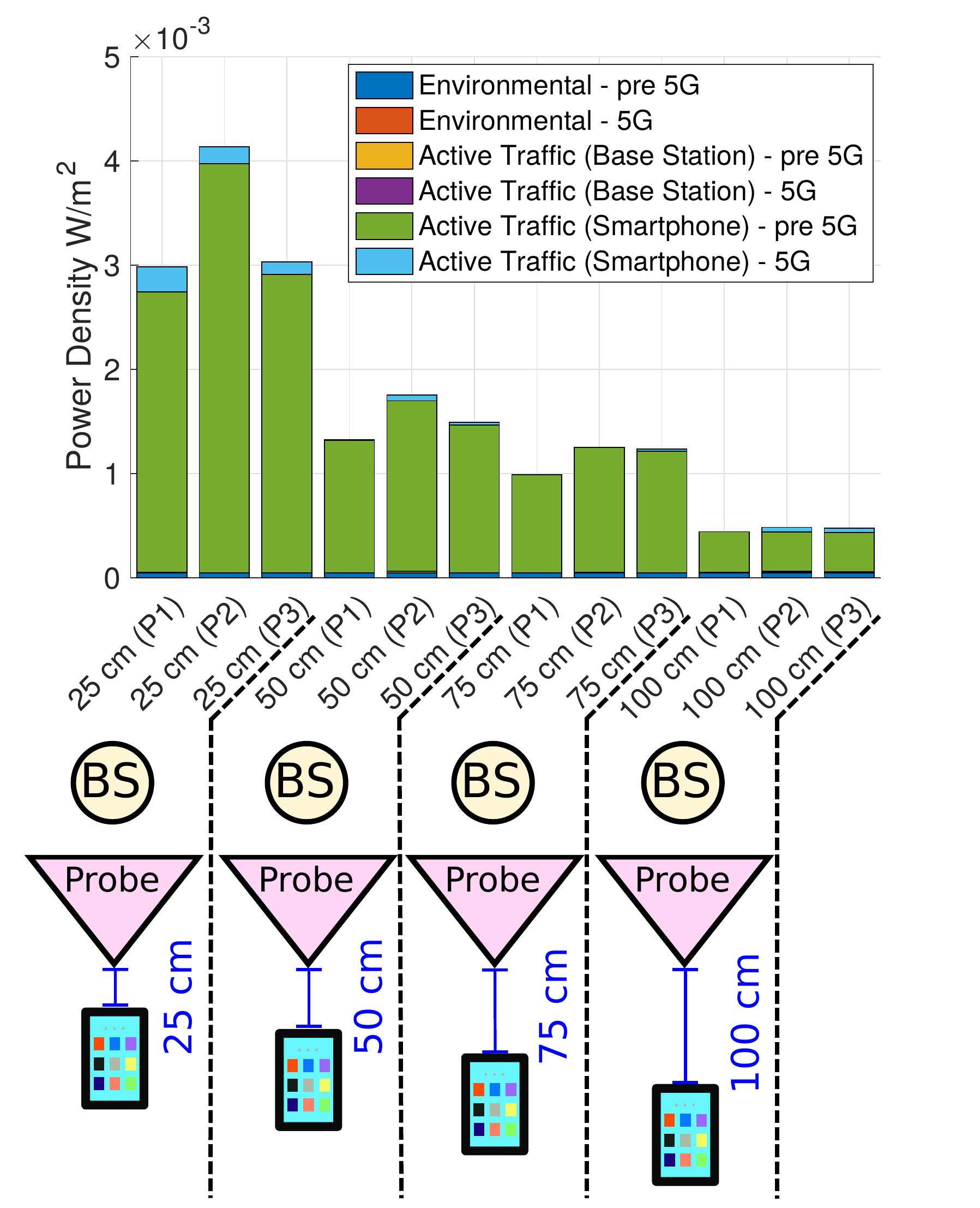}
\label{fig:var_distance}
}
\caption{Impact of smartphone orientation and EMF evaluation distance from the smartphone  - LOS location with maximum UL traffic setting.}.
\label{fig:var_distance_orientation}
\vspace{-3mm}
\end{figure}

Eventually, Fig.~\ref{fig:uni_test_emf_bit_throughput} reports the total exposure-per-Mbps vs. throughput for the two locations and the \ac{UL}/\ac{DL} tests. Astonishingly, the inversely proportional law between exposure-per-Mbps and throughput clearly emerges in each location and in each direction. Given a location, the total exposure-per-Mbps is lower for the \ac{DL} tests compared to the \ac{UL} ones. Morever, given a direction of traffic generation, the total exposure-per-Mbps in \ac{NLOS} tends to be higher than the one recorded in the \ac{LOS} condition.  

In the final part of our work, we evaluate the impact of changing the orientation and relative positioninig of the smartphone. Fig.~\ref{fig:var_orientation} reports the exposure breakdown for each smartphone orientation setting reported on the bottom of the figure, corresponding to a smartphone rotation of $0^\circ$, $90^\circ$, $270^\circ$ in clockwise direction of the horizontal plane. For each angle, we perform three independent assessment with the maximum \ac{UL} traffic. Interestingly, the dominance of smartphone exposure is evident in all the experiments. However, the 
$270^\circ$ rotation generally results in an higher exposure from the smartphone than the other angles. We argue that this phenomenon is due to the positioning of the smartphone {main antenna} on {bottom right part} the smartphone - opposite to the \ac{RBS} when an angle of $270^\circ$ is set, which results in worse propagation conditions than the other settings.

Finally, we analyze the impact of increasing the distance of the exposure evaluation point w.r.t. the smartphone, as shown in Fig.~\ref{fig:var_distance}. Starting from the default value of 0.25~[m], we increase the distance up to 1~[m]. {As expected}, the exposure experiences a sharp decrease when the distance is increased. However, we point out that the smartphone exposure is still the dominant component even when the distance is set to the maximum value of 1~[m].    

\section{Conclusions and Future Work}
\label{sec:conclusions}

We have {investigated} the problem of exposure assessment in a commercial 5G network, by {evaluating} the impact of user generated traffic on the exposure from the \ac{RBS} and the smartphone. 
To solve the complex and innovative measurement requirements - which include several aspects related to 5G implementation and its inter-viewing with legacy 4G networks - we have designed an innovative framework, called \textsc{5G-EA}. Our framework splits the complex problem into a set of procedures, which are tailored to a specific measurement goal (environmental vs. active traffic assessment). In addition, \textsc{5G-EA} relies on a completely softwarized approach, in which the measurement algorithm is run on a general purpose machine that controls the programmable spectrum analyzer.

We have then performed an extensive set of assessments in both outdoor and indoor locations. Interestingly, our results demonstrate that the smartphone contribution largely dominates over the other exposure components, particularly when \ac{UL} traffic is injected. However, the largest contribution is due to pre-5G technologies, while 5G always constitute a small share (up to {38\%}) {out of the total one that is radiated by the smartphone}. {In addition, the total exposure dramatically decreases when outdoor \ac{LOS} conditions are experienced, and in general when the exposure from the \ac{RBS} becomes detectable by the \ac{SAN}. Moreover, we have designed and evaluated an exposure estimator based on the maximum \ac{UL} traffic that is achieved by \texttt{iPerf} in the measurement location.  Eventually, the exposure tends to increase in indoor locations when passing from \ac{LOS} to \ac{NLOS} condition, for a given level of \ac{UL} traffic that is set towards the smartphone. Finally,  we have demonstrated that the measured exposure levels are influenced by key parameters, like \ac{DL} vs. \ac{UL} direction, smartphone orientation and relative distance of the smartphone w.r.t. the measurement antenna.}

We believe that our work paves the way for future research in the field. {First}, the application of \text{5G-EA} in other deployments (e.g., subject to different exposure regulations and/or different radio configurations) is an interesting step. {Second}, the evaluation of exposure should be extended by considering multiple \ac{UE} models{/types}{, locations in balconies/terraces in close proximity to the serving \ac{RBS}, and additional sources like WiFi}. {Third}, the assessment of exposure in 5G deployments including mm-Wave frequencies is another line of research. {Fourth}, we plan to perform extensive assessment by running commonly used smartphone applications {(social media, video streaming, online conference, etc.).} {Fifth, the decrease of exposure observed in \ac{LOS} locations suggest that deploying a dense 5G network, in which most of territory is in LOS w.r.t. the serving RBS, is the best solution to reduce the exposure from the terminals. This goal could be alternatively achieved by installing intelligent surfaces (active or passive), to improve the signal coverage over the territory. The evaluation of exposure in such innovative deployments is therefore a future activity.}

\section*{Acknowledgements}

This work was supported by the PLAN-EMF Project (KAUST-CNIT) under Award OSR-2020-CRG9-4377.

\bibliographystyle{ieeetr}

\begin{IEEEbiography}[{\includegraphics[width=1in,height=1.25in,clip,keepaspectratio]{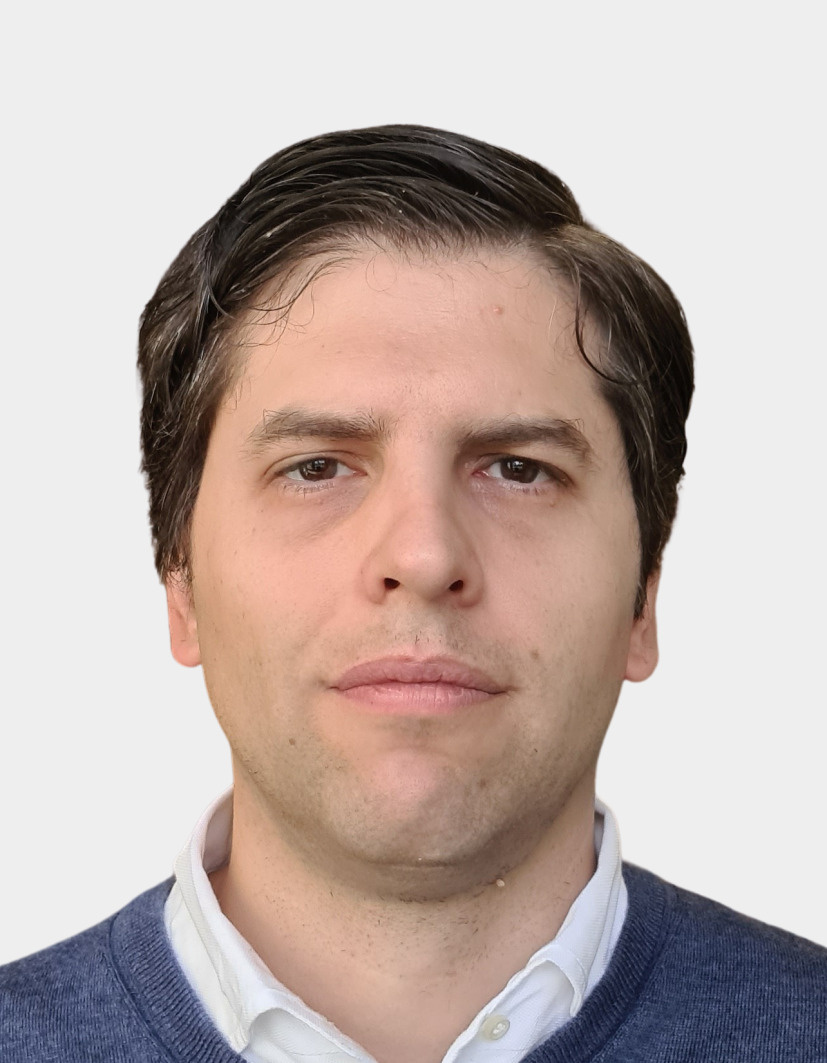}}]{Luca Chiaraviglio} (M'09-SM'16) received the Ph.D. degree in telecommunication and electronics engineering from the Polytechnic Unvisersity of Turin, Italy. He is currently an Associate Professor with the University of Rome ``Tor Vergata'', Italy, teaching the courses of ``Internet of Things'' and ``Networking and Internet''. He has co-authored more than 160 articles published in international journals, books, and conferences. His current research interests include 6G networks, electromagnetic fields assessments of 5G networks, and health risks assessment of 5G communications. He received the Best Paper Award at the IEEE Vehicular Technology Conference (VTC Spring), in 2020 and 2016, and at the Conference on Innovation in Clouds, Internet and Networks (ICIN), in 2018, all of them appearing as the first author. Some of his papers are listed as the Best Readings on Green Communications by the IEEE. Moreover, he has been recognized as an author in the top 1\% most highly cited papers in the Information and Communication Technology (ICT) field worldwide and top 2\% world scientists according to the 2021 and 2022 updates of the science-wide author databases of standardized citation indicators.
\end{IEEEbiography}

\begin{IEEEbiography}[{\includegraphics[width=1in,height=1.25in,clip,keepaspectratio]{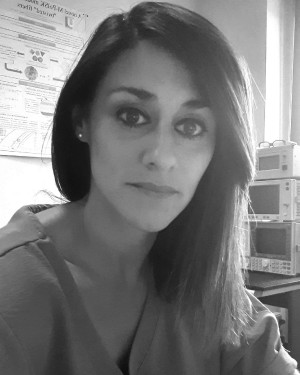}}]{Chiara Lodovisi} received the Ph.D. degree in engineering electronics from the University of Rome Tor Vergata, Italy. She worked for five years as a RF Engineer Consultant for H3G mobile operator and then as a Research Associate at the University of Rome Tor Vergata,  focusing on  optical communications,  submarine and satellite optical links, and radio over fiber. She is currently a Researcher with CNIT, Italy. Her research interests include 5G networks, health risk assessment of 5G communications, interoperability over fiber between TETRA/LTE systems, and 5G networks.
\end{IEEEbiography}

\begin{IEEEbiography}[{\includegraphics[width=1in,height=1.25in,clip,keepaspectratio]{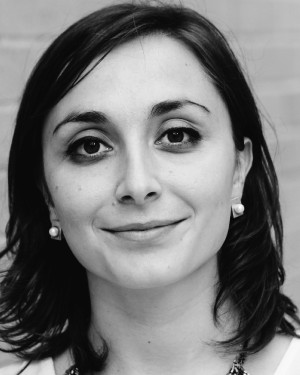}}]{Stefania Bartoletti} received the Laurea degree (summa cum laude) in electronics and telecommunications engineering and the Ph.D. degree in information engineering from the University of Ferrara, Ferrara, Italy, in 2011 and 2015, respectively. She is currently Assistant Professor at the University of Rome ``Tor Vergata'', Italy. She was a Marie Sklodowska-Curie Global Fellow within the Horizon 2020 European Framework for a research project with the Wireless Information and Network Sciences Laboratory, Massachusetts Institute of Technology (MIT), Cambridge, MA, USA, and the University of Ferrara during 2016-2019. Her research interests include theory and experimentation of wireless networks for passive localization and physical behavior analysis. Dr. Bartoletti was the recipient of the 2016 Paul Baran Young Scholar Award of the Marconi Society. She served as a Chair of the TPC for the IEEE ICC and Globecom Workshops on Advances in Network Localization and Navigation (ANLN) from 2017 to 2021, and as reviewer for numerous IEEE journals and international conferences. She is Editor of the IEEE Communications Letters.
\end{IEEEbiography}

\begin{IEEEbiography}[{\includegraphics[width=1in,height=1.25in,clip,keepaspectratio]{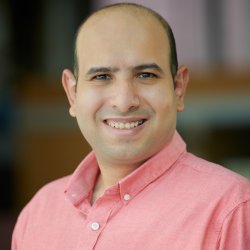}}]{Ahmed Elzanaty} (Senior Member, IEEE) received the Ph.D. degree (cum laude) in electronics, telecommunications, and information technology from the University of Bologna, Italy, in 2018. He was a Research Fellow with the University of Bologna from 2017 to 2019. He was a Post-Doctoral Fellow with the King Abdullah University of Science and Technology, Saudi Arabia. He is currently a Lecturer (Assistant Professor) with the Institute for Communication Systems, University of Surrey, U.K. He has participated in several national and European projects, such as GRETA and EuroCPS. His research interests include cellular networks design with EMF constraints, coded modulation, compressive sensing, and distributed training of neural networks.
\end{IEEEbiography}

\begin{IEEEbiography}[{\includegraphics[width=1in,height=1.25in,clip,keepaspectratio]{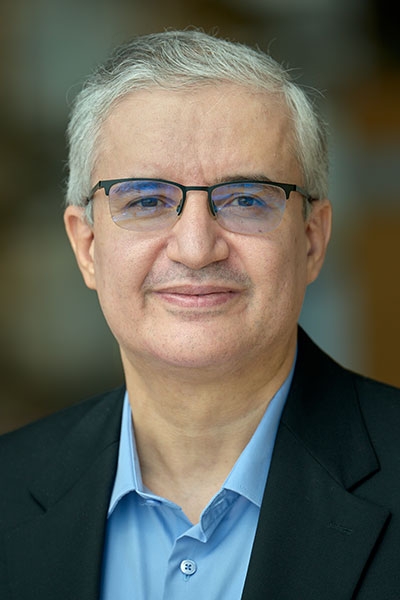}}]{Mohamed Slim Alouini} was born in Tunis, Tunisia. He received the Ph.D. degree in Electrical Engineering from the California Institute of Technology (Caltech) in 1998.  He served as a faculty member at the University of Minnesota then in the Texas A\&M University at Qatar before joining in 2009 the King Abdullah University of Science and Technology (KAUST) where he is now a Distinguished Professor of Electrical and Computer Engineering.  Prof. Alouini is a Fellow of the IEEE and OPTICA (Formerly the Optical Society of America (OSA)). He is currently particularly interested in addressing the technical challenges associated with the uneven distribution, access to, and use of information and communication technologies in rural,  low-income, disaster, and/or hard-to-reach areas. 
\end{IEEEbiography}

\end{document}